\documentclass[conference]{IEEEtran}

\usepackage{cite}
\usepackage{amsmath,amssymb,amsfonts}
\usepackage{algorithm}
\usepackage{algorithmic}
\usepackage{graphicx}
\usepackage{textcomp}
\usepackage{xcolor}
\usepackage[utf8]{inputenc}
\usepackage{subcaption} 
\usepackage{booktabs}  
\usepackage{multirow}
\usepackage{siunitx}
\graphicspath{ {./} } 
\usepackage[hyphens]{url}
\usepackage{tabularx}
\usepackage{comment}

\def\BibTeX{{\rm B\kern-.05em{\sc i\kern-.025em b}\kern-.08em
T\kern-.1667em\lower.7ex\hbox{E}\kern-.125emX}}

\begin{document}

\title{DEEP-GAP: Deep-learning Evaluation of Execution Parallelism in GPU Architectural Performance}

\author{\IEEEauthorblockN{Kathiravan Palaniappan}
\IEEEauthorblockA{\textit{Independent Researcher (University of Colorado Colorado Springs Alumni)} \\
kpalania@uccs.edu}
}

\maketitle

\begin{abstract}

Modern datacenters increasingly rely on low-power, single-slot inference accelerators
to balance performance, energy efficiency, and rack density constraints. The NVIDIA T4
GPU has become one of the most widely deployed inference accelerators in production
environments due to its favorable performance-per-watt characteristics, mature software
ecosystem, and cost efficiency \cite{t4_datasheet}. Its architectural successor, the
NVIDIA L4 GPU, introduces significant advancements, including enhanced Tensor Core
throughput, expanded cache capacity, higher memory bandwidth, and improved parallel
execution capability \cite{l4_datasheet,ada_whitepaper}. However, limited empirical
evidence exists quantifying the real-world inference performance gap between these two
generations under controlled and reproducible conditions \cite{mlperf}.

In this work, we introduce \textbf{DEEP-GAP}, a systematic benchmarking study that extends the GDEV-AI \cite{gdevai} methodology to GPU-based inference, directly evaluating the performance gap between the widely deployed T4 GPU and the more recent L4 GPU across multiple precision modes (FP32, FP16, INT8). Using identical model configurations, workload parameters, and measurement methodology, we evaluate inference performance for ResNet18, ResNet50, and ResNet101 using both PyTorch eager execution and TensorRT optimization pipelines \cite{tensorrt_docs,tensorrt_guide}.

Results show that reduced precision significantly enhances inference performance,
with INT8 achieving up to \textbf{58× throughput improvement over CPU baselines}.
L4 achieves up to \textbf{4.4× higher throughput than T4} while reaching peak efficiency
at smaller batch sizes (B=16–32), thereby improving latency–throughput tradeoff characteristics for latency-sensitive workloads. This is consistent with well-established system-level tradeoffs between response time and processing efficiency \cite{latency_throughput}. However, T4 remains competitive for large-batch,
throughput-oriented workloads and environments where cost efficiency, power constraints,
or existing deployment compatibility are primary considerations.

By maintaining methodological continuity with previous CPU benchmarking, DEEP-GAP establishes a controlled empirical basis for understanding the transition from CPU-bound inference to contemporary GPU acceleration. Our findings provide practical guidance for selecting precision modes, batch sizes, and accelerator generations in accordance with workload requirements and deployment constraints.
\end{abstract}


\section{Introduction}

The rapid growth of machine learning inference workloads in production datacenters
has increased the demand for accelerators capable of delivering high throughput under strict power and density constraints. Single-slot, low-power GPUs have emerged as attractive solutions for scalable inference deployment.
\begin{figure}[t]
  \centering
  \includegraphics[width=\linewidth]{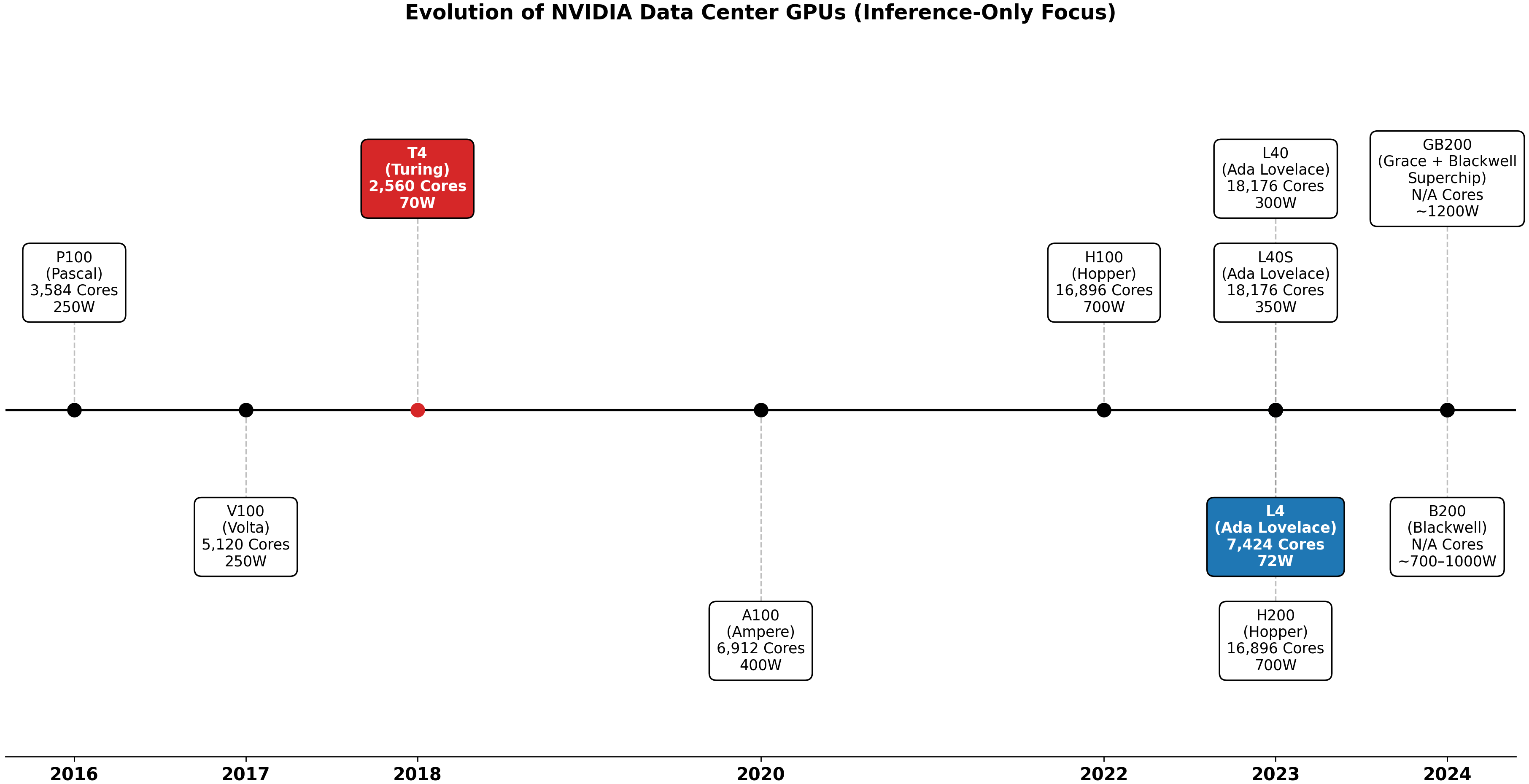}
  \caption{Evolution of NVIDIA datacenter GPUs across architectural generations,
  highlighting inference-optimized accelerators (T4 and L4) within the broader
  compute-class roadmap. The figure contextualizes the GPUs evaluated in this
  study relative to contemporary training-oriented accelerators.}
  \label{fig:gpu_timeline}
\end{figure}

While large language models (LLMs) dominate many current AI applications, we
select ResNet architectures for this study due to their well-established,
reproducible benchmark \cite{he2016resnet} for evaluating compute-bound convolutional workloads \cite{krizhevsky2012imagenet}. Compared to transformer-based models, which often introduce additional variability due to attention mechanisms and sequence-dependent computation \cite{transformer_cost}, ResNet models provide a controlled setting for analyzing the impact of numerical precision and batch-level parallelism \cite{deng2009imagenet} on Tensor Core utilization.

Deep neural networks have become the dominant approach for a wide range of applications,
including computer vision and natural language processing \cite{dnn_survey}.

This allows us to isolate architectural performance characteristics related to
parallel execution efficiency, arithmetic throughput, and scaling behavior across
precision modes. These factors are central to understanding the generational
performance differences between Turing-based T4 GPUs \cite{turing_whitepaper} and Ada Lovelace–based L4 GPUs \cite{ada_whitepaper} in real-world datacenter inference deployments.

While the majority of NVIDIA datacenter GPUs are designed to prioritize high-throughput 
training performance, the T4 and L4 stand out as purposefully inference-optimized 
accelerators. Their architectural characteristics and low TDP profiles reflect 
a design focus on efficient deployment rather than maximum compute density.

The T4 GPU has been widely adopted due to its 70W power envelope and compatibility with 
standard rack servers. The more recent L4 GPU introduces architectural enhancements \cite{ada_whitepaper} including improved Tensor Core throughput and increased memory bandwidth, promising performance improvements while preserving deployment flexibility.

Despite these advancements, there remains a lack of controlled benchmarking studies that isolate the true performance gap between these two accelerators under identical runtime configurations. This paper aims to provide a rigorous, empirical comparison.

\subsection{Motivation: Inference-Centric Datacenter Demand}

Modern datacenter deployments increasingly rely on deep learning inference \cite{barroso2013datacenter}
to support real-time applications that demand predictable latency, high throughput,
and efficient resource utilization. Selecting appropriate precision modes
(FP32, FP16, INT8) and hardware platforms directly impacts system performance,
cost efficiency, and scalability.

Although recent GPU architectural advancements have primarily focused on
training workloads, many production environments require sustained inference
performance under power and deployment constraints. Despite the widespread
adoption of inference-optimized accelerators such as T4 and L4, publicly
available empirical comparisons across precision modes remain limited.

This work provides a systematic evaluation of inference performance across
GPU generations and precision configurations, enabling clearer understanding
of how architectural improvements translate into practical deployment benefits.

\begin{table*}[t]
\centering
\caption{Peak Inference Throughput: Granite Rapids CPU vs NVIDIA T4 and L4 across Precisions. Speedup values for ResNet-101 are marked N/A because CPU experiments were not performed
due to substantially higher computational cost and impractically long runtime on the
Granite Rapids baseline configuration.}
\label{tab:cpu_vs_t4_l4_all}
\begin{tabular}{llccc}
\toprule
\textbf{Platform} & \textbf{Model} & \textbf{Peak Batch (B)} &
\textbf{Peak Throughput (images/sec)} & \textbf{Speedup vs CPU} \\
\midrule

\multirow{2}{*}{Granite Rapids (24 Threads)}
& ResNet-18 & 8  & \num{670}  & 1.0$\times$ \\
& ResNet-50 & 8  & \num{230}  & 1.0$\times$ \\

\midrule

\multirow{3}{*}{NVIDIA T4 (FP32)}
& ResNet-18 & 384 & \num{1289} & 1.92$\times$ \\
& ResNet-50 & 256 & \num{382}  & 1.66$\times$ \\
& ResNet-101 & 256  & \num{226}  & N/A \\

\midrule

\multirow{3}{*}{NVIDIA T4 (FP16)}
& ResNet-18 & 512 & \num{2569} & 3.83$\times$ \\
& ResNet-50 & 384  & \num{849} & 3.69$\times$ \\
& ResNet-101 & 384  & \num{512}  & N/A \\

\midrule

\multirow{3}{*}{NVIDIA T4 (INT8 TensorRT)}
& ResNet-18 & 512 & \num{8837} & 13.19$\times$ \\
& ResNet-50 & 384 & \num{5066} & 22.03$\times$ \\
& ResNet-101 & 256  & \num{3125}  & N/A \\

\midrule

\multirow{3}{*}{NVIDIA L4 (FP32)}
& ResNet-18 & 8 & \num{3483} & 5.20$\times$ \\
& ResNet-50 & 8 & \num{1068} & 4.64$\times$ \\
& ResNet-101 & 8  & \num{687}  & N/A \\

\midrule

\multirow{3}{*}{NVIDIA L4 (FP16)}
& ResNet-18 & 16 & \num{5923} & 8.84$\times$ \\
& ResNet-50 & 8  & \num{1928}  & 8.38$\times$ \\
& ResNet-101 & 16  & \num{1206}  & N/A \\

\midrule

\multirow{3}{*}{NVIDIA L4 (INT8 TensorRT)}
& ResNet-18 & 32 & \num{38932} & 58.11$\times$ \\
& ResNet-50 & 32 & \num{13388} & 58.21$\times$ \\
& ResNet-101 & 32  & \num{8026}  & N/A \\

\bottomrule
\end{tabular}
\end{table*}

Table~\ref{tab:cpu_vs_t4_l4_all} summarizes peak inference throughput for
Granite Rapids CPU, NVIDIA T4, and NVIDIA L4 across FP32, FP16, and INT8 precision modes.

Across all configurations, GPU-based inference significantly outperforms the CPU baseline in performance.
Even at FP32 precision, GPUs demonstrate clear advantages due to higher parallelism.
The NVIDIA T4 achieves up to 1.92$\times$ speedup for ResNet-18 and 1.66$\times$ for ResNet-50 compared to the CPU baseline.

The use of reduced precision significantly increases throughput.
With FP16 precision, the T4 achieves 3.83$\times$ speedup for ResNet-18 and 3.69$\times$ for ResNet-50.
Further improvements are observed with INT8 TensorRT optimization, reaching
13.19$\times$ and 22.03$\times$ speedup respectively.

The NVIDIA L4 demonstrates significantly stronger performance improvements over both
the CPU and T4 across all precision modes.
At FP32 precision, L4 achieves 5.20$\times$ speedup for ResNet-18 and 4.64$\times$ for ResNet-50.
FP16 precision further improves throughput, reaching 8.84$\times$ and 8.38$\times$ speedup.

INT8 TensorRT produces the highest performance gains, with L4 achieving a
58.11$\times$ speedup for ResNet-18 and a 58.21$\times$ speedup for ResNet-50 relative to CPU execution.
These results highlight the effectiveness of reduced precision inference
and specialized Tensor Core hardware \cite{jouppi2017tpu} in modern GPU architectures.

Overall, the results demonstrate a consistent trend:
lower numerical precision enables significantly higher throughput,
while newer GPU architectures further amplify these gains through
improved parallel execution capability and memory subsystem efficiency.

\begin{figure}[t]
    \centering
    \includegraphics[width=\linewidth]{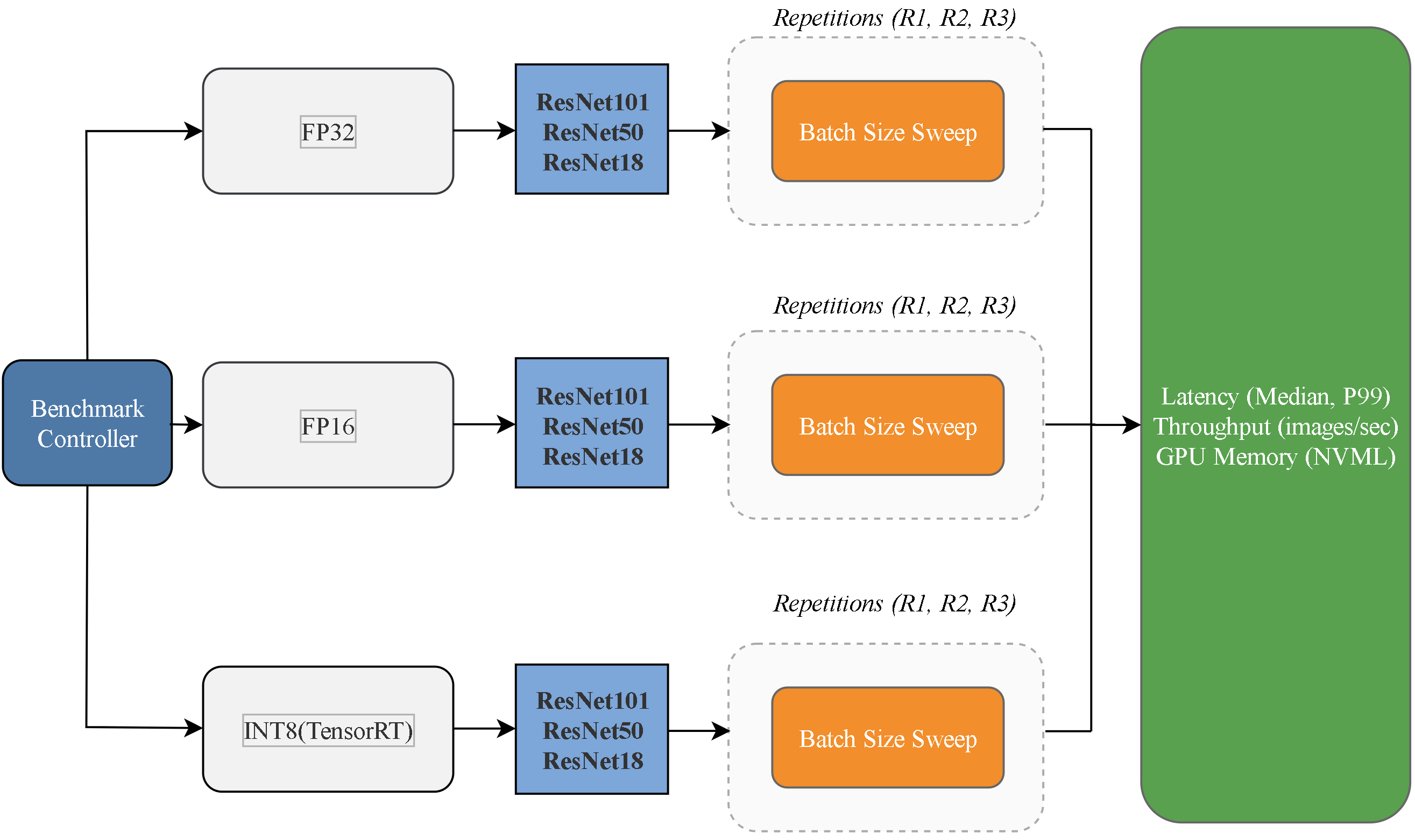}
    \caption{
Benchmark execution structure. For each model (ResNet101, ResNet50, ResNet18),
we perform a sweep over batch sizes under multiple precision modes (FP32, FP16, INT8),
executing each configuration across repeated runs (R1--R3) to capture run-to-run variability.
Collected metrics include median latency, P99 latency, throughput (images/sec),
and GPU memory utilization (NVML).
}
    \label{fig:thread-sweep-diagram}
\end{figure}

\section{Background and Architectural Overview}
\label{sec:background}
\subsection{Evolution of Inference-Oriented GPUs}

Modern GPU architectures have evolved significantly to improve deep learning performance,
with enhancements in Tensor Core design, memory bandwidth, and cache hierarchy \cite{ampere_whitepaper}.

This study investigates how architectural evolution in inference-oriented GPUs 
translates into measurable performance gains under real-world deployment constraints. 
Specifically, we quantify the generational performance gap between NVIDIA T4 and L4 
accelerators using controlled benchmarking across precision modes (FP32, FP16, INT8) 
and batch sizes.

In contrast to training-oriented GPUs, which are optimized for peak throughput and multi-GPU scaling, 
inference accelerators are designed for sustained performance-per-watt, rack-level 
compatibility, and deployment flexibility. These constraints are critical in 
datacenter environments where power, cooling, and form factor limit hardware upgrades \cite{barroso2013datacenter}. 
By focusing on low-profile, single-slot GPUs, this work isolates architectural 
improvements within a consistent deployment class, enabling a fair and practical 
comparison for real-world upgrade scenarios.

\subsubsection{T4 as Baseline Inference Accelerator}

In this work, the T4 GPU serves as the baseline inference accelerator \cite{turing_whitepaper}, representing 
widely deployed infrastructure across cloud and enterprise datacenters. Its 70W power 
envelope, single-slot low-profile design, and PCIe-based compatibility allow seamless 
integration into existing 1U and 2U servers without specialized cooling or power 
requirements.

Architecturally, T4 introduces Tensor Cores optimized for mixed-precision inference, 
supporting FP16 and INT8 acceleration. These capabilities establish the baseline 
performance characteristics against which generational improvements are measured. 
In the DEEP-GAP analysis, T4 defines the lower bound of throughput and latency 
performance under identical system configurations.

\subsubsection{L4 as Generational Successor}

The L4 GPU represents the next-generation accelerator within the same deployment 
class. While maintaining similar power and form factor constraints, L4 introduces 
substantial architectural enhancements, including increased CUDA core counts, 
expanded L2 cache capacity, higher memory bandwidth, and improved Tensor Core 
throughput.

In this study, L4 is evaluated as the target platform for generational comparison. 
We hypothesize that these architectural improvements will yield disproportionate 
performance gains under mixed-precision inference workloads, particularly at higher 
batch sizes. This hypothesis is evaluated through systematic benchmarking in 
Section~\ref{sec:results}.

\subsubsection{Why Deployment-Constrained GPUs Matter}

Datacenter operators must scale inference capacity within strict constraints on 
rack density, power consumption, and cooling infrastructure. Large multi-slot GPUs 
often require specialized chassis and higher operational costs, limiting their 
adoption in existing deployments.

This work focuses on low-profile, low-power GPUs that can be deployed incrementally 
within current infrastructure. By evaluating T4 and L4 within the same deployment 
constraints, DEEP-GAP provides insights into achievable performance gains without 
requiring major hardware redesigns. This perspective is critical for practitioners 
making cost-performance trade-offs in production environments.

\subsection{Microarchitectural Differences}

L4 introduces architectural advancements over T4, including a higher CUDA core count,
increased memory bandwidth, larger L2 cache, and enhanced Tensor Core performance.
These changes are expected to improve throughput scaling and mixed-precision efficiency,
particularly for FP16 and INT8 inference workloads.

\section{DEEP-GAP Benchmarking Framework}
\label{sec:methodology}

To systematically evaluate GPU-based inference performance, we extend the 
GDEV-AI CPU benchmarking framework to a GPU-centric methodology that captures 
throughput, latency (including tail latency), and memory utilization under 
different numerical precision modes, as illustrated in Figure~\ref{fig:thread-sweep-diagram}.

Unlike CPU benchmarking, which primarily explores thread-level parallelism, 
GPU inference is driven by massive data parallelism, where batch size and 
precision play dominant roles in performance scaling.

\subsection{Benchmark Design}

We evaluate three precision modes commonly used in production inference: 
\textbf{FP32} (full precision) using PyTorch CUDA execution, 
\textbf{FP16} (half precision) using mixed-precision acceleration, and 
\textbf{INT8} (quantized inference) using TensorRT-optimized execution.

The benchmark is performed across multiple convolutional neural networks, 
including ResNet-18, ResNet-50, and ResNet-101.

Deep neural networks exhibit varying levels of compute intensity and memory access patterns,
which significantly influence hardware utilization and performance \cite{dnn_analysis}.

We systematically sweep across increasing batch sizes to identify peak throughput, saturation points, and latency-scaling characteristics, since batching plays a central role in GPU scheduling efficiency and parallel execution behavior \cite{batching_latency}.

\subsection{Measurement Strategy}

Each experiment follows a structured execution pipeline consisting of warm-up iterations to stabilize GPU execution, followed by timed inference runs for latency measurement. Multiple repeats are performed to ensure statistical reliability, and NVML-based sampling is used to capture total GPU memory usage throughout execution.

For each configuration, we record median latency, mean latency, standard deviation, and P99 latency \cite{tail_at_scale} to capture both average and tail behavior. In addition, we measure throughput in images per second and track peak GPU memory usage using NVML.

\subsection{Benchmark Procedure}

The DEEP-GAP benchmarking procedure is designed to ensure fair, reproducible, 
and consistent evaluation across all configurations. Each experiment follows a 
standardized execution pipeline that isolates GPU performance while minimizing 
external variability.

For each combination of model, precision mode, and batch size, the model is first 
loaded onto the target device and configured for inference execution. Gradient 
computation is disabled to avoid unnecessary overhead. 

For GPU experiments, host-side parallelism is constrained by fixing the PyTorch 
inter-op thread count to one (torch.set\_num\_interop\_threads(1)), ensuring that 
measured performance reflects GPU execution characteristics rather than variability 
introduced by CPU-side scheduling effects.

For CPU experiments, thread counts are systematically varied across available cores to evaluate scaling behavior.

Each configuration begins with a warm-up phase to stabilize execution by triggering 
kernel initialization, memory allocation, and cache population. Following warm-up, 
timed inference iterations are executed, during which latency and throughput metrics 
are recorded. Median latency and tail latency (p99) are computed to capture both 
typical and worst-case performance behavior.

Batch sizes are systematically increased to evaluate scaling characteristics, 
allowing identification of saturation points where throughput plateaus or latency 
begins to degrade. This enables differentiation between compute-bound and 
memory-bound regimes for each architecture.

INT8 inference is executed using TensorRT-optimized engines to reflect
production deployment conditions. FP32 and FP16 experiments are executed
using PyTorch CUDA eager mode, providing a consistent comparison across
precision modes. All configurations are evaluated under identical system
settings, with GPU memory usage monitored via NVML.

\begin{algorithm}[t]
\caption{DEEP-GAP GPU Inference Benchmark Procedure}
\label{alg:deepgap_benchmark}
\begin{algorithmic}[1]

\STATE Disable gradient computation
\STATE Set \texttt{torch.set\_num\_interop\_threads(1)}

\FOR{precision $\in \{\text{FP32}, \text{FP16}, \text{INT8-TensorRT}\}$}
    \FOR{model $\in \{\text{ResNet101}, \text{ResNet50}, \text{ResNet18}\}$}
    
        \IF{precision == INT8-TensorRT}
            \STATE Export model to ONNX format
            \STATE Build TensorRT engine with optimization profile
            \STATE Perform INT8 calibration and cache calibration data
        \ENDIF
        
        \FOR{sweep $= 1$ to $10$}
            \FOR{batch size $\in \{1,2,4,8,16,32,64,128,256,384,512\}$}
                \FOR{repeat $= 1$ to $3$}
                
                    \STATE Allocate GPU input/output tensors
                    \STATE Run warm-up iterations
                    
                    \FOR{iteration $= 1$ to $100$}
                        \STATE Execute inference
                        \STATE Record latency
                        \STATE Sample NVML total GPU memory usage
                    \ENDFOR
                    
                    \STATE Compute median, mean, std, and P99 latency
                    \STATE Compute throughput (images/sec)
                    \STATE Track peak memory usage
                    
                \ENDFOR
            \ENDFOR
        \ENDFOR        
    \ENDFOR
\ENDFOR

\end{algorithmic}
\end{algorithm}

\subsection{Experimental Parameters}

To ensure reproducibility and statistical stability, all experiments are conducted using a fixed benchmarking protocol, aligning with established best practices for reproducible machine learning experimentation \cite{reproducibility}.

The methodology follows the structured benchmarking framework introduced in GDEV-AI \cite{gdevai}, extended here to GPU-based inference workloads.

Each measurement includes an initial warm-up phase of 20 iterations to eliminate cold-start effects such as kernel initialization, cache population, and memory allocation overhead. Performance metrics are then collected over 100 timed iterations, providing stable estimates of steady-state inference latency and throughput.

To account for runtime variability, each configuration is repeated 3 times, and the maximum observed resource usage (e.g., GPU memory utilization reported by NVML) is recorded across repeats. 

For each model and precision mode, batch sizes are systematically swept across 10 configurations to evaluate performance scaling behavior under increasing workload intensity.

This experimental design enables reliable estimation of median latency, tail latency (p99), and throughput, while minimizing measurement noise caused by transient system effects.

\subsection{Experimental Scope}

The DEEP-GAP study evaluates inference performance across multiple dimensions, 
including model complexity, numerical precision, and batch size scaling.

Experiments are conducted using ResNet-18, ResNet-50, and ResNet-101 to represent 
increasing computational and memory demands. Each model is evaluated under FP32, 
FP16, and INT8 precision modes to capture the impact of reduced numerical precision 
on performance.

Batch sizes are systematically varied to analyze scaling behavior, identifying compute-bound and memory-bound regimes for each architecture, as larger batch sizes typically improve hardware utilization by exposing greater parallel execution opportunities \cite{volkov2010gpu}. All experiments are 
performed under identical system configurations to isolate the effect of GPU 
microarchitectural differences between T4 and L4.

\subsection{Impact of Numerical Precision on Parallelism}
\label{sec:precision_parallelism}

\begin{figure*}[t]
\centering
\includegraphics[width=0.95\textwidth]{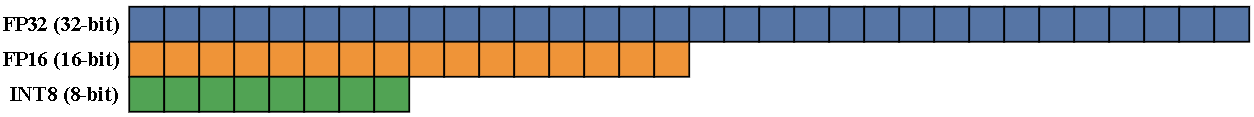}
\caption{Effect of numerical precision on data representation and parallel execution. 
Lower precision reduces data size, enabling more parallel operations per compute unit. 
FP32 uses higher precision with larger data representation, while FP16 and INT8 progressively 
reduce precision and increase parallelism.}
\label{fig:precision_data}
\end{figure*}

\begin{figure*}[t]
\centering
\includegraphics[width=0.95\textwidth]{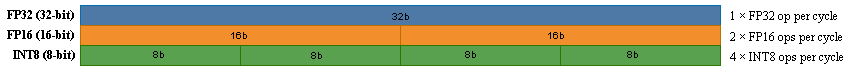}
\caption{Compute efficiency across precision modes. Lower precision enables more computations 
per cycle due to reduced data width. FP32 performs one operation per cycle, FP16 increases throughput, 
and INT8 maximizes parallel execution using specialized hardware units.}
\label{fig:precision_cycles}
\end{figure*}

Modern GPUs exploit reduced numerical precision to improve computational throughput, 
as illustrated in Figure~\ref{fig:precision_data} and Figure~\ref{fig:precision_cycles}. 
Lower precision formats require fewer bits to represent each value, reducing memory 
bandwidth requirements and enabling more data elements to be processed in parallel.

As shown in Figure~\ref{fig:precision_data}, FP32 (32-bit) provides high numerical accuracy 
but incurs larger data movement and lower parallelism. FP16 (16-bit) reduces data size 
and improves throughput through Tensor Core acceleration \cite{micikevicius2018mixedprecision}, offering a balance between 
accuracy and performance. INT8 (8-bit) further compresses data representation \cite{jacob2018quantization}, enabling 
significantly higher parallel execution and maximizing hardware utilization.

Conceptually, as illustrated in Figure~\ref{fig:precision_cycles}, lower precision allows 
more values to be packed into the same compute unit (e.g., conceptually increases the number of operations per cycle), thereby increasing the number of operations 
that can be executed per cycle \cite{quant_whitepaper}.

As precision decreases, the data footprint becomes smaller, more values fit within 
the same hardware resources, and the number of operations per cycle increases. 
This creates a trade-off between numerical accuracy and performance, where FP32 
offers fine-grained resolution, FP16 provides a balanced middle ground, and INT8 
achieves the highest throughput at reduced precision.

This relationship motivates the DEEP-GAP evaluation across all three precision 
modes, enabling us to quantify how reduced precision translates into real-world 
inference performance gains.

The evaluation spans three orthogonal dimensions: three models \cite{he2016resnet} (ResNet-18, 
ResNet-50, and ResNet-101), three precision modes (FP32, FP16, and INT8 using 
TensorRT), and two GPU architectures (NVIDIA T4 and NVIDIA L4). This results in 
18 distinct configurations, each evaluated across multiple batch sizes and 
repeated runs to ensure statistical reliability.

\subsection{Key Differences from CPU Benchmarking}

Compared to the GDEV-AI CPU methodology, DEEP-GAP shifts from thread-level scaling 
to batch-driven parallelism, where performance is primarily governed by batch size 
rather than thread count. It introduces precision-aware execution, where FP32, FP16, 
and INT8 exhibit significantly different performance characteristics, and leverages 
hardware acceleration through CUDA cores and Tensor Cores. Additionally, INT8 
execution utilizes TensorRT for optimized inference, and memory usage is measured 
at the system level using NVML rather than framework-level allocation.

This methodology enables a controlled and reproducible comparison of inference 
efficiency across GPU architectures and precision modes.

\section{Experimental Setup}
\label{sec:experimental_methodology}
To ensure reproducibility and isolate GPU architectural effects, all experiments
were conducted under controlled and documented system configurations. The
methodology follows the structured benchmarking framework introduced in
GDEV-AI \cite{gdevai}, extended here to GPU-based inference.
To facilitate transparency and independent verification of results, the complete
benchmarking framework, experiment scripts are
publicly available as an open-source repository \cite{deepgap_repo}.

\subsection{Hardware Configuration}

All GPU experiments were conducted on the host platform previously
characterized in GDEV-AI \cite{gdevai}.

For DEEP-GAP, both T4 and L4 GPUs were deployed on the same Granite
Rapids-based system shown in Table~\ref{tab:hardware_specs} to minimize
host-side bottlenecks and ensure that measured performance differences
reflect GPU architectural characteristics rather than CPU limitations.

\begin{table}[htbp]
\caption{Host System Configuration}
\label{tab:hardware_specs}
\centering
\small
\begin{tabular}{p{3.3cm} p{4.7cm}}
\hline
\textbf{Component} & \textbf{Specification} \\
\hline
CPU & Intel Xeon 6 Performance (Granite Rapids), 24 cores / 48 threads \\
Memory & 32\,GB DDR5-6400 \\
Storage & NVMe SSD \\
Operating System & Ubuntu 22.04 LTS \\
GPU Driver / CUDA & Fixed NVIDIA driver and CUDA toolkit across all experiments \\
\hline
\end{tabular}
\end{table}

\subsubsection{PCIe Configuration}

Both GPUs were installed in the same physical PCIe x16 slot to ensure consistent 
host platform characteristics. The T4 operated over PCIe Gen3, while the L4 
utilized PCIe Gen4, consistent with their respective architectural specifications.

Input tensors were allocated directly on the GPU prior to timing measurements, 
and performance metrics were collected exclusively for model forward-pass execution. 
Host-to-device data transfer latency was therefore excluded from reported results. 
Under these conditions, inference performance is primarily determined by on-device 
compute throughput, memory hierarchy efficiency, and Tensor Core utilization, 
rather than PCIe bandwidth differences.

While PCIe Gen4 provides higher theoretical transfer bandwidth than Gen3, its 
impact on measured performance is expected to be limited in this study due to the 
absence of runtime data transfers within the timed region.

\subsubsection{Thermal and Power Stability}

To prevent performance variability due to thermal throttling, GPU temperature, 
clock frequency, and power draw were monitored throughout execution. All reported 
measurements were collected after reaching steady-state thermal conditions.

Power consumption was recorded using device telemetry via \texttt{nvidia-smi}, 
enabling evaluation of performance-per-watt across precision modes.

\begin{figure*}[t]
\centering

\begin{subfigure}{0.32\textwidth}
\centering
\includegraphics[width=\linewidth]{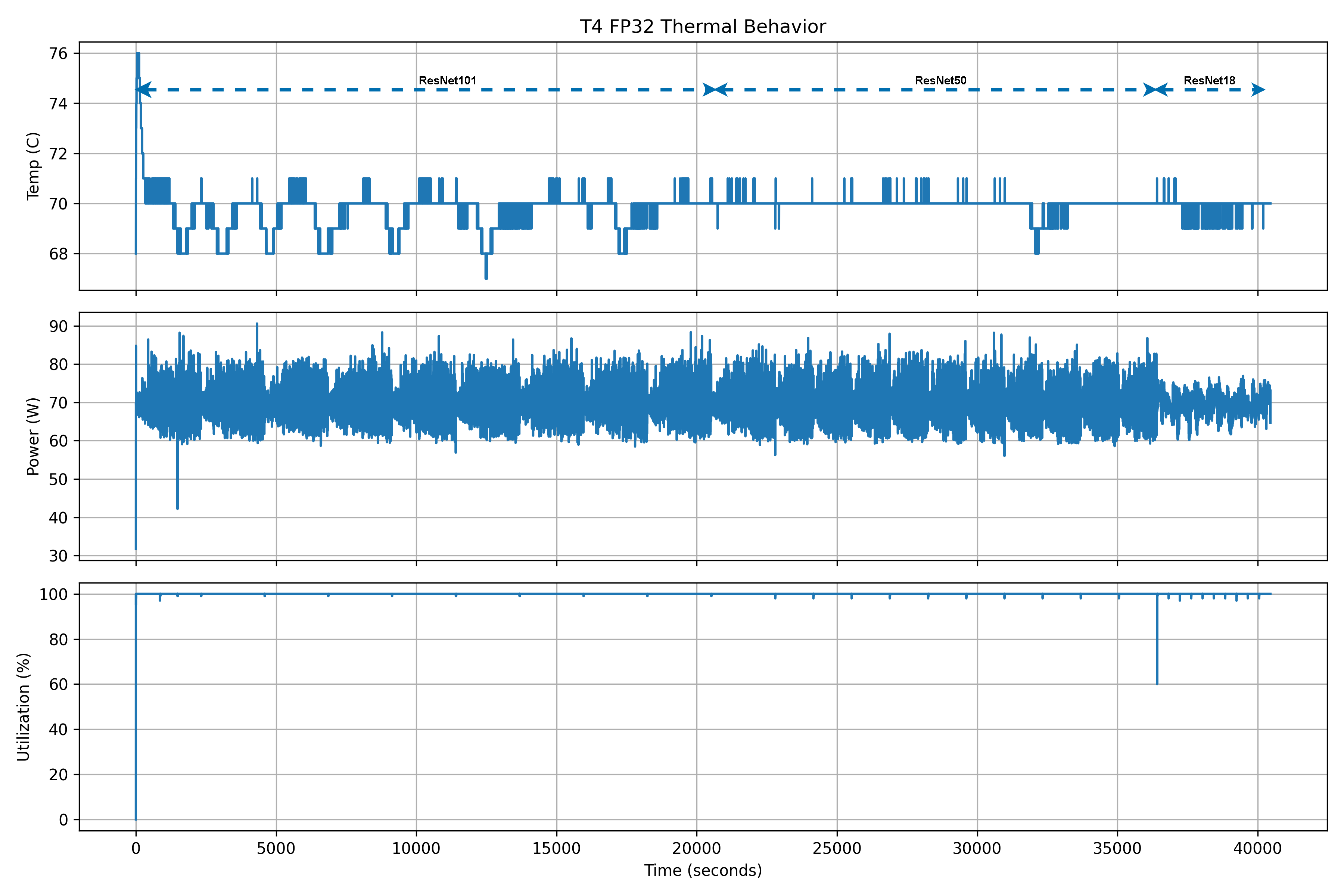}
\caption{T4 FP32}
\end{subfigure}
\hfill
\begin{subfigure}{0.32\textwidth}
\centering
\includegraphics[width=\linewidth]{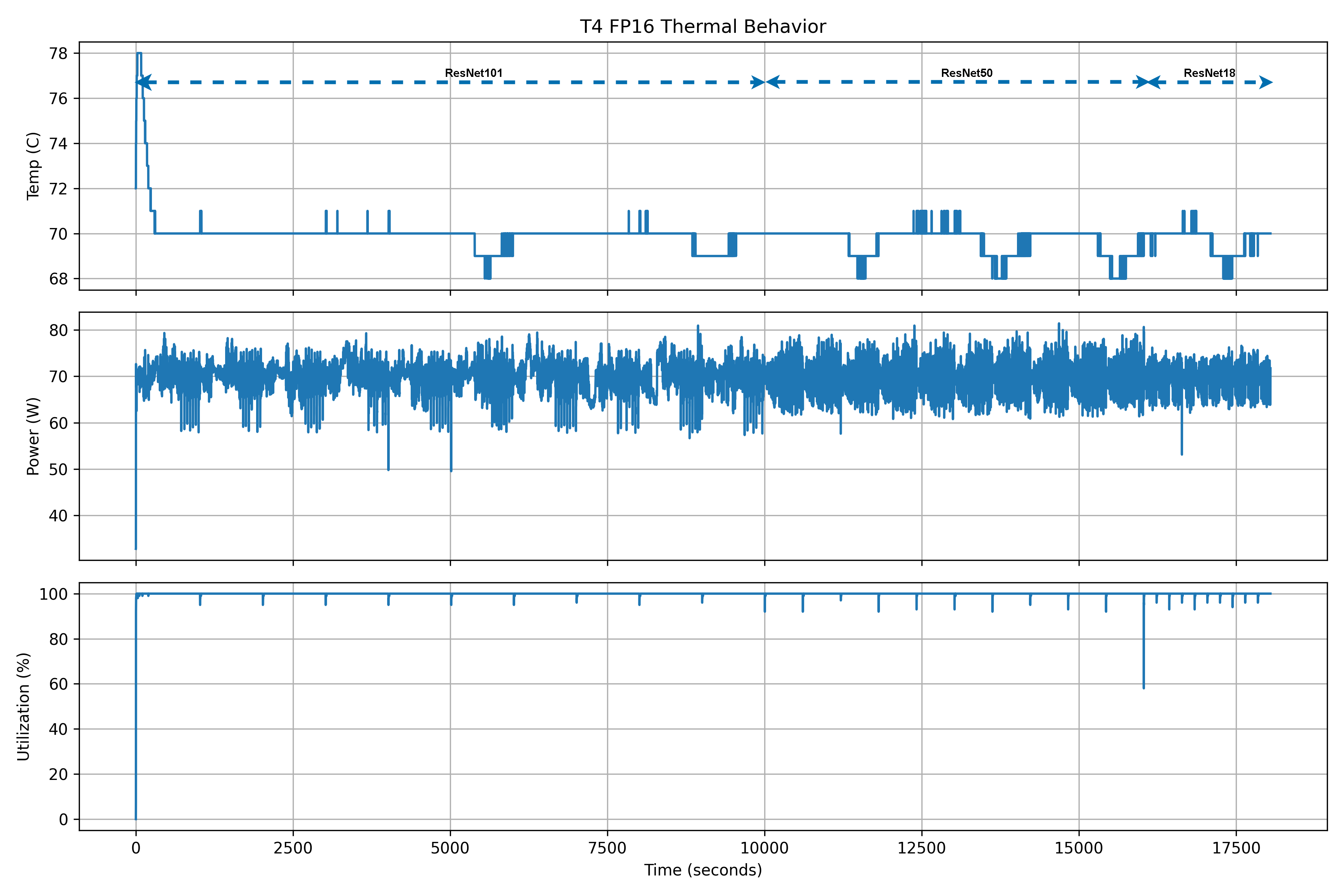}
\caption{T4 FP16}
\end{subfigure}
\hfill
\begin{subfigure}{0.32\textwidth}
\centering
\includegraphics[width=\linewidth]{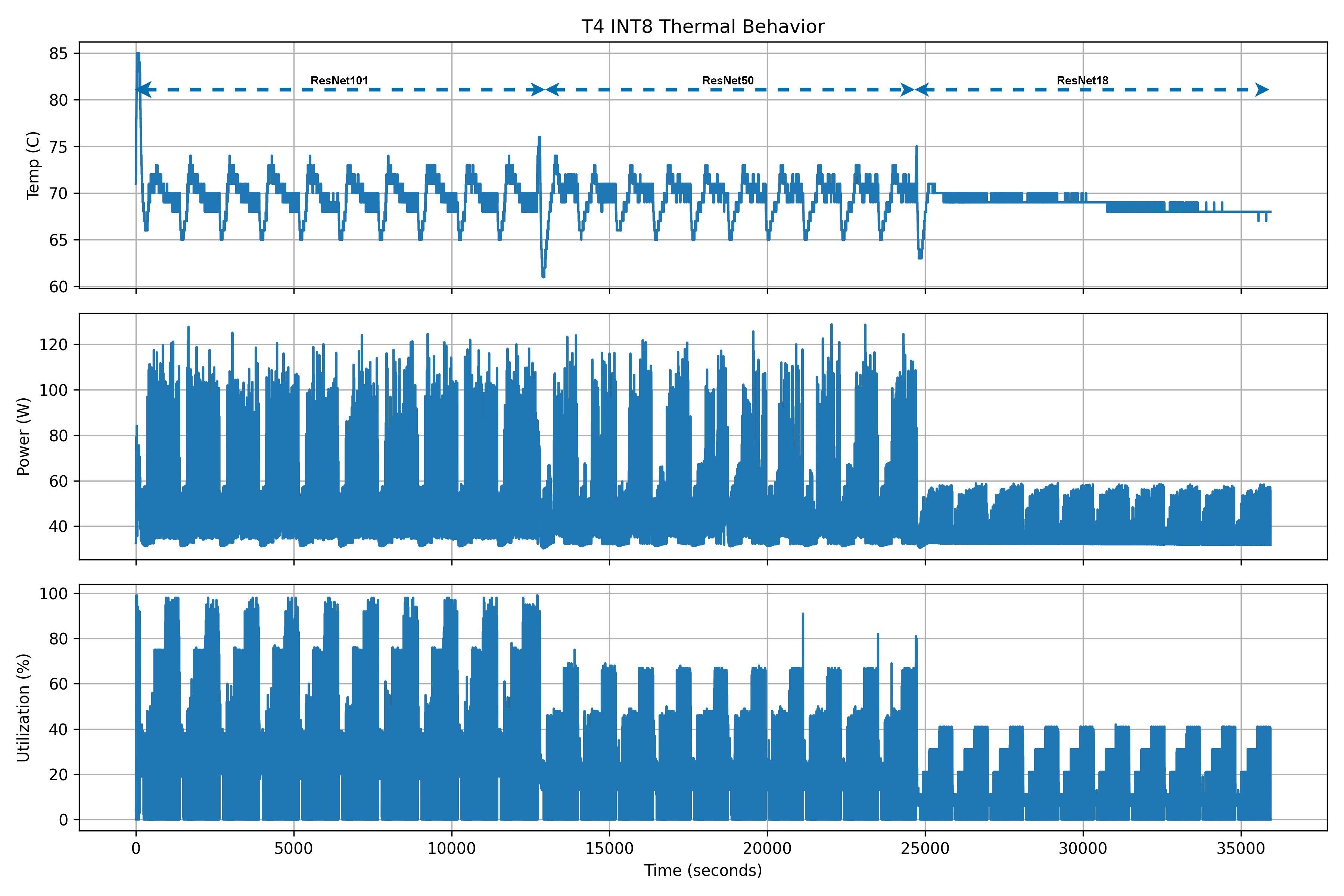}
\caption{T4 INT8}
\end{subfigure}

\vspace{0.5em}

\begin{subfigure}{0.32\textwidth}
\centering
\includegraphics[width=\linewidth]{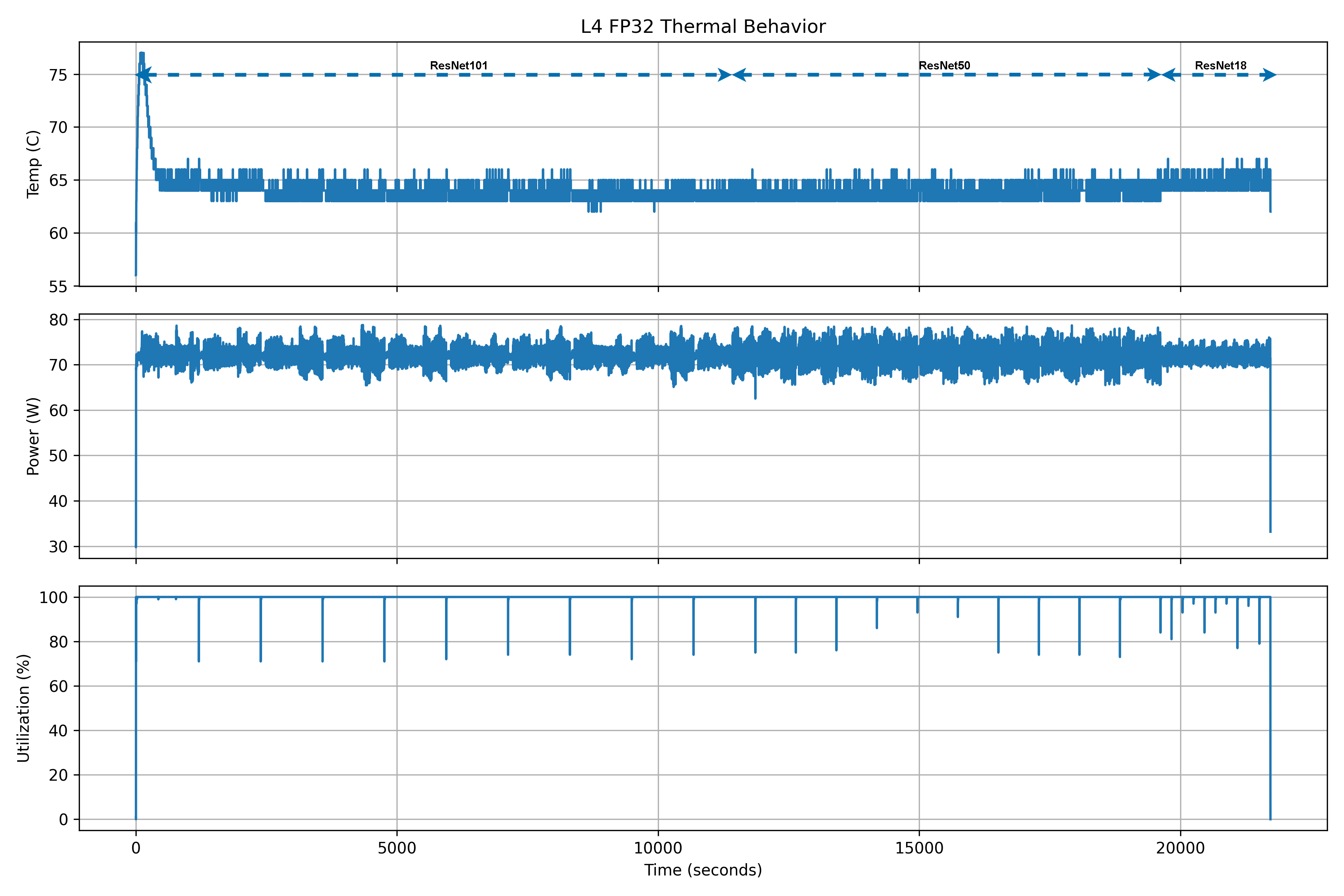}
\caption{L4 FP32}
\end{subfigure}
\hfill
\begin{subfigure}{0.32\textwidth}
\centering
\includegraphics[width=\linewidth]{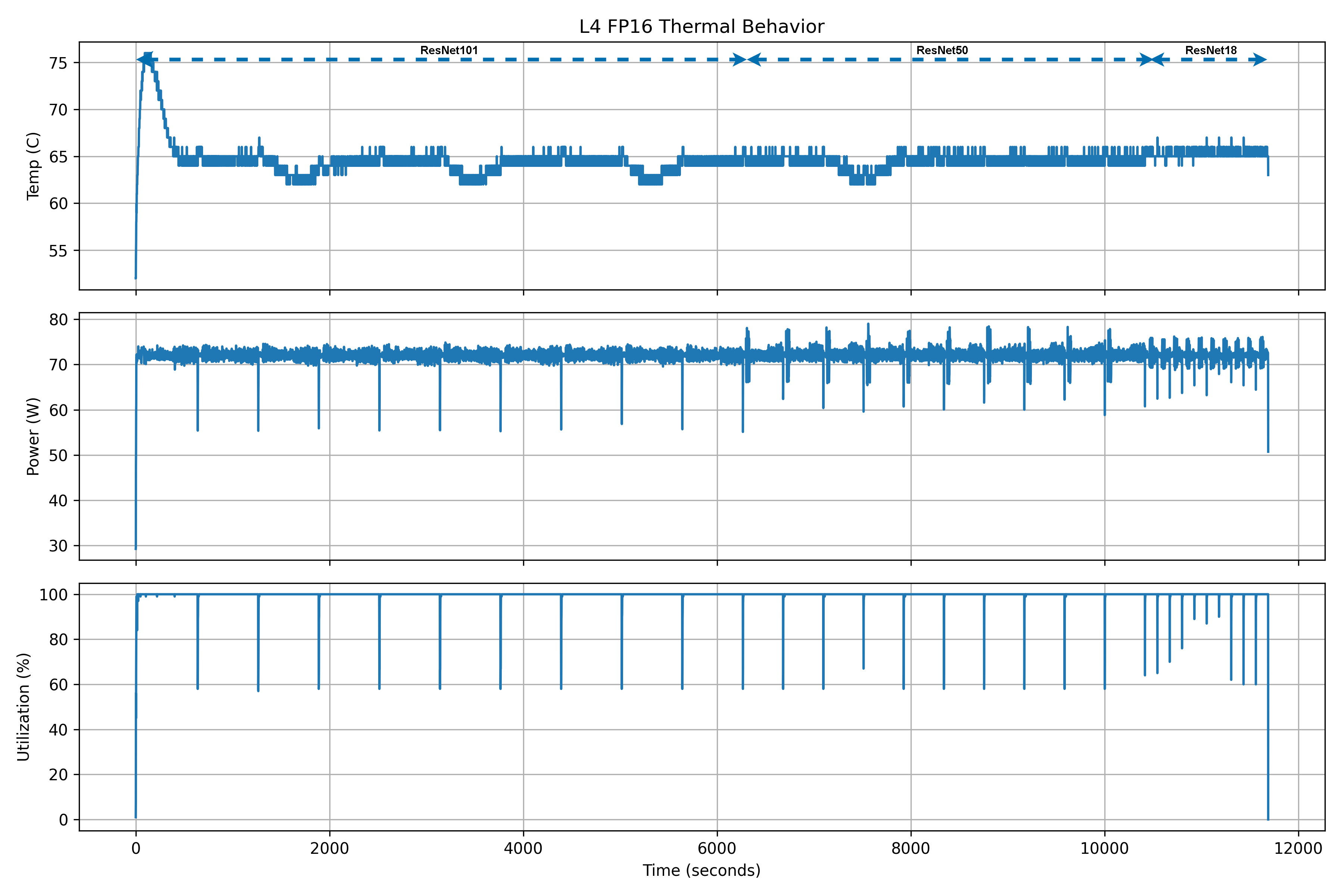}
\caption{L4 FP16}
\end{subfigure}
\hfill
\begin{subfigure}{0.32\textwidth}
\centering
\includegraphics[width=\linewidth]{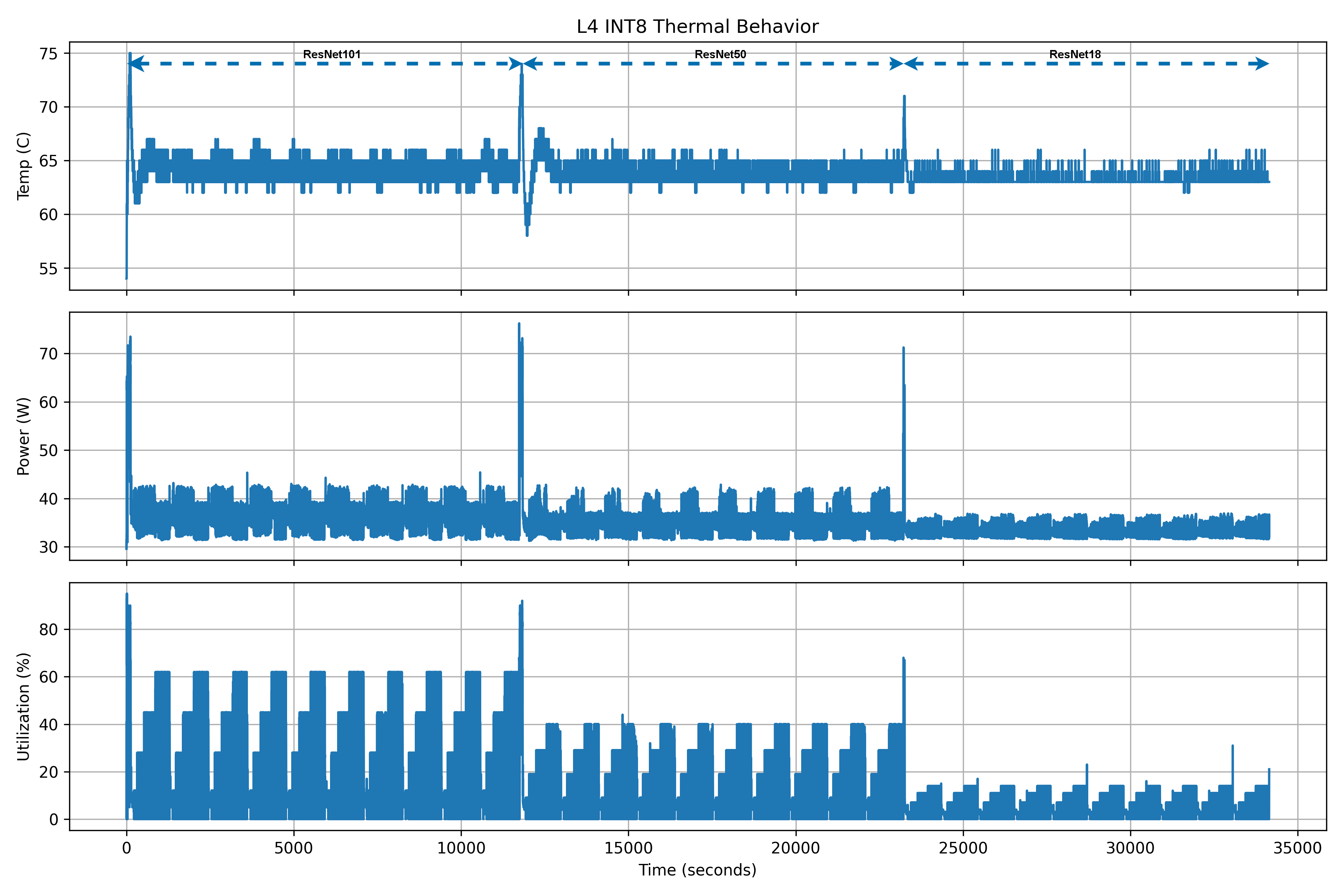}
\caption{L4 INT8}
\end{subfigure}

\caption{
Thermal behavior, power consumption, and utilization across NVIDIA T4 (top row) and NVIDIA L4 (bottom row)
during DEEP-GAP benchmark execution under FP32, FP16, and INT8 precision modes.
Each subplot shows temperature (°C), power draw (W), and GPU utilization (\%) over time for ResNet101,
ResNet50, and ResNet18 workloads executed sequentially.
Both GPUs reach stable steady-state operating conditions with no sustained evidence of thermal throttling.
Observed temperature ranges remain consistent with expected datacenter operating envelopes
(~65–72°C), indicating that reported throughput and latency results primarily reflect architectural
characteristics rather than temperature-induced performance variability.
INT8 workloads show periodic utilization patterns due to TensorRT engine execution phases and batch-level scheduling behavior.
}
\label{fig:thermal_behavior_t4_l4}
\end{figure*}

To ensure measurement stability, GPU temperature, power draw, and utilization were monitored throughout execution using NVML telemetry.
Figure~\ref{fig:thermal_behavior_t4_l4} shows thermal behavior across FP32, FP16, and INT8 precision modes for both NVIDIA T4 and NVIDIA L4.
Across all evaluated workloads, both GPUs reached stable steady-state operating conditions without evidence of sustained thermal throttling.
This confirms that the reported throughput and latency measurements reflect architectural behavior and precision effects rather than temperature-induced performance degradation.

\subsection{Empirical Throughput vs Theoretical Peak Compute}

To contextualize the observed inference throughput, we compare empirical performance trends with the theoretical peak compute capability of the Turing-based NVIDIA T4 and Ada Lovelace–based NVIDIA L4 architectures.

Table~\ref{tab:theoretical_compute} summarizes peak theoretical compute throughput across precision modes. These values represent hardware upper bounds assuming ideal compute saturation and full Tensor Core utilization.

\begin{table}[h]
\centering
\caption{Theoretical Peak Compute Capability}
\begin{tabular}{lccc}
\toprule
GPU & FP32 TFLOPS & FP16 TFLOPS & INT8 TOPS \\
\midrule
T4 (Turing) & 8.1 & 65 & 130 \\
L4 (Ada) & 30.3 & 121 & 242 \\
\bottomrule
\end{tabular}
\label{tab:theoretical_compute}
\end{table}

Real-world inference workloads typically achieve lower utilization due to memory bandwidth constraints, kernel launch overhead, synchronization cost, and non-arithmetic operations, consistent with Roofline-style performance limits that balance compute throughput and memory bandwidth \cite{roofline}.

Nevertheless, empirical results follow the same directional scaling trend as theoretical compute capability. The L4 demonstrates substantially higher throughput compared to T4 across all precision modes, consistent with its increased CUDA core count, expanded L2 cache capacity, higher memory bandwidth, and improved Tensor Core throughput.

Additionally, L4 reaches peak throughput at smaller batch sizes compared to T4, indicating improved architectural efficiency in scheduling parallel workloads and maintaining high Tensor Core utilization even under latency-sensitive execution regimes.

\subsection{Performance per Watt Analysis}

Energy efficiency is a critical consideration for datacenter deployment, where power constraints directly influence rack density, cooling requirements, and operational cost.

Using NVML telemetry, we compute performance per watt (PPW) as:

\begin{equation}
PPW = \frac{Throughput}{Power}
\end{equation}

where Throughput denotes inference throughput (images/sec) and Power denotes average power consumption (W).

Average power consumption is measured during the same benchmark runs used to determine peak throughput for each model, precision mode, and GPU architecture.

Results show that reduced numerical precision substantially improves energy efficiency. INT8 TensorRT execution achieves the highest performance per watt across both architectures due to increased Tensor Core utilization and reduced data movement.

Across all evaluated workloads, L4 demonstrates significantly higher performance per watt compared to T4 while operating within a similar power envelope. This improvement reflects architectural enhancements in Ada Lovelace, including improved Tensor Core throughput, higher memory bandwidth, and increased execution parallelism.

L4 provides higher inference capacity per watt, making it well suited for power-constrained deployments.

\subsection{Benchmark Procedure Differences from CPU-Only Experiments}

The CPU-only methodology explored scaling across 
host thread counts and small batch sizes. In contrast, GPU inference shifts the 
primary source of parallelism to device-side execution, making CPU thread scaling 
less relevant.

Accordingly, GPU experiments use a fixed host thread configuration and instead 
focus on batch-size scaling to expose GPU utilization and saturation behavior. 
Batch sizes were swept across 
$\{1,2,4,8,16,32,64,128,256,384,512\}$ 
to evaluate throughput scaling under increasing arithmetic intensity and kernel 
occupancy.

Timing is collected using CUDA events with explicit synchronization to eliminate 
host-side measurement artifacts. Each configuration includes a warm-up phase to 
reach steady state, followed by repeated timed iterations to compute median and 
p99 latency.

GPU memory usage is recorded using NVML to identify saturation points and 
out-of-memory boundaries.

\subsection{Software Stack}

All experiments were conducted under a fixed software environment to ensure 
consistency across configurations.

\subsubsection{Operating System}

Benchmarking was performed on Ubuntu 22.04 LTS with background services minimized 
to reduce scheduling interference.

\subsubsection{NVIDIA Driver and CUDA Toolkit}

A consistent NVIDIA driver version was used across both GPUs. CUDA Toolkit support 
was verified for compatibility with both Turing (T4) and Ada Lovelace (L4)  \cite{l4_datasheet,ada_whitepaper}
architectures. Identical runtime libraries were used to avoid version-induced 
performance variation.

\subsubsection{PyTorch Runtime}

FP32 and FP16 experiments were executed using PyTorch with CUDA acceleration. 
The same PyTorch version was used across both GPUs to ensure consistent operator 
implementations and kernel selection.

\subsubsection{TensorRT Optimization}

INT8 experiments were conducted using NVIDIA TensorRT to leverage optimized 
execution, including kernel fusion and quantization-aware acceleration. Engines 
were built separately for each GPU to enable architecture-specific optimization 
while maintaining identical model configurations.

All runtime parameters, including precision modes and batch sizes, were 
programmatically controlled to ensure consistency across runs.

\subsection{Models Evaluated}

To maintain methodological continuity with GDEV-AI \cite{gdevai}, this study 
evaluates three widely used convolutional neural network architectures: 
ResNet-18, ResNet-50, and ResNet-101. These models are selected to represent 
progressively increasing computational complexity and arithmetic intensity, 
enabling analysis of how architectural differences between T4 and L4 manifest 
under varying workload intensities.

Collectively, these models span a range of compute characteristics from 
relatively lightweight inference workloads (ResNet-18) to high-intensity 
execution regimes (ResNet-101), providing a structured basis for evaluating 
throughput scaling, latency behavior, and Tensor Core utilization across 
GPU generations.

All models are instantiated using standard implementations provided by the PyTorch
Torchvision library \cite{torchvision_resnet}.

\subsubsection{ResNet-18}

ResNet-18 represents a lightweight convolutional architecture with relatively 
low computational and memory requirements. It is well-suited for latency-sensitive 
inference scenarios and smaller batch sizes, where kernel launch overheads, 
memory access latency, and PCIe transfer costs can have a more pronounced impact 
on performance.

In the context of DEEP-GAP, ResNet-18 is used to evaluate GPU behavior under 
lower compute intensity, helping identify scenarios where architectural 
improvements may be limited by underutilization of available compute resources. 
This provides insight into performance scaling in latency-bound regimes, as 
analyzed in Section~\ref{sec:results}.

\subsubsection{ResNet-50}

ResNet-50 is a deeper and more computationally intensive architecture, with 
significantly higher parameter count and arithmetic complexity compared to 
ResNet-18. It places greater demand on GPU compute units, memory bandwidth, 
and Tensor Core throughput.

Within DEEP-GAP, ResNet-50 serves as a representative high-intensity workload, 
enabling evaluation of how architectural enhancements in L4 translate into 
measurable gains under sustained parallel execution. In particular, it allows 
analysis of scaling behavior at larger batch sizes, where workloads transition 
toward compute-bound or memory-bound regimes.

Comparing ResNet-18 and ResNet-50 enables a structured analysis of how workload 
intensity influences the observed performance gap between T4 and L4, providing 
a more comprehensive understanding of generational improvements across different 
deployment scenarios.

\subsubsection{ResNet-101}

ResNet-101 further increases model depth and computational complexity, representing 
a high-intensity inference workload. Its larger parameter count and deeper network 
structure significantly increase arithmetic demand and memory traffic, placing 
sustained pressure on GPU compute resources and memory hierarchy.

In DEEP-GAP, ResNet-101 is used to evaluate performance under fully utilized 
execution conditions, where GPUs operate closer to their peak throughput limits. 
This model is particularly useful for exposing differences in memory bandwidth, 
cache efficiency, and Tensor Core utilization between T4 and L4.

By including ResNet-101, the study captures behavior in high compute intensity 
regimes, where architectural improvements are expected to have the most pronounced 
impact. The resulting performance trends are analyzed in Section~\ref{sec:throughput_scaling}, 
with particular focus on throughput saturation and latency stability.

\paragraph{Discussion}

Together, ResNet-18, ResNet-50, and ResNet-101 provide a structured evaluation 
across low, medium, and high workload intensities. This progression enables a 
comprehensive analysis of how workload characteristics influence the observed 
performance gap between T4 and L4, and how architectural improvements scale 
with increasing computational demand.

\subsubsection{Rationale for Model Selection}

ResNet architectures are selected due to their widespread adoption in both 
research literature and production inference environments. Their consistent 
structure and extensive prior evaluation make them suitable baselines for 
architectural performance analysis and enable direct comparison with the 
CPU-based results reported in GDEV-AI \cite{gdevai}.

The combination of ResNet-18, ResNet-50, and ResNet-101 spans a range of 
computational intensities, allowing evaluation of GPU performance across 
varying workload scales. Increasing network depth introduces higher arithmetic 
demand and memory traffic, enabling observation of how architectural differences 
between T4 and L4 manifest under progressively more demanding inference conditions.

These models are primarily composed of convolutional operations, which are well 
suited to GPU execution and benefit from mixed-precision acceleration using FP16 
and INT8 Tensor Core pathways. This makes them effective workloads for analyzing 
throughput scaling, latency characteristics, and precision-dependent performance behavior.

All experiments are conducted in inference mode (\texttt{model.eval()}) with gradient 
computation disabled, ensuring that measurements reflect forward-pass execution only.

\section{Results}
\label{sec:results}
\subsection{Throughput Scaling Across Batch Sizes}
\label{sec:throughput_scaling}
\begin{figure*}[t]
    \centering


    \begin{subfigure}[b]{0.32\textwidth}
        \centering
        \includegraphics[width=\textwidth]{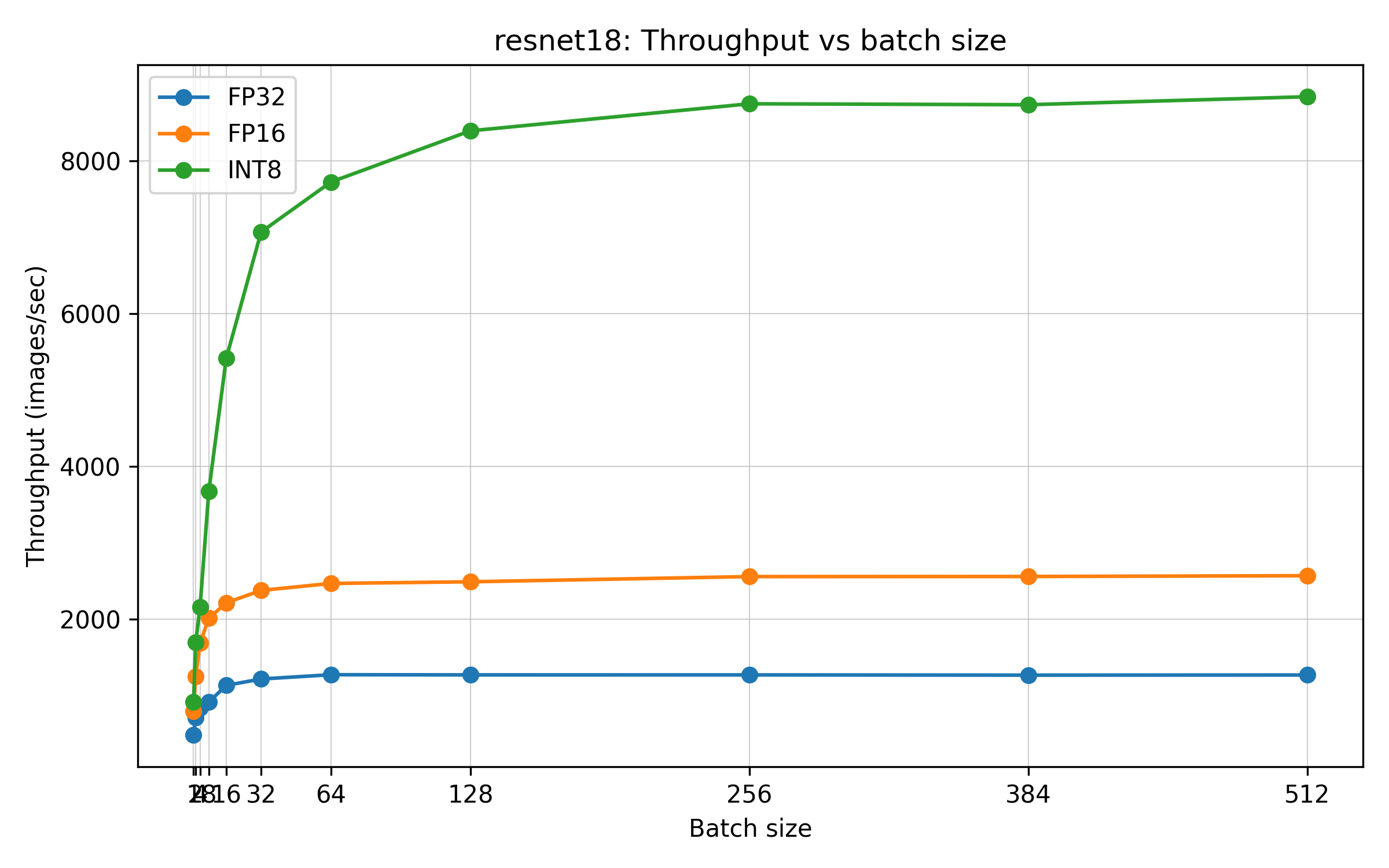}
        \caption{ResNet18}
        \label{fig:r18_t4_throughput}
    \end{subfigure}
    \hfill
    \begin{subfigure}[b]{0.32\textwidth}
        \centering
        \includegraphics[width=\textwidth]{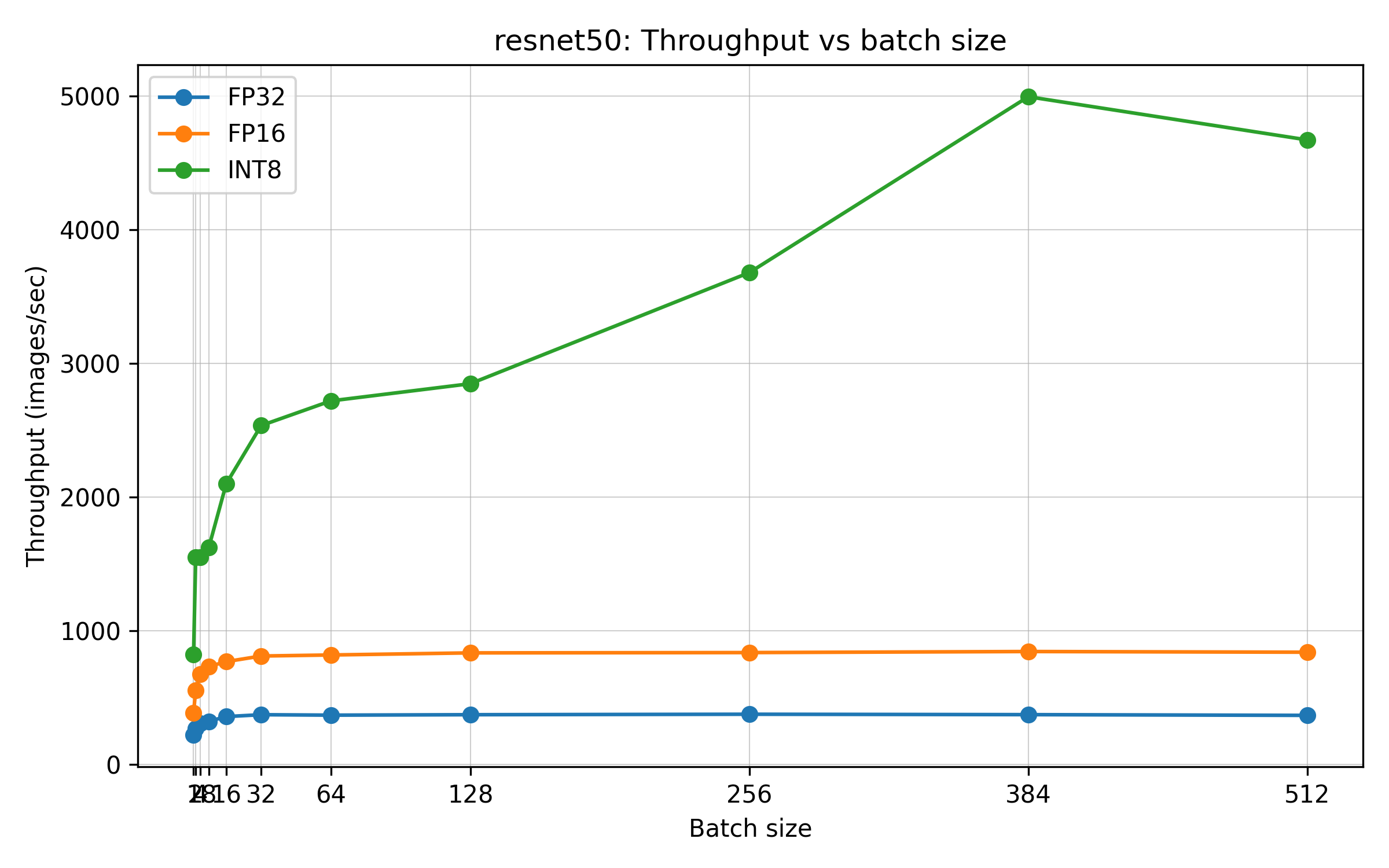}
        \caption{ResNet50}
        \label{fig:r50_t4_throughput}
    \end{subfigure}
    \hfill
    \begin{subfigure}[b]{0.32\textwidth}
        \centering
        \includegraphics[width=\textwidth]{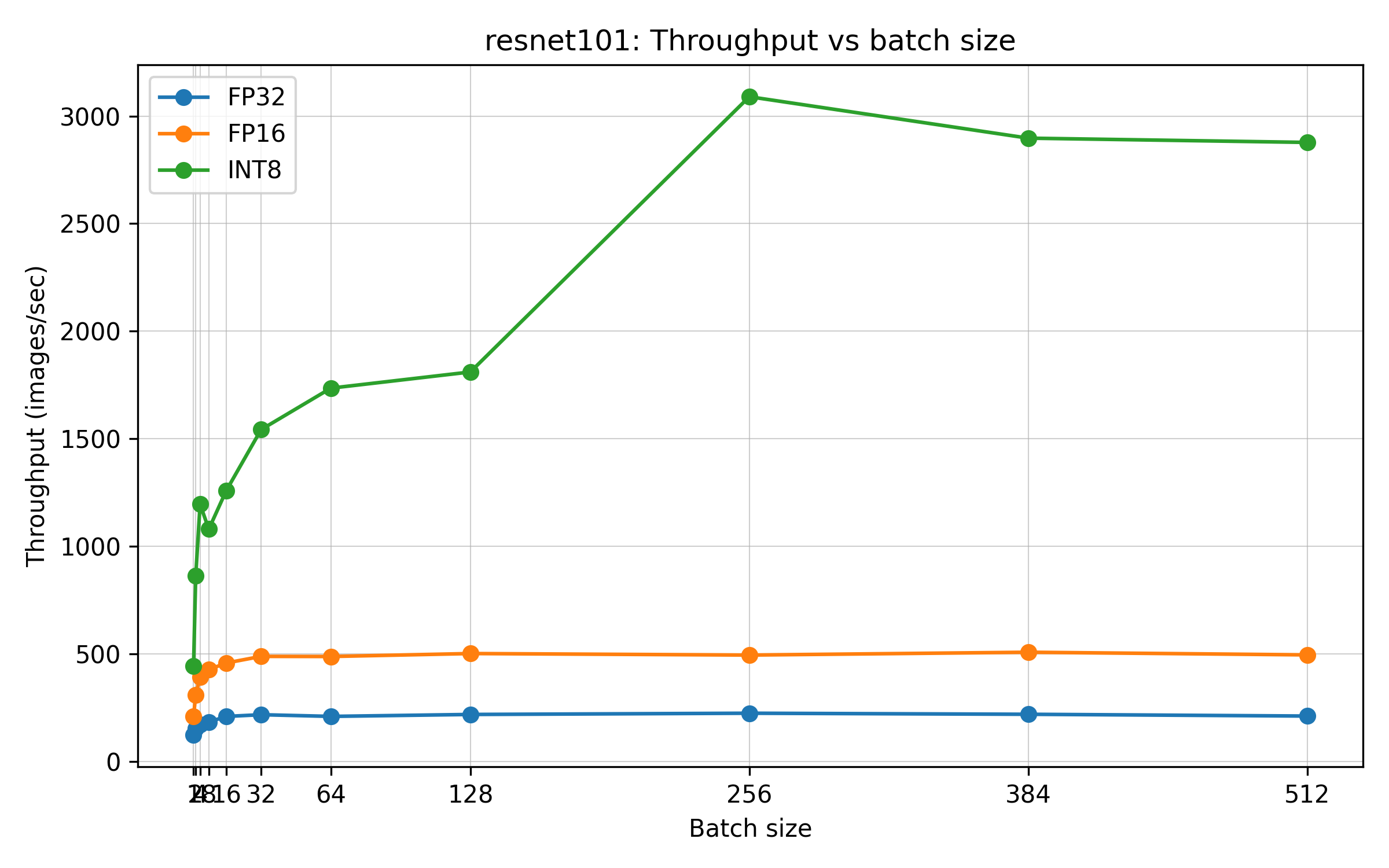}
        \caption{ResNet101}
        \label{fig:r101_t4_throughput}
    \end{subfigure}

    \vspace{12pt}


    \begin{subfigure}[b]{0.32\textwidth}
        \centering
        \includegraphics[width=\textwidth]{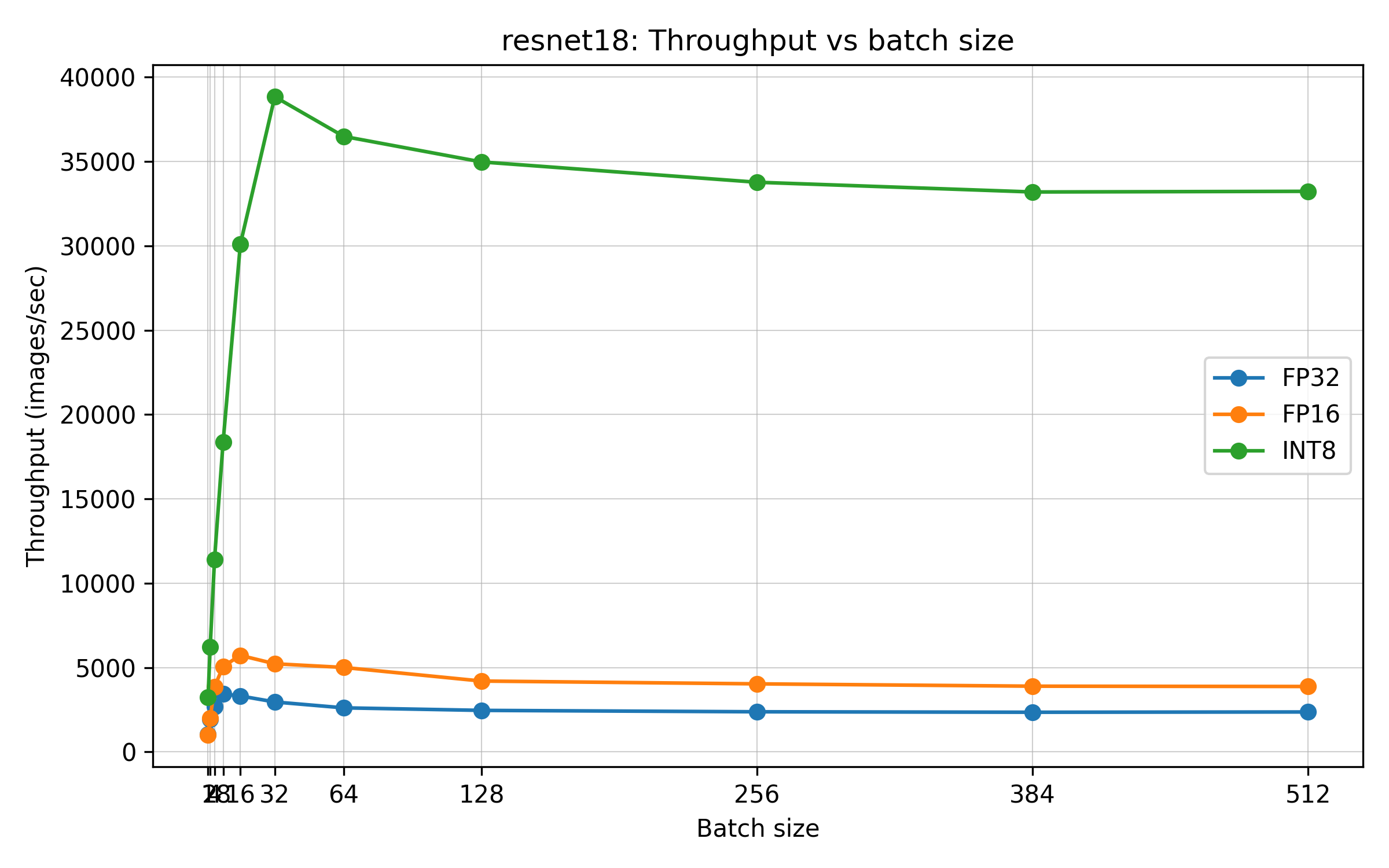}
        \caption{ResNet18}
        \label{fig:r18_l4_throughput}
    \end{subfigure}
    \hfill
    \begin{subfigure}[b]{0.32\textwidth}
        \centering
        \includegraphics[width=\textwidth]{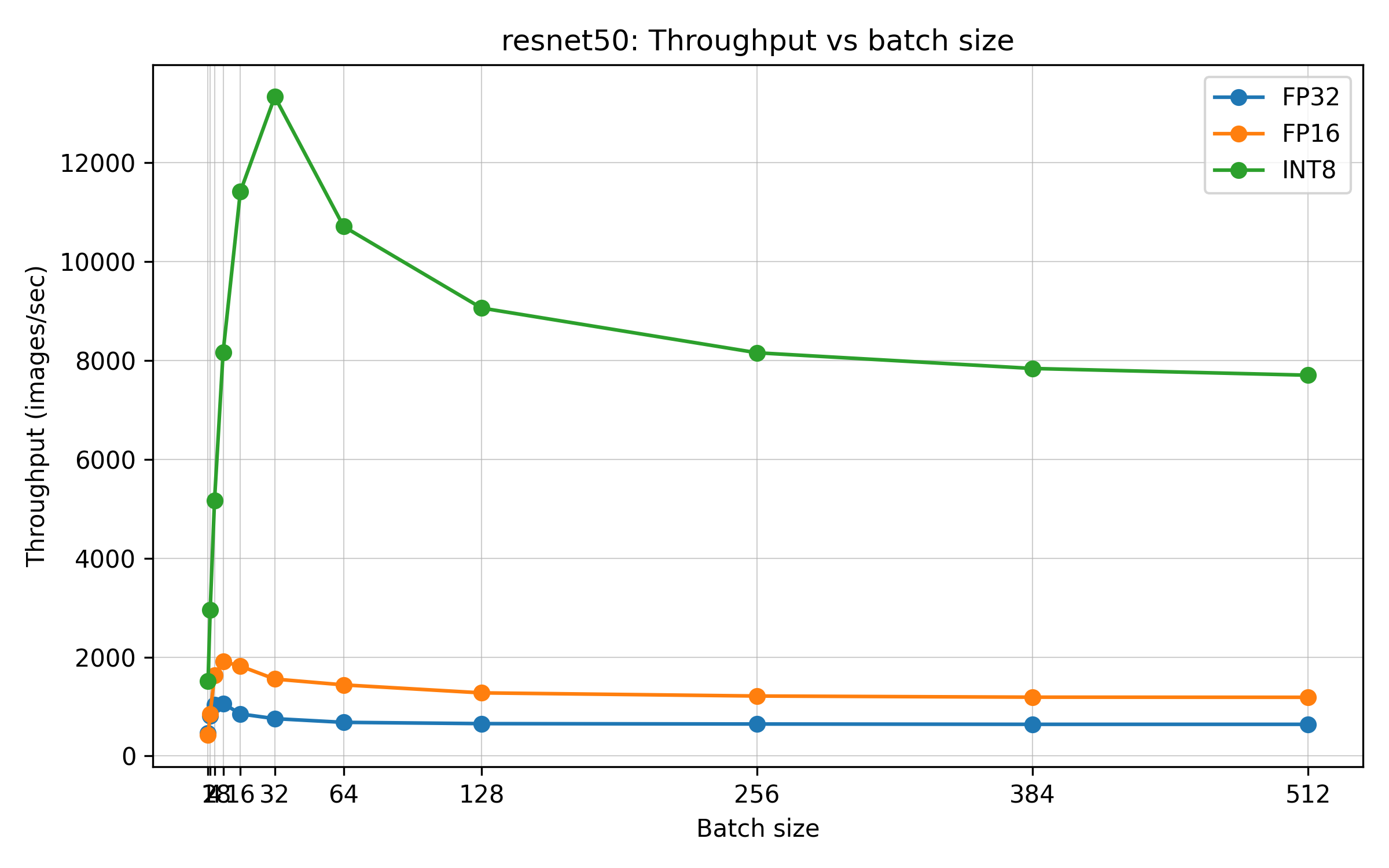}
        \caption{ResNet50}
        \label{fig:r50_l4_throughput}
    \end{subfigure}
    \hfill
    \begin{subfigure}[b]{0.32\textwidth}
        \centering
        \includegraphics[width=\textwidth]{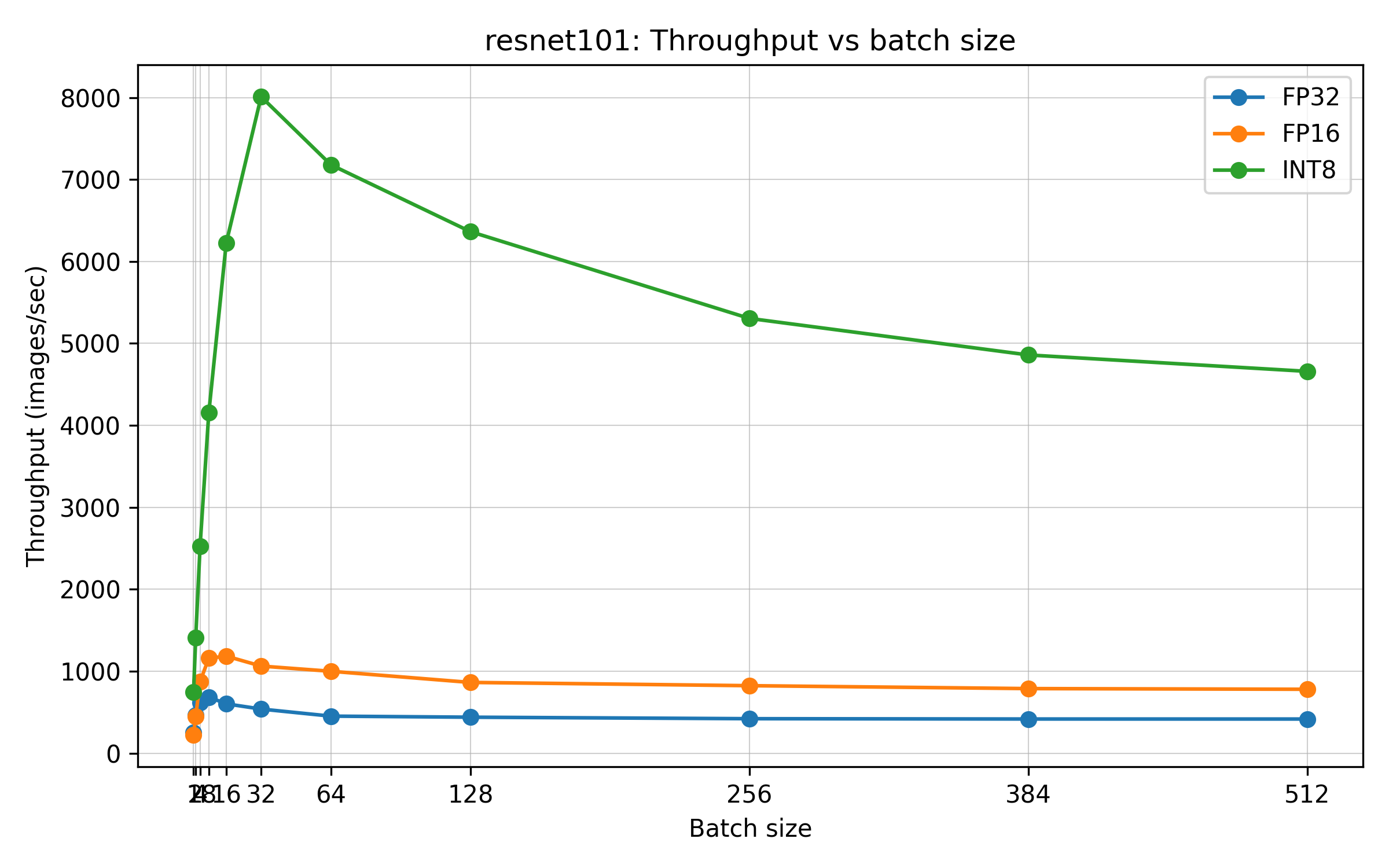}
        \caption{ResNet101}
        \label{fig:r101_l4_throughput}
    \end{subfigure}

    \caption{
    Throughput scaling across batch sizes for ResNet18, ResNet50, and ResNet101.
    Top row shows NVIDIA T4 results, bottom row shows NVIDIA L4 results.
    INT8 achieves highest throughput due to reduced numerical precision and
    improved Tensor Core utilization. L4 demonstrates significantly higher
    throughput and improved scaling efficiency compared to T4.
    }

    \label{fig:throughput_t4_vs_l4}

\end{figure*}

We first analyze inference throughput scaling across increasing batch sizes
for ResNet18, ResNet50, and ResNet101 across NVIDIA T4 and NVIDIA L4 GPUs.
Figure~\ref{fig:throughput_t4_vs_l4} shows throughput (images/sec)
across FP32, FP16, and INT8 precision modes.

Across all models and architectures, throughput increases rapidly as batch size
grows from 1 to 32, demonstrating the transition from underutilized execution
to effective GPU parallelism. Beyond moderate batch sizes, throughput begins to
plateau, indicating saturation of available compute resources, memory bandwidth,
or kernel execution efficiency.

\textbf{CPU baseline.}
On the Granite Rapids CPU platform (24 threads), peak throughput occurs at
batch size $B=8$, reaching 670 images/sec for ResNet18 and 230 images/sec for
ResNet50. Throughput does not significantly improve beyond this point,
suggesting early saturation due to limited data parallelism and memory bandwidth.

\textbf{T4 scaling behavior.}
On NVIDIA T4, FP32 throughput improves modestly over CPU baselines.
ResNet18 reaches 1289 images/sec at $B=384$, providing a 1.92$\times$
speedup over CPU. Similarly, ResNet50 reaches 382 images/sec at $B=256$
(1.66$\times$ speedup).

Reduced precision significantly increases throughput.
Using FP16, ResNet18 achieves 2569 images/sec at $B=512$
(3.83$\times$ CPU speedup), while ResNet50 reaches 849 images/sec at $B=384$
(3.69$\times$ speedup).

INT8 TensorRT demonstrates substantially higher efficiency.
ResNet18 achieves 8837 images/sec at $B=512$, representing a 13.19$\times$
speedup relative to CPU. ResNet50 achieves 5066 images/sec at $B=384$
(22.03$\times$ speedup), while ResNet101 reaches 3125 images/sec at $B=256$.

These results indicate that T4 benefits significantly from reduced precision,
but requires relatively large batch sizes to fully utilize Tensor Core parallelism.

\textbf{L4 scaling behavior.}
NVIDIA L4 demonstrates substantially improved scaling efficiency compared to T4.
Even at small batch sizes, L4 achieves strong performance gains.
Using FP32 precision at $B=8$, ResNet18 achieves 3483 images/sec,
representing a 5.20$\times$ speedup over CPU and a 2.70$\times$ improvement
over T4 FP32 peak throughput.

FP16 further increases throughput efficiency.
ResNet18 achieves 5923 images/sec at $B=16$ (8.84$\times$ CPU speedup),
while ResNet50 achieves 1928 images/sec at $B=8$ (8.38$\times$ speedup).

INT8 TensorRT provides the highest throughput across all configurations.
ResNet18 achieves 38,932 images/sec at $B=32$, corresponding to a 58.11$\times$
speedup over CPU and approximately 4.4$\times$ improvement over T4 INT8.
ResNet50 achieves 13,388 images/sec at $B=32$ (58.21$\times$ speedup),
while ResNet101 achieves 8026 images/sec at $B=32$.

\begin{figure*}[t]
    \centering

    \begin{subfigure}[b]{0.48\textwidth}
        \centering
        \includegraphics[width=\textwidth]{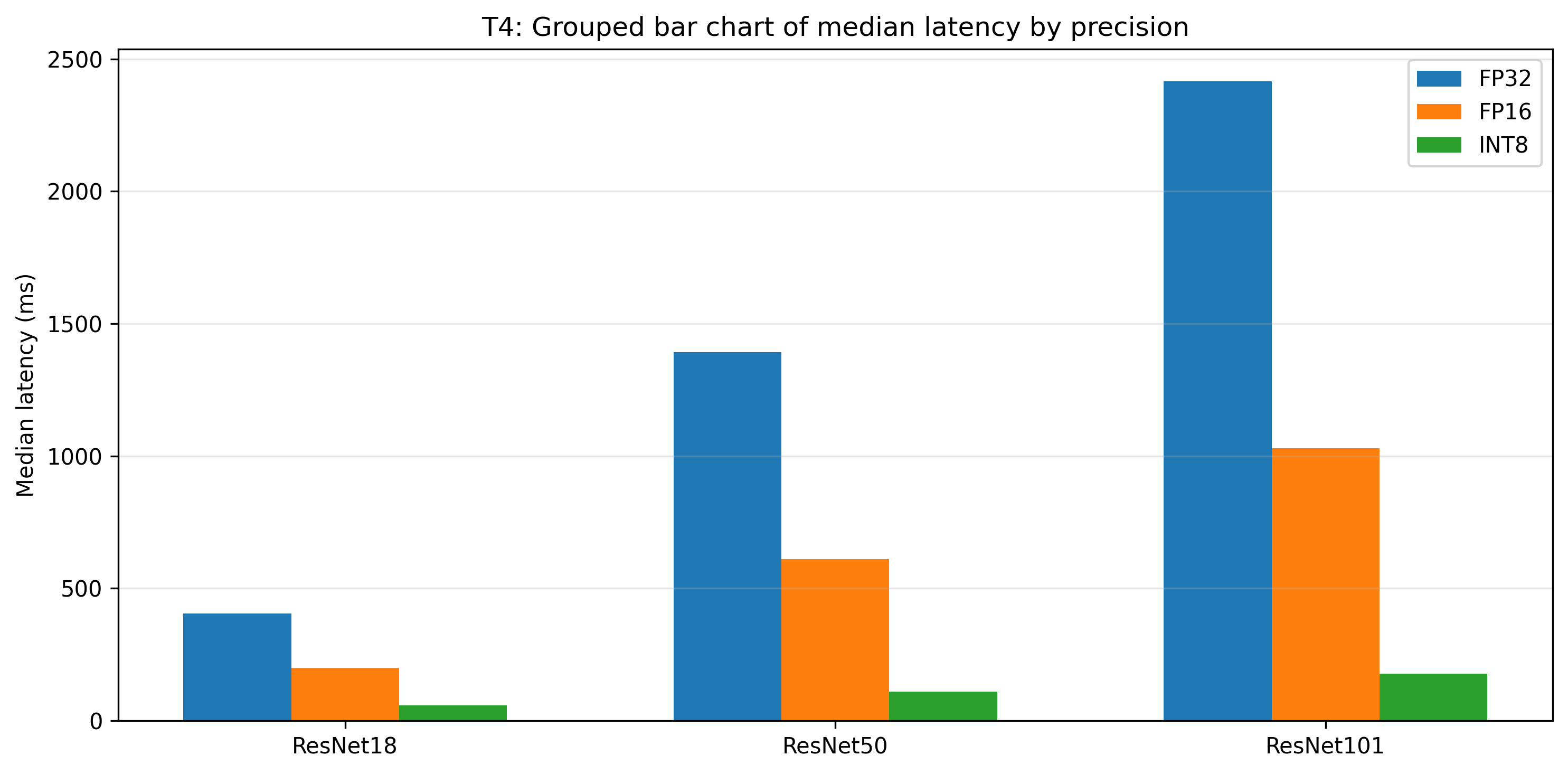}
        \caption{T4 median latency comparison}
        \label{fig:t4_latency_summary}
    \end{subfigure}
    \hfill
    \begin{subfigure}[b]{0.48\textwidth}
        \centering
        \includegraphics[width=\textwidth]{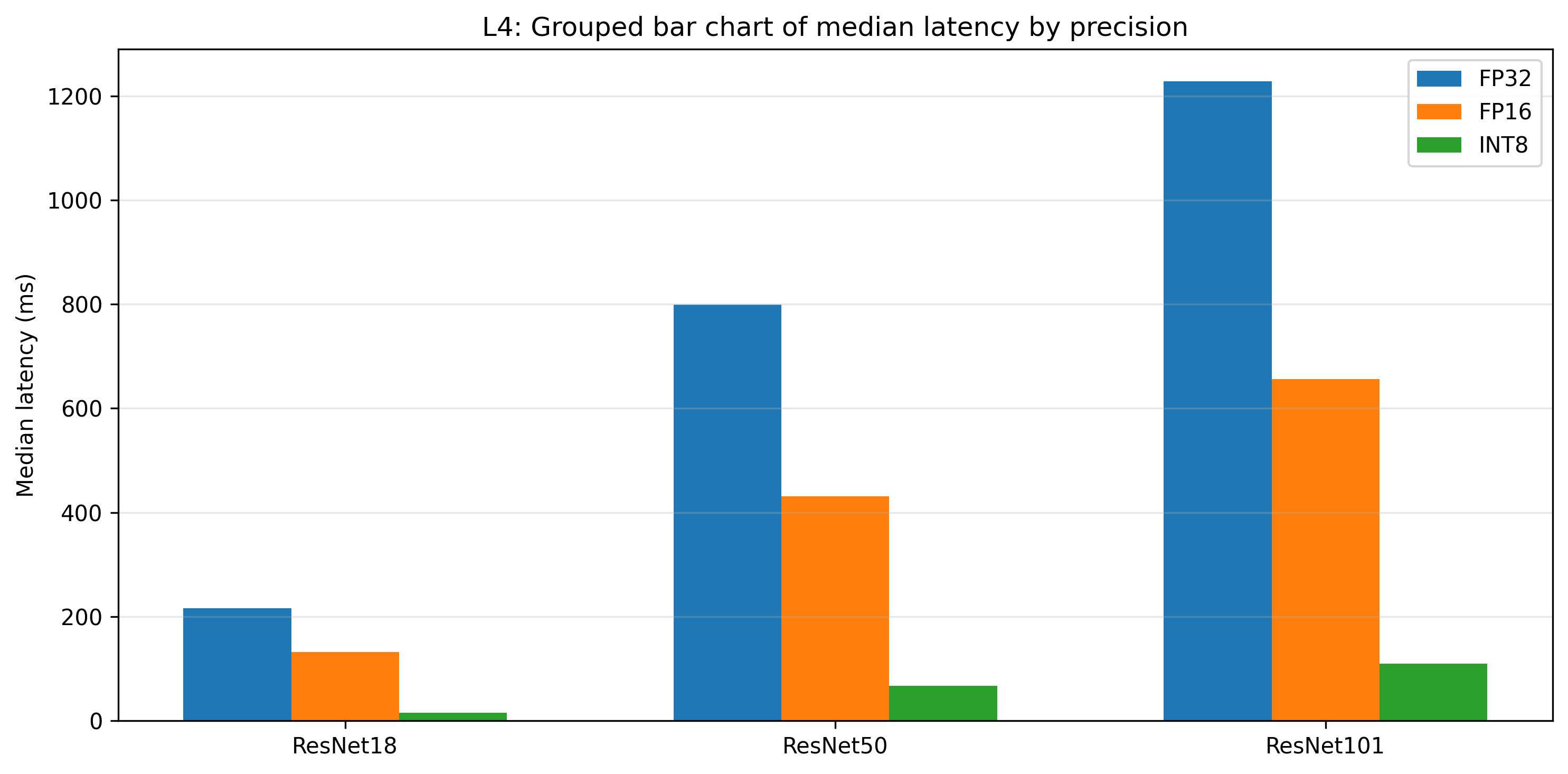}
        \caption{L4 median latency comparison}
        \label{fig:l4_latency_summary}
    \end{subfigure}

    \caption{
    Median latency comparison across ResNet models and precision modes.
    INT8 consistently achieves the lowest latency, followed by FP16 and FP32.
    Larger models (ResNet101) show higher latency due to increased computational complexity.
    L4 demonstrates significantly lower latency than T4 across all precision modes,
    highlighting architectural improvements in Tensor Core efficiency and memory bandwidth.
    }
    \label{fig:latency_summary_t4_l4}

\end{figure*}

\begin{figure*}[t]
    \centering


    \begin{subfigure}[b]{0.32\textwidth}
        \centering
        \includegraphics[width=\textwidth]{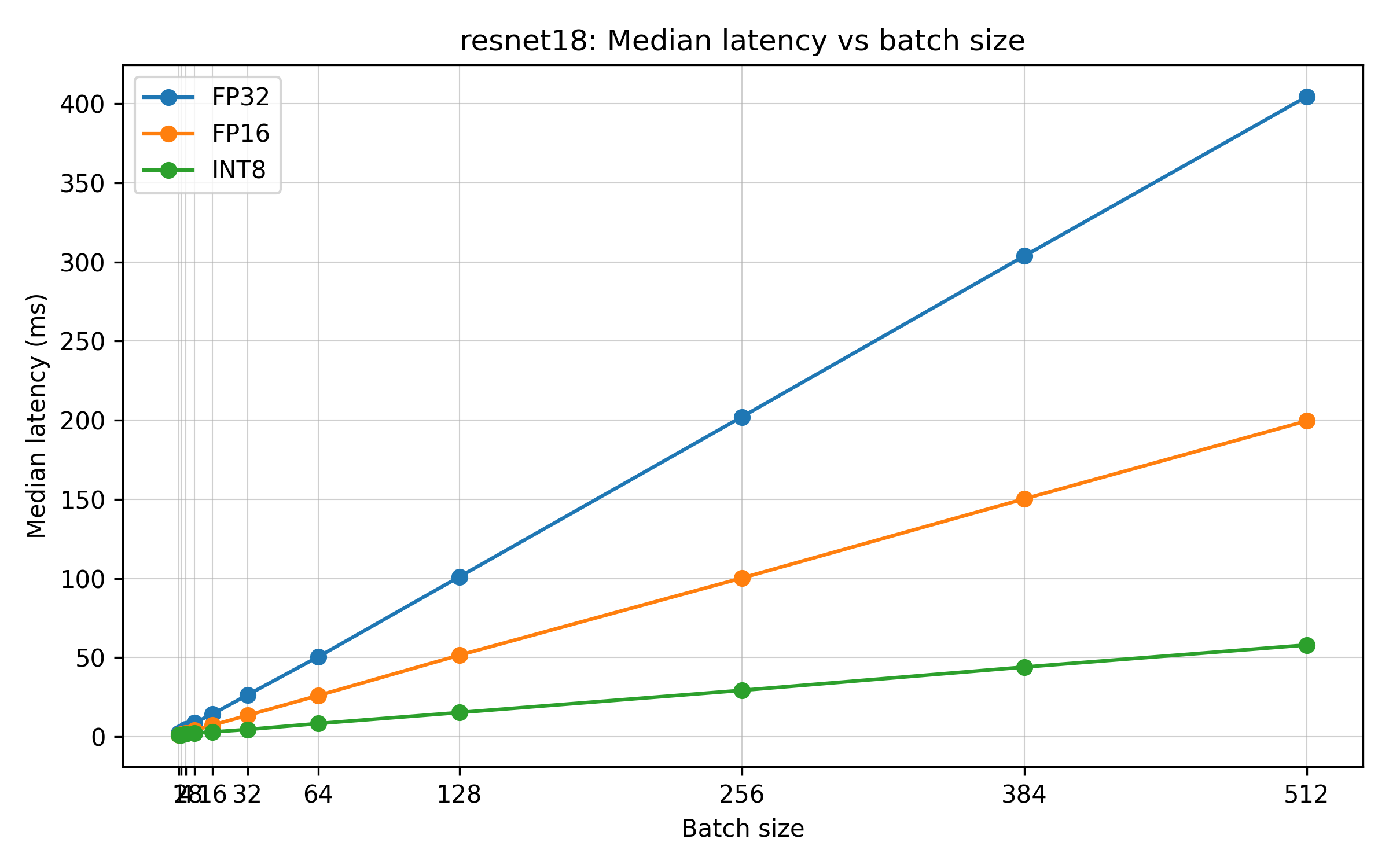}
        \caption{ResNet18}
        \label{fig:r18_t4_latency}
    \end{subfigure}
    \hfill
    \begin{subfigure}[b]{0.32\textwidth}
        \centering
        \includegraphics[width=\textwidth]{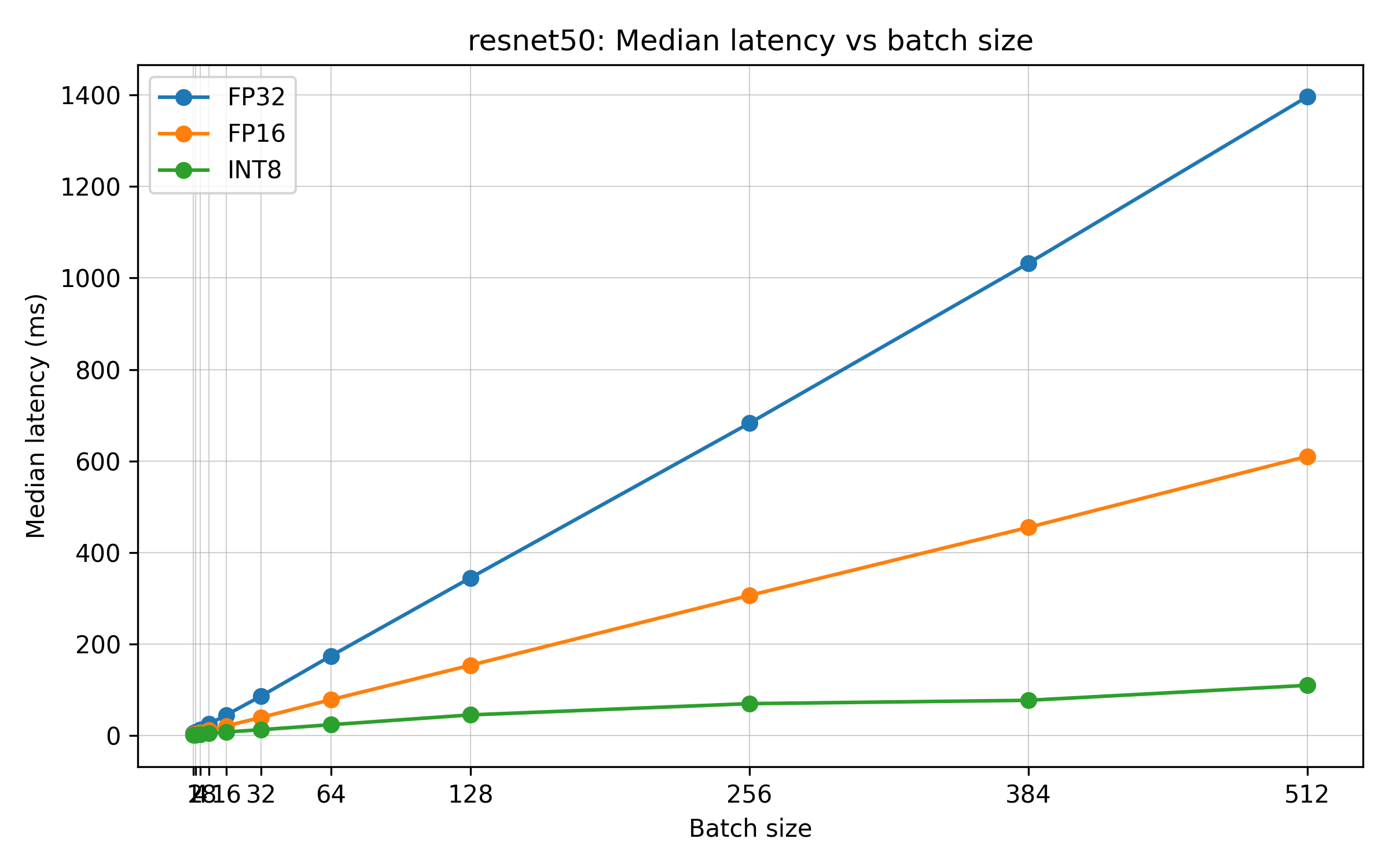}
        \caption{ResNet50}
        \label{fig:r50_t4_latency}
    \end{subfigure}
    \hfill
    \begin{subfigure}[b]{0.32\textwidth}
        \centering
        \includegraphics[width=\textwidth]{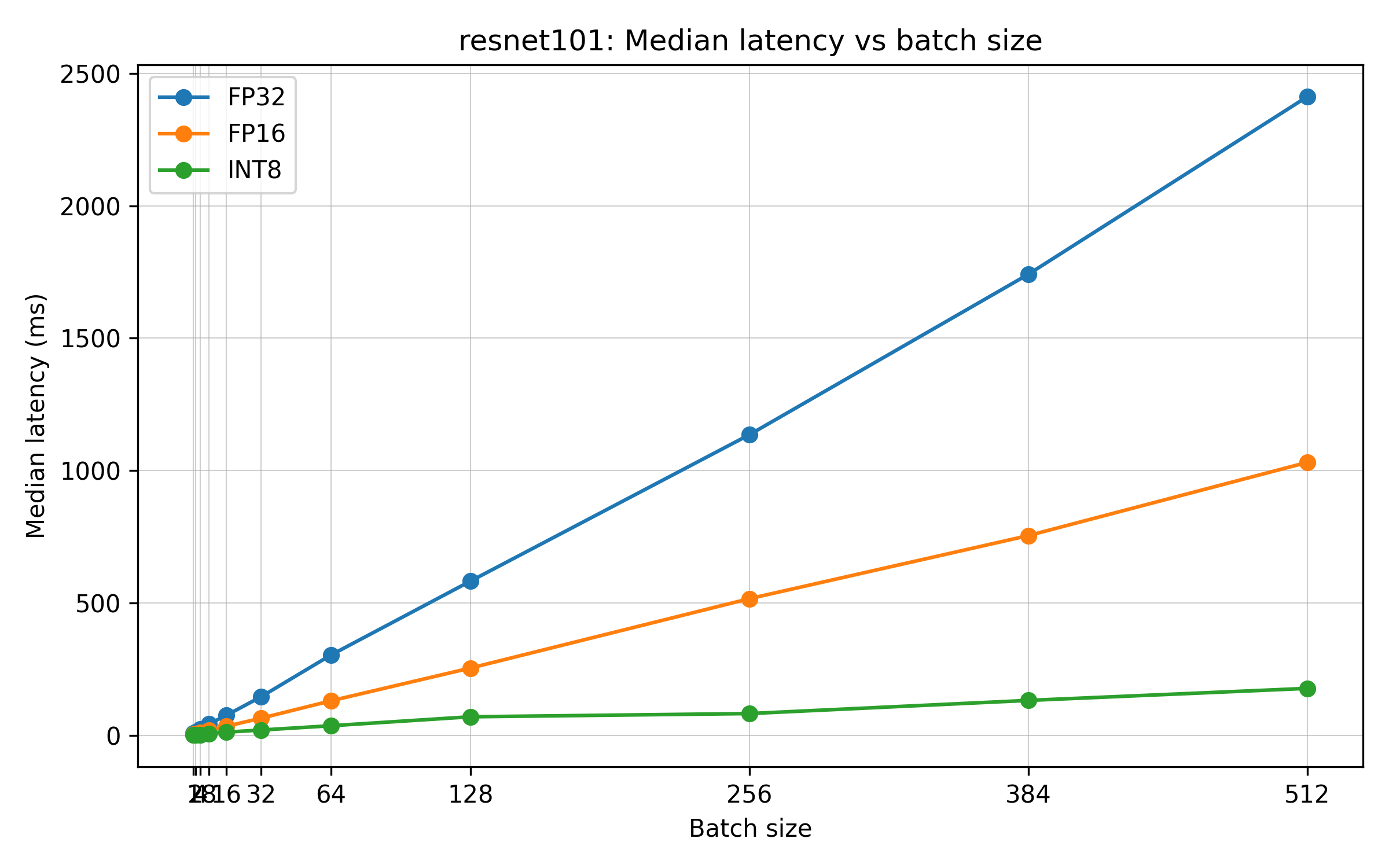}
        \caption{ResNet101}
        \label{fig:r101_t4_latency}
    \end{subfigure}

    \vspace{12pt}


    \begin{subfigure}[b]{0.32\textwidth}
        \centering
        \includegraphics[width=\textwidth]{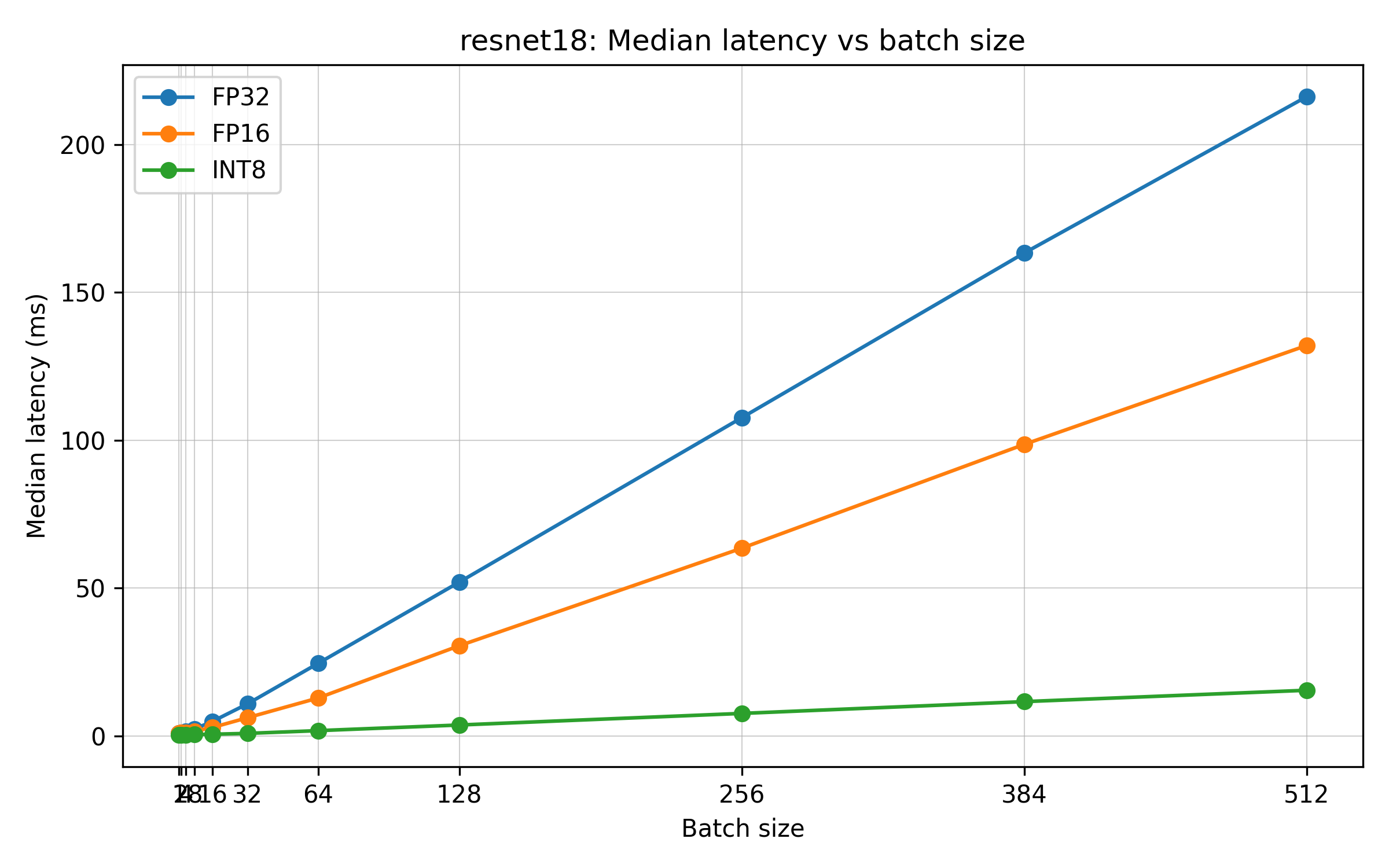}
        \caption{ResNet18}
        \label{fig:r18_l4_latency}
    \end{subfigure}
    \hfill
    \begin{subfigure}[b]{0.32\textwidth}
        \centering
        \includegraphics[width=\textwidth]{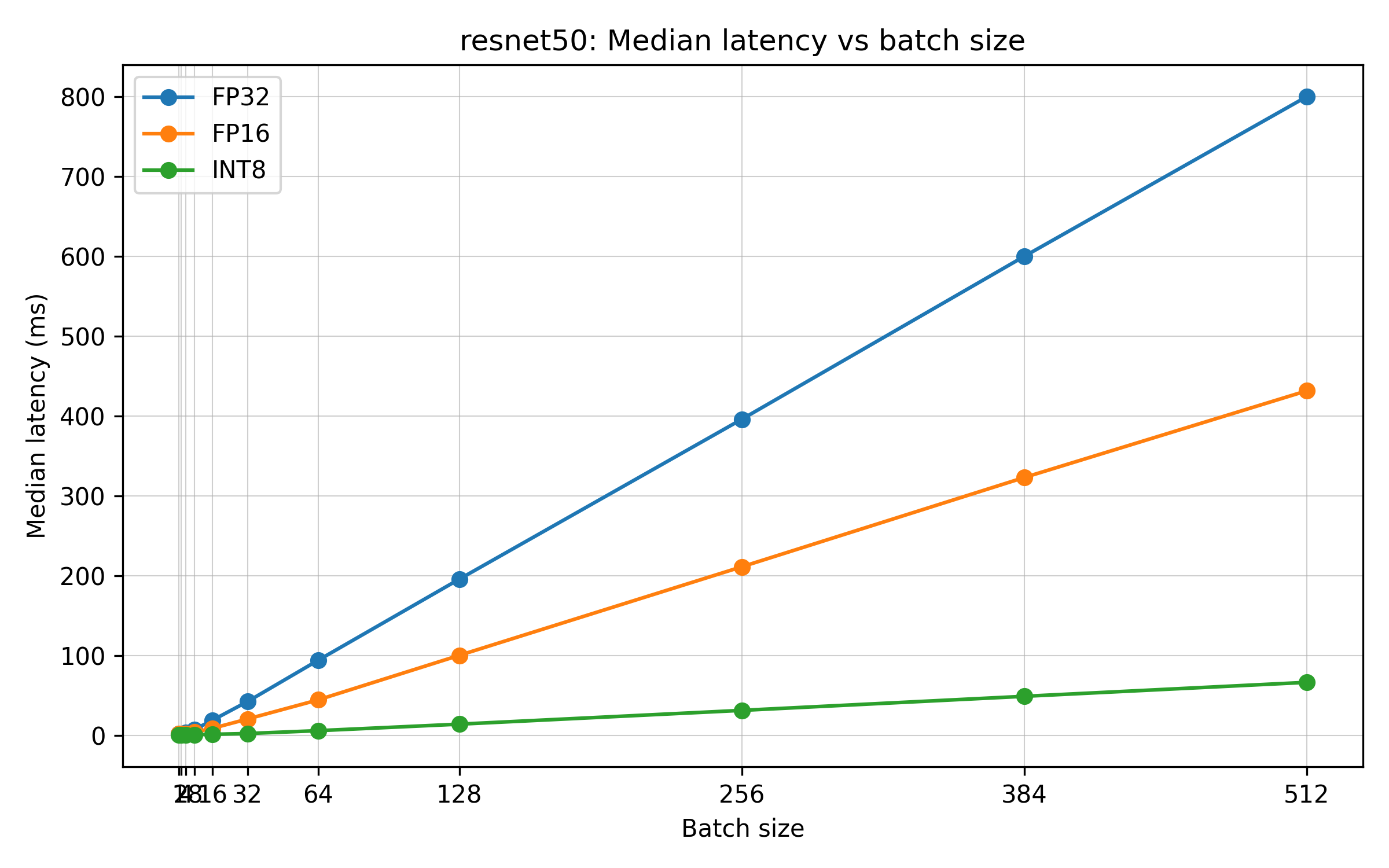}
        \caption{ResNet50}
        \label{fig:r50_l4_latency}
    \end{subfigure}
    \hfill
    \begin{subfigure}[b]{0.32\textwidth}
        \centering
        \includegraphics[width=\textwidth]{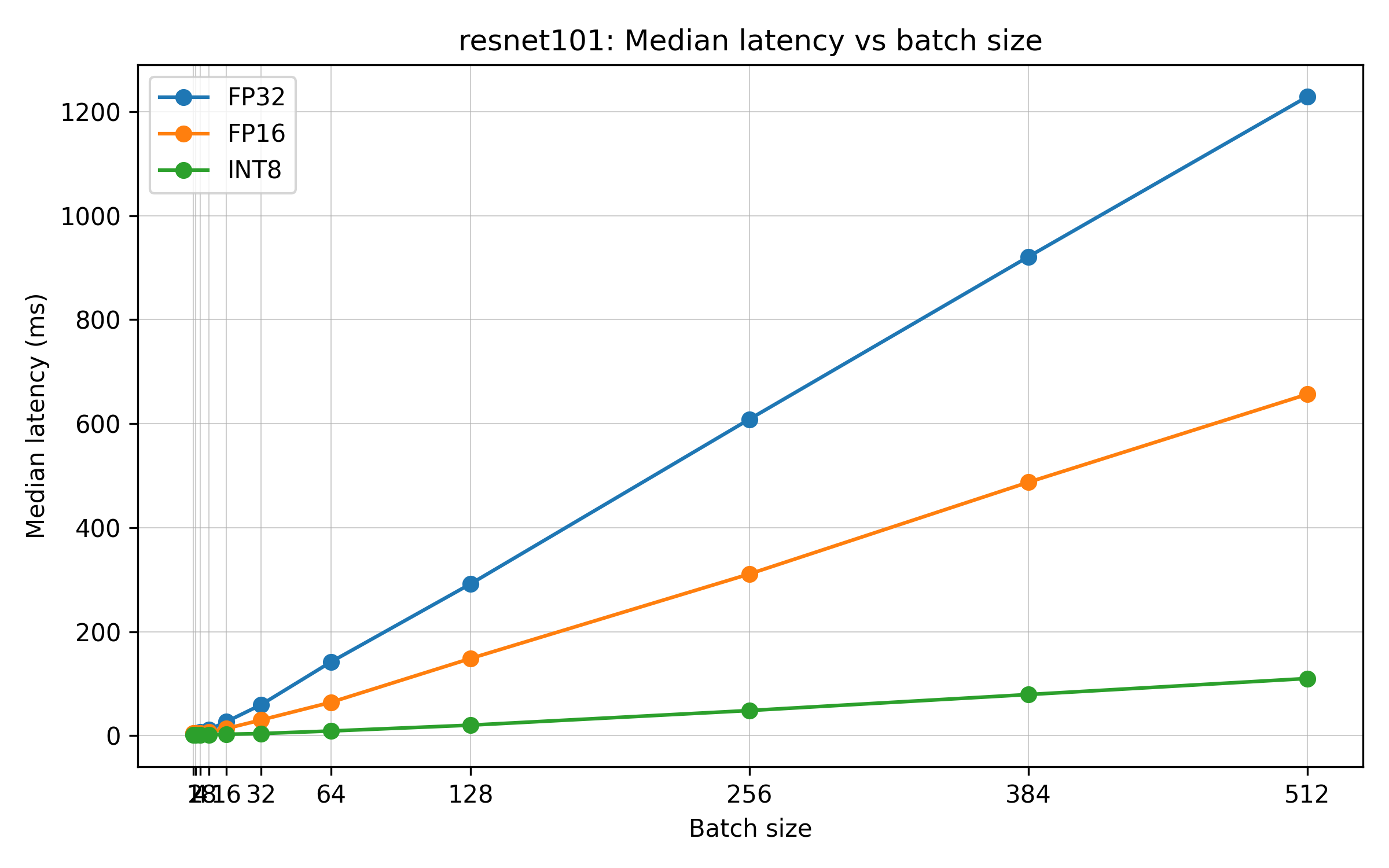}
        \caption{ResNet101}
        \label{fig:r101_l4_latency}
    \end{subfigure}

    \caption{
    Median latency scaling across batch sizes for ResNet18, ResNet50, and ResNet101.
    Top row shows NVIDIA T4 results and bottom row shows NVIDIA L4 results.
    Latency increases with batch size due to larger workload processing per inference call.
    Reduced precision modes (FP16 and INT8) significantly reduce latency due to
    lower computational cost and improved Tensor Core utilization.
    L4 consistently demonstrates lower latency compared to T4 across all models.
    }

    \label{fig:latency_t4_vs_l4}

\end{figure*}

Several clear scaling patterns emerge from the experimental results. First, the NVIDIA T4 requires relatively larger batch sizes ($B \geq 256$) to reach peak throughput, indicating that its Tensor Core resources require higher levels of data parallelism to achieve full utilization. This behavior suggests that the T4 architecture benefits most in throughput-oriented scenarios where large batch sizes can be sustained.

In contrast, the NVIDIA L4 achieves peak throughput at significantly smaller batch sizes ($B=16$--$32$), demonstrating improved architectural efficiency and more effective utilization of parallel execution units. This indicates that the L4 architecture is capable of reaching high compute efficiency even under moderate batching conditions, which is advantageous for latency-sensitive inference workloads.

Across all evaluated models, INT8 precision consistently delivers the highest throughput, highlighting the strong impact of reduced numerical precision on arithmetic intensity, memory bandwidth utilization, and Tensor Core efficiency. Reduced precision enables more operations to be executed per cycle, allowing both architectures to achieve substantially higher inference throughput compared to FP32 and FP16 modes.

As model complexity increases, peak throughput decreases due to higher computational demand and increased memory movement requirements. For example, L4 INT8 throughput decreases from 38,932 images/sec for ResNet18 to 8,026 images/sec for ResNet101, reflecting the additional arithmetic workload and deeper network structure associated with larger models.

Overall, the results demonstrate that batch size and numerical precision strongly influence GPU inference efficiency. Modern architectures such as L4 achieve high throughput at smaller batch sizes, making them well suited for latency-sensitive datacenter workloads where very large batching levels may not be practical.

\subsection{Latency Results}

Figure~\ref{fig:latency_t4_vs_l4} shows median latency scaling across batch sizes
for both NVIDIA T4 and NVIDIA L4 GPUs across FP32, FP16, and INT8 precision modes.

Across all models, latency increases approximately linearly with batch size,
reflecting the increased computational workload required to process larger input batches.
This behavior is expected because GPU kernels process more data per inference call,
increasing execution time even though throughput improves.

\begin{figure*}[t]
    \centering


    \begin{subfigure}[b]{0.32\textwidth}
        \centering
        \includegraphics[width=\textwidth]{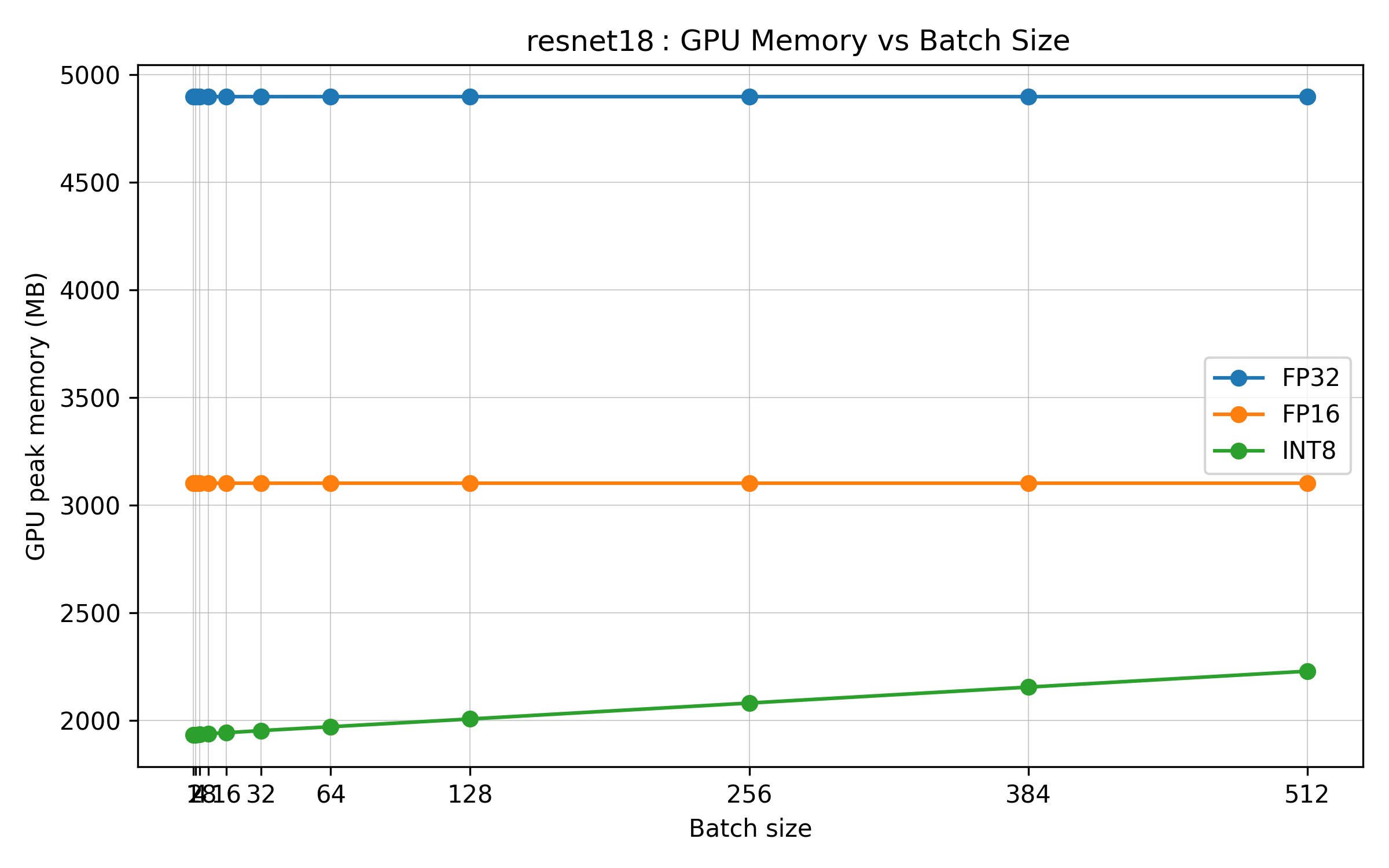}
        \caption{ResNet18}
        \label{fig:r18_t4_mem}
    \end{subfigure}
    \hfill
    \begin{subfigure}[b]{0.32\textwidth}
        \centering
        \includegraphics[width=\textwidth]{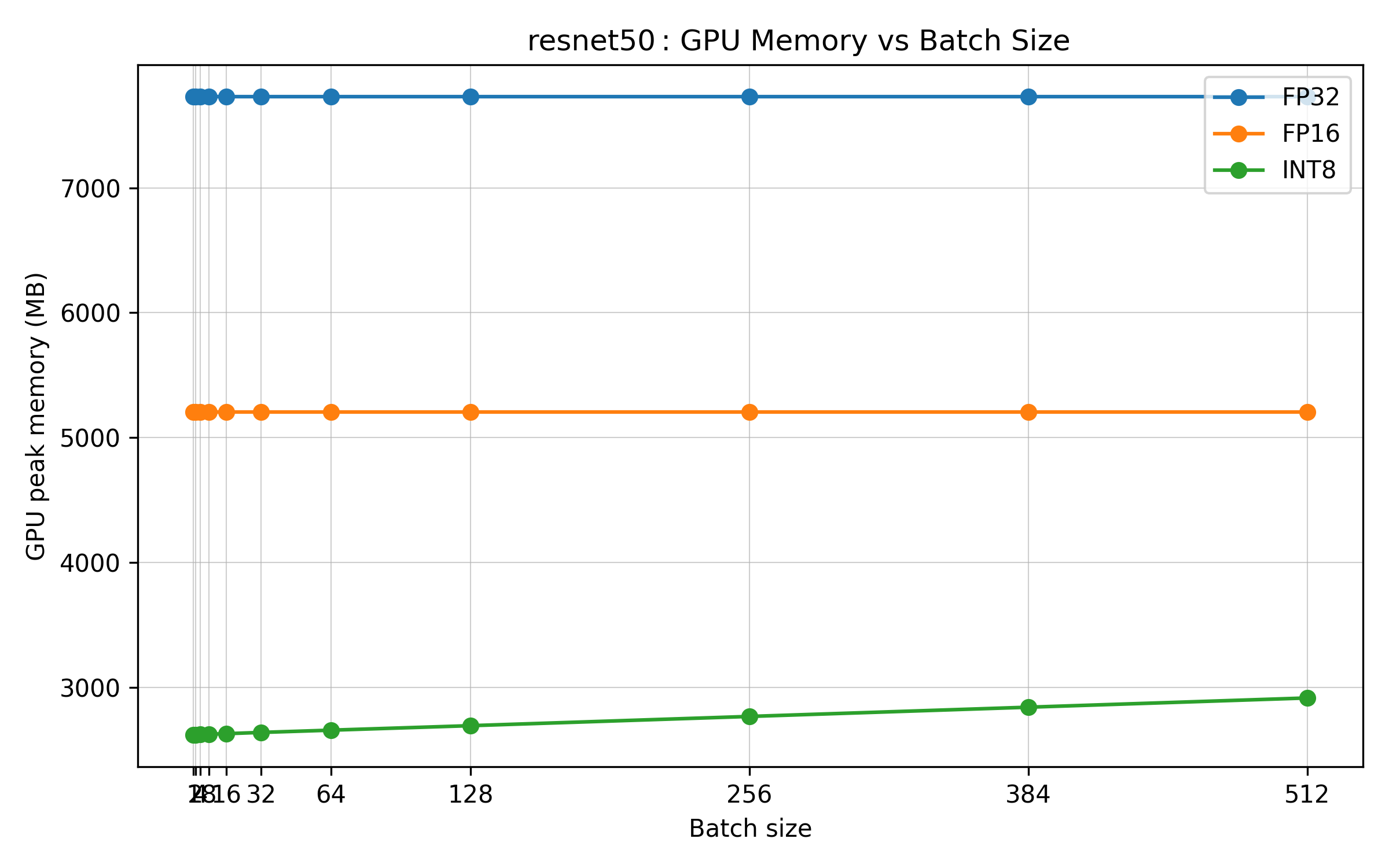}
        \caption{ResNet50}
        \label{fig:r50_t4_mem}
    \end{subfigure}
    \hfill
    \begin{subfigure}[b]{0.32\textwidth}
        \centering
        \includegraphics[width=\textwidth]{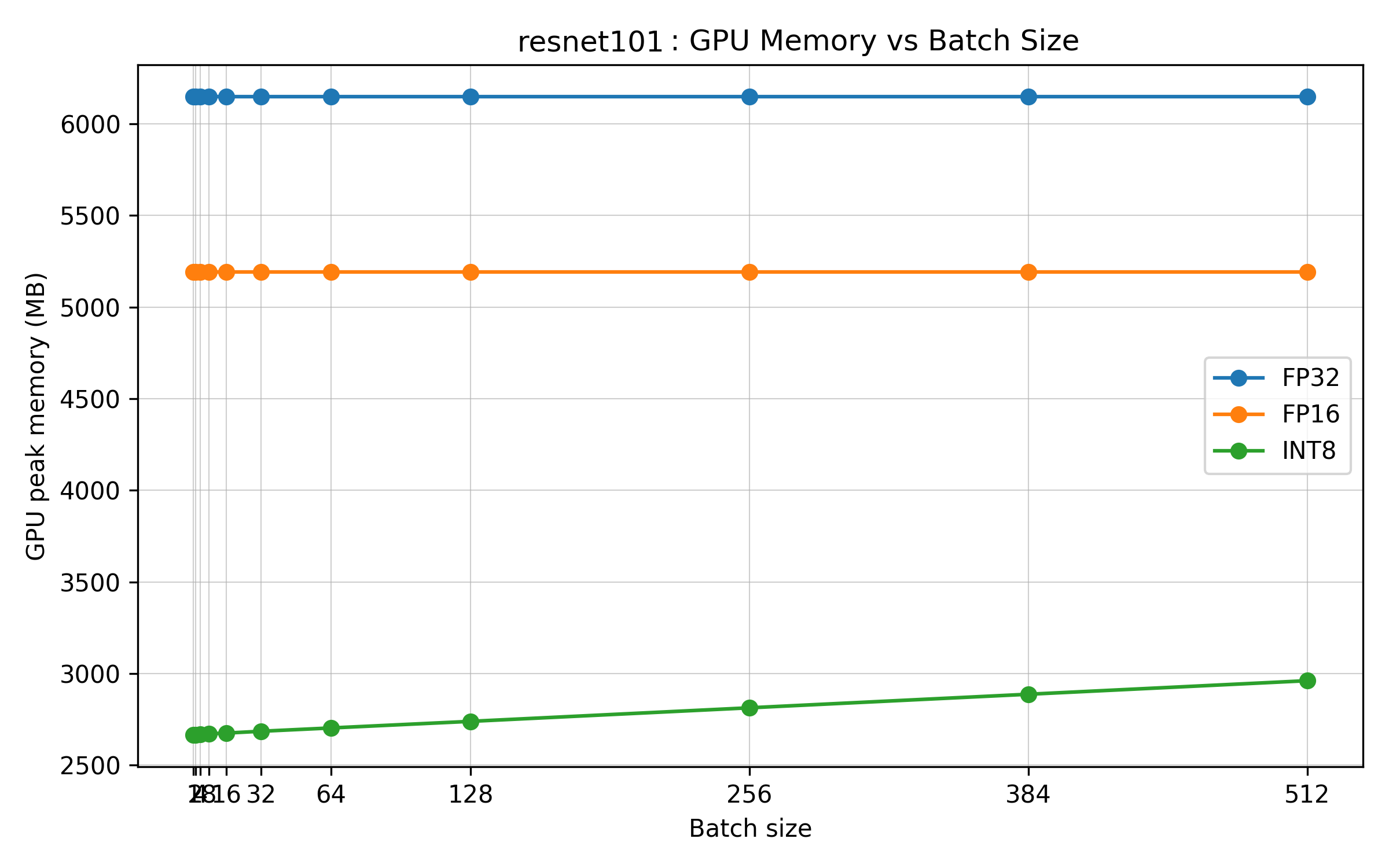}
        \caption{ResNet101}
        \label{fig:r101_t4_mem}
    \end{subfigure}

    \vspace{12pt}


    \begin{subfigure}[b]{0.32\textwidth}
        \centering
        \includegraphics[width=\textwidth]{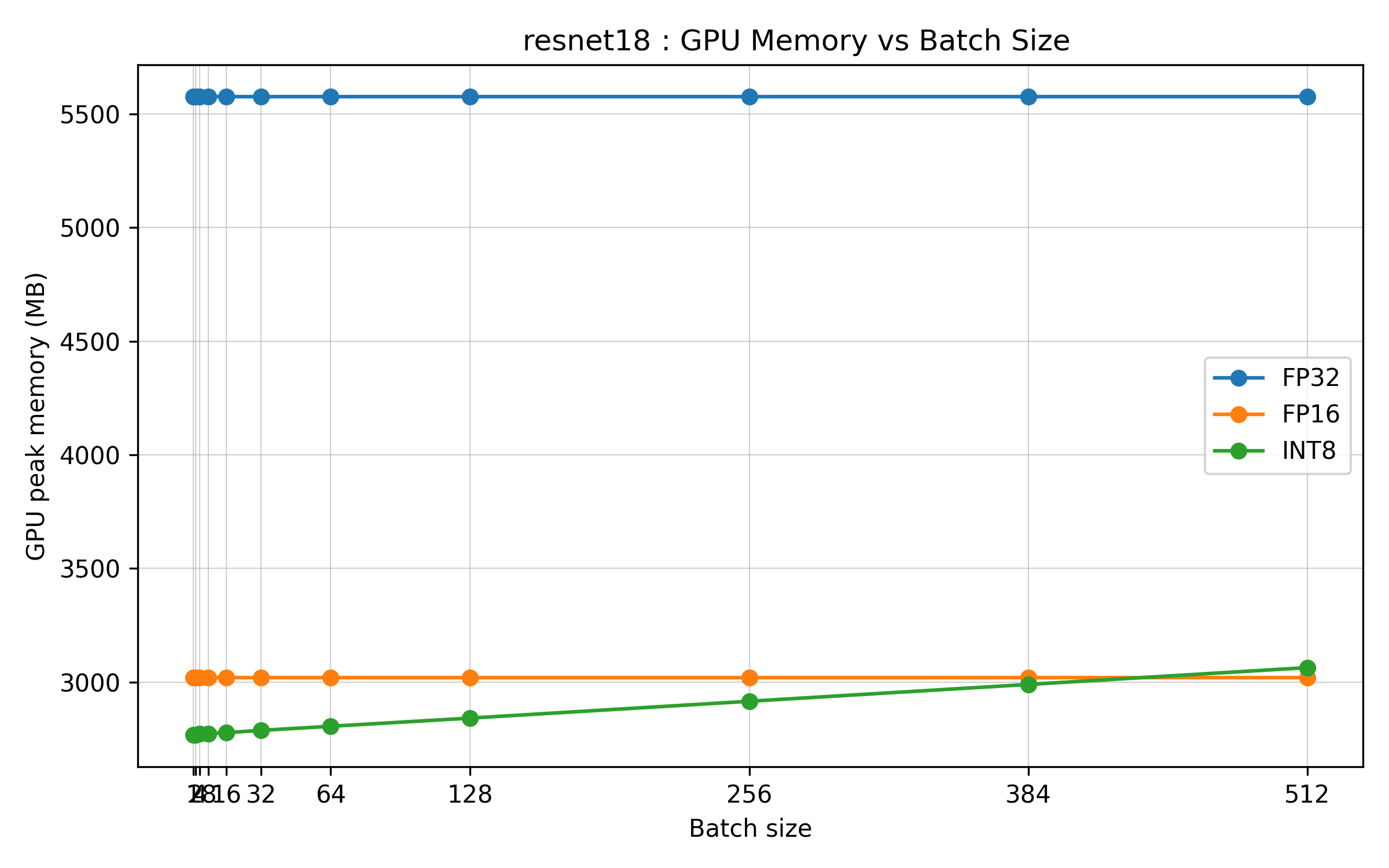}
        \caption{ResNet18}
        \label{fig:r18_l4_mem}
    \end{subfigure}
    \hfill
    \begin{subfigure}[b]{0.32\textwidth}
        \centering
        \includegraphics[width=\textwidth]{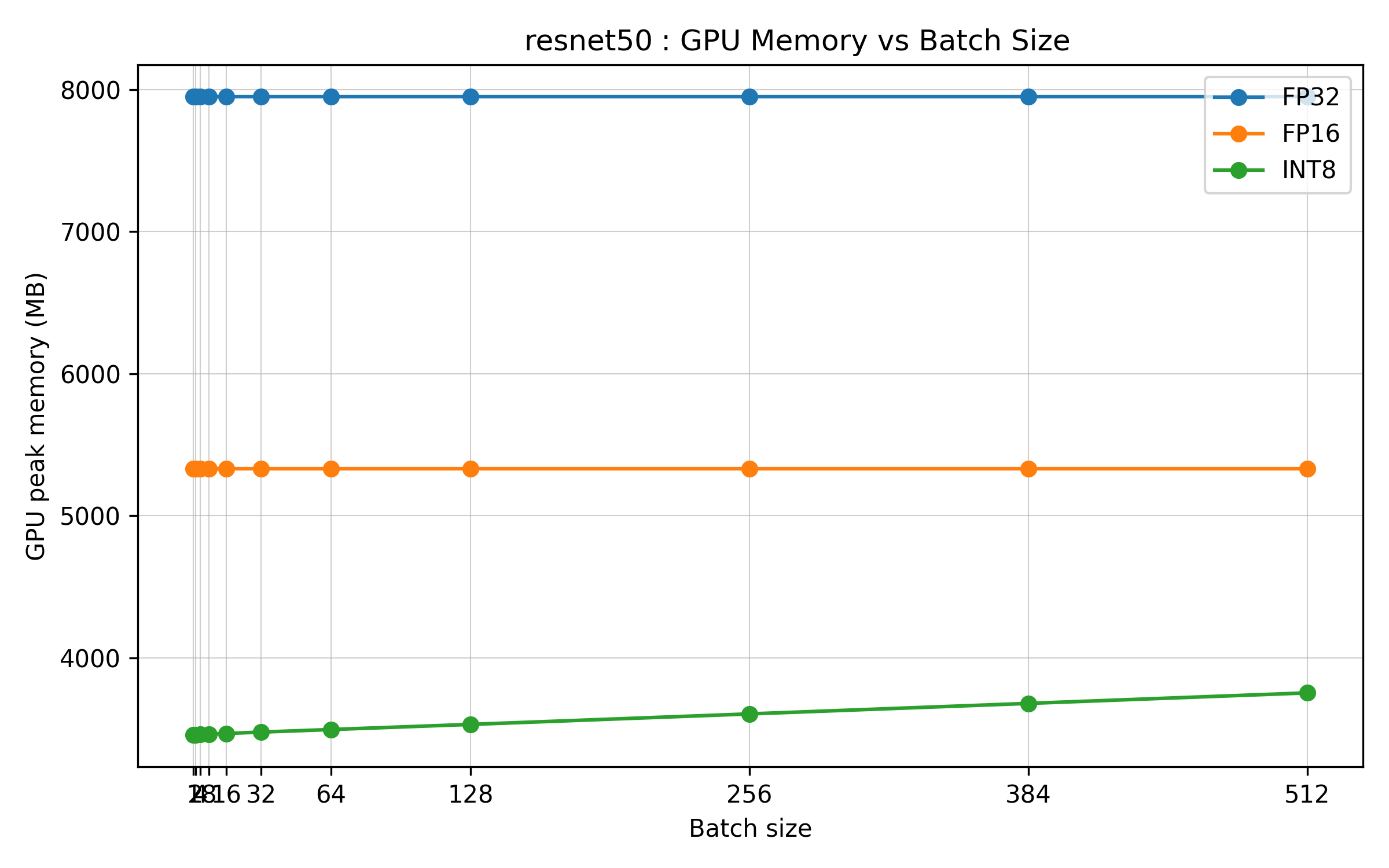}
        \caption{ResNet50}
        \label{fig:r50_l4_mem}
    \end{subfigure}
    \hfill
    \begin{subfigure}[b]{0.32\textwidth}
        \centering
        \includegraphics[width=\textwidth]{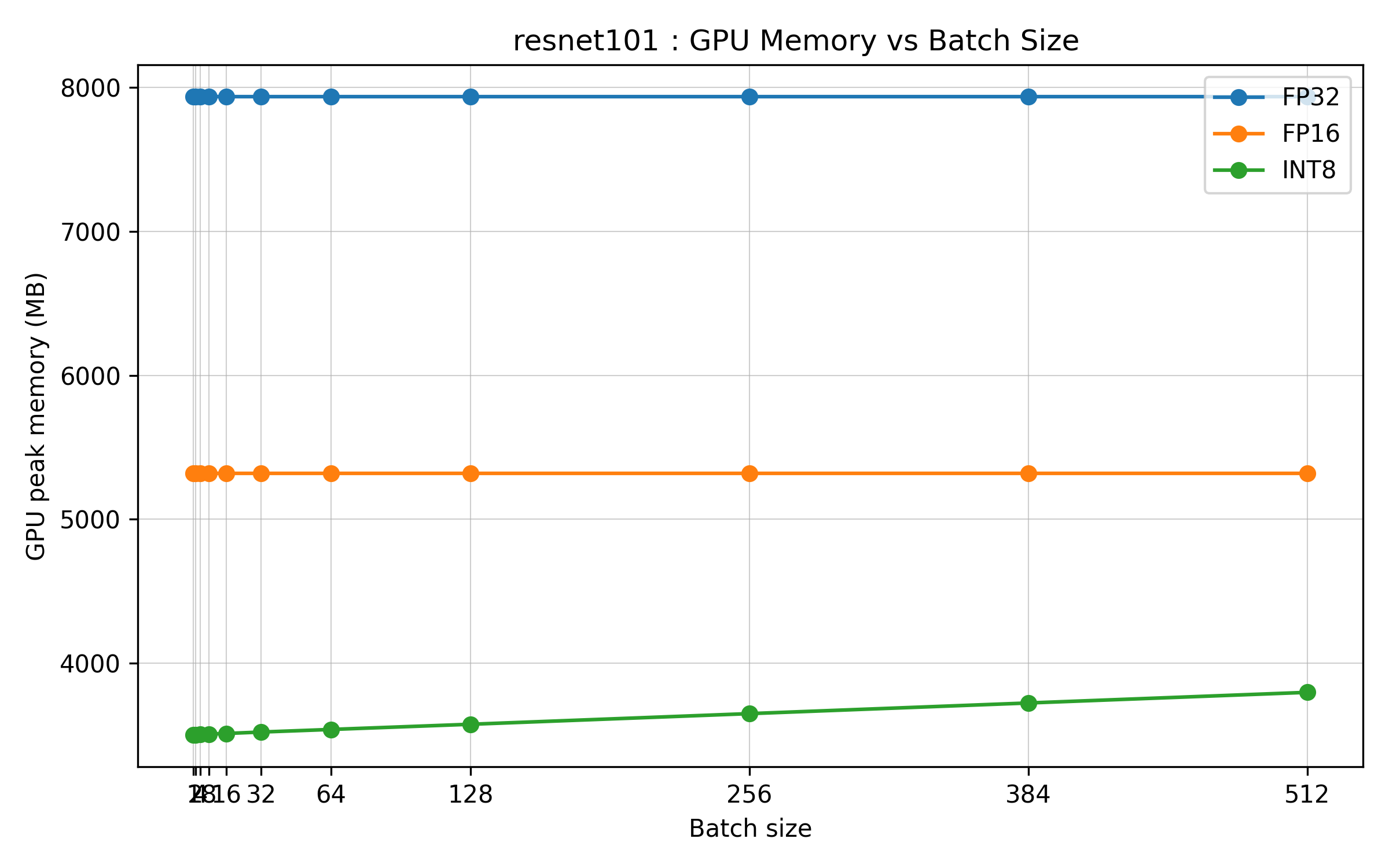}
        \caption{ResNet101}
        \label{fig:r101_l4_mem}
    \end{subfigure}

    \caption{
    GPU memory usage measured via NVML across batch sizes.
    Top row shows NVIDIA T4 results, bottom row shows NVIDIA L4 results.
    FP32 requires the highest memory allocation due to larger tensor representation,
    followed by FP16 and INT8.
    Memory usage remains relatively stable across batch sizes,
    indicating that model weights dominate memory consumption,
    while activation memory increases gradually with batch size.
    L4 shows slightly higher baseline allocation due to architectural differences
    and runtime memory management behavior.
    }

    \label{fig:nvml_memory_t4_vs_l4}

\end{figure*}
\textbf{T4 latency behavior.}
On NVIDIA T4, FP32 latency increases significantly as batch size grows.
For ResNet18, median latency increases from approximately 6 ms at $B=1$
to over 400 ms at $B=512$.
ResNet50 latency increases from roughly 10 ms at $B=1$ to approximately 1400 ms at $B=512$,
while ResNet101 increases from approximately 18 ms to over 2400 ms across the same range.

Reduced precision significantly lowers latency.
Using FP16 precision, ResNet18 latency at $B=512$ decreases to approximately 200 ms,
representing roughly a 2$\times$ improvement over FP32.
Similarly, ResNet50 FP16 latency reaches approximately 600 ms at $B=512$,
while ResNet101 reaches approximately 1000 ms.

INT8 TensorRT demonstrates substantially lower latency across all batch sizes.
For example, ResNet18 INT8 latency at $B=512$ is approximately 60 ms,
while ResNet50 reaches roughly 110 ms and ResNet101 approximately 170 ms.
These results indicate that reduced numerical precision significantly improves
execution efficiency by reducing memory movement and increasing Tensor Core utilization.

\textbf{L4 latency behavior.}
NVIDIA L4 demonstrates consistently lower latency compared to T4 across all models
and precision modes.
For ResNet18, FP32 latency increases from approximately 4 ms at $B=1$
to approximately 220 ms at $B=512$, representing roughly a 1.8$\times$ reduction
compared to T4 FP32 latency at large batch sizes.

FP16 further improves response time efficiency.
ResNet18 FP16 latency reaches approximately 130 ms at $B=512$,
while ResNet50 reaches approximately 430 ms and ResNet101 approximately 660 ms.

INT8 provides the lowest latency across all configurations.
For ResNet18, INT8 latency remains below 20 ms even at $B=512$,
demonstrating strong execution efficiency.
ResNet50 INT8 latency reaches approximately 65 ms at $B=512$,
while ResNet101 remains near 105 ms.

Several important latency trends emerge from the experimental results. Latency increases approximately linearly with batch size across all evaluated models, indicating predictable scaling behavior as inference workloads grow. This linear relationship suggests that the GPUs maintain stable execution characteristics even as the number of input samples processed simultaneously increases.

INT8 precision consistently provides the lowest latency across all models and batch sizes, demonstrating the efficiency of reduced precision execution on Tensor Core architectures. Lower numerical precision reduces arithmetic complexity and improves computational density, allowing faster kernel execution compared to FP32 and FP16 modes.

Across all tested configurations, the NVIDIA L4 achieves significantly lower latency than the NVIDIA T4, indicating improvements in kernel execution efficiency, memory bandwidth utilization, and architectural parallelism. These architectural enhancements enable the L4 to process inference workloads more efficiently while maintaining stable execution characteristics across different batch sizes.

As expected, larger models exhibit higher latency due to increased computational complexity and deeper network structures. For example, FP32 latency at $B=512$ increases from approximately 400 ms for ResNet18 to approximately 2400 ms for ResNet101 on T4, reflecting the greater arithmetic workload required for larger convolutional networks.

Overall, the results demonstrate that modern GPU architectures such as L4 improve both throughput and latency simultaneously, enabling higher request processing capacity without proportional increases in response time. Tail latency (P99) measurements show similar scaling trends and remain stable across batch sizes, indicating predictable execution behavior suitable for real-time inference deployment scenarios. These characteristics make L4 particularly well suited for latency-sensitive inference workloads commonly encountered in production datacenter environments.

\begin{figure*}[t]
    \centering


    \begin{subfigure}[b]{0.32\textwidth}
        \centering
        \includegraphics[width=\textwidth]{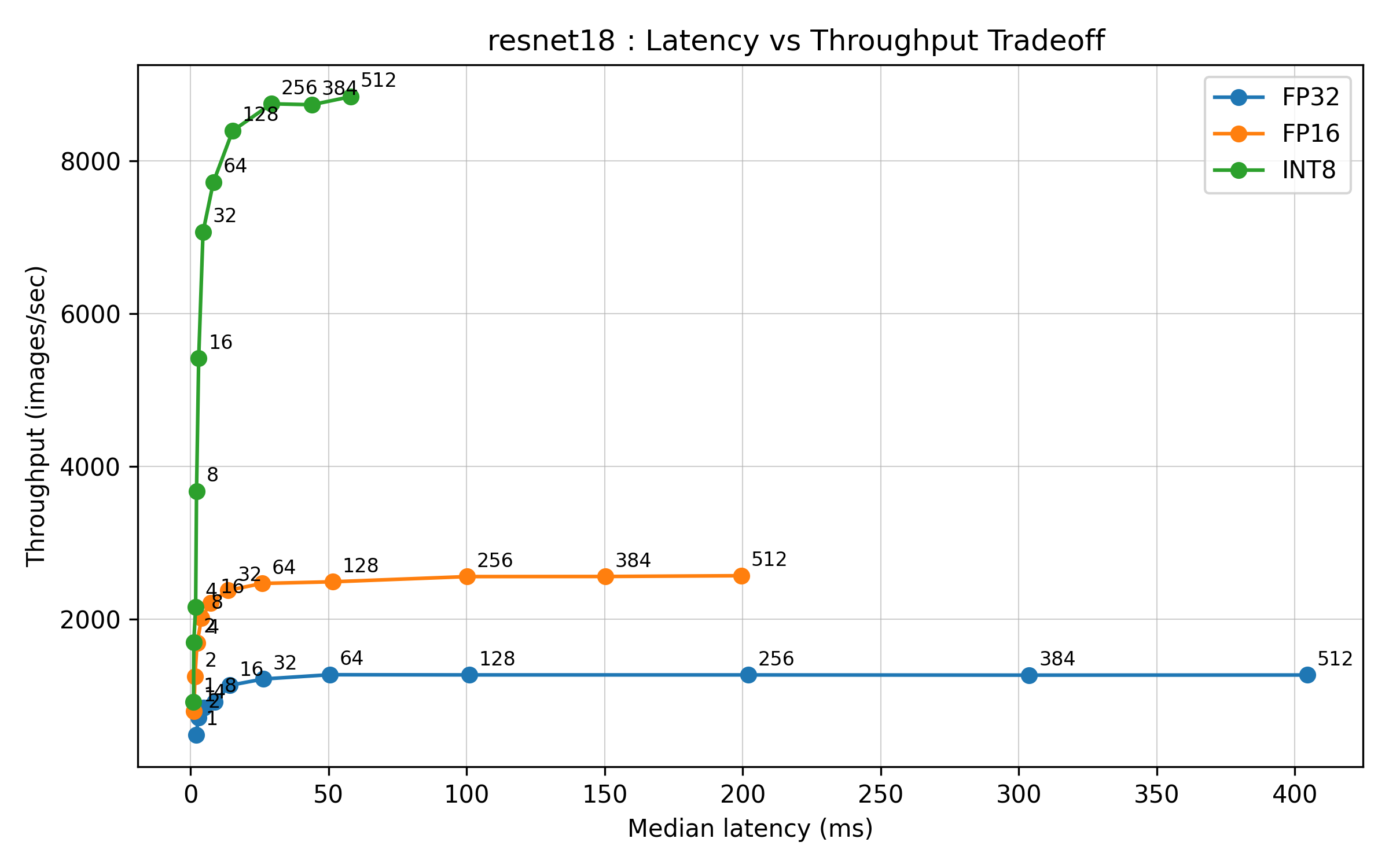}
        \caption{ResNet18}
        \label{fig:r18_t4_pareto}
    \end{subfigure}
    \hfill
    \begin{subfigure}[b]{0.32\textwidth}
        \centering
        \includegraphics[width=\textwidth]{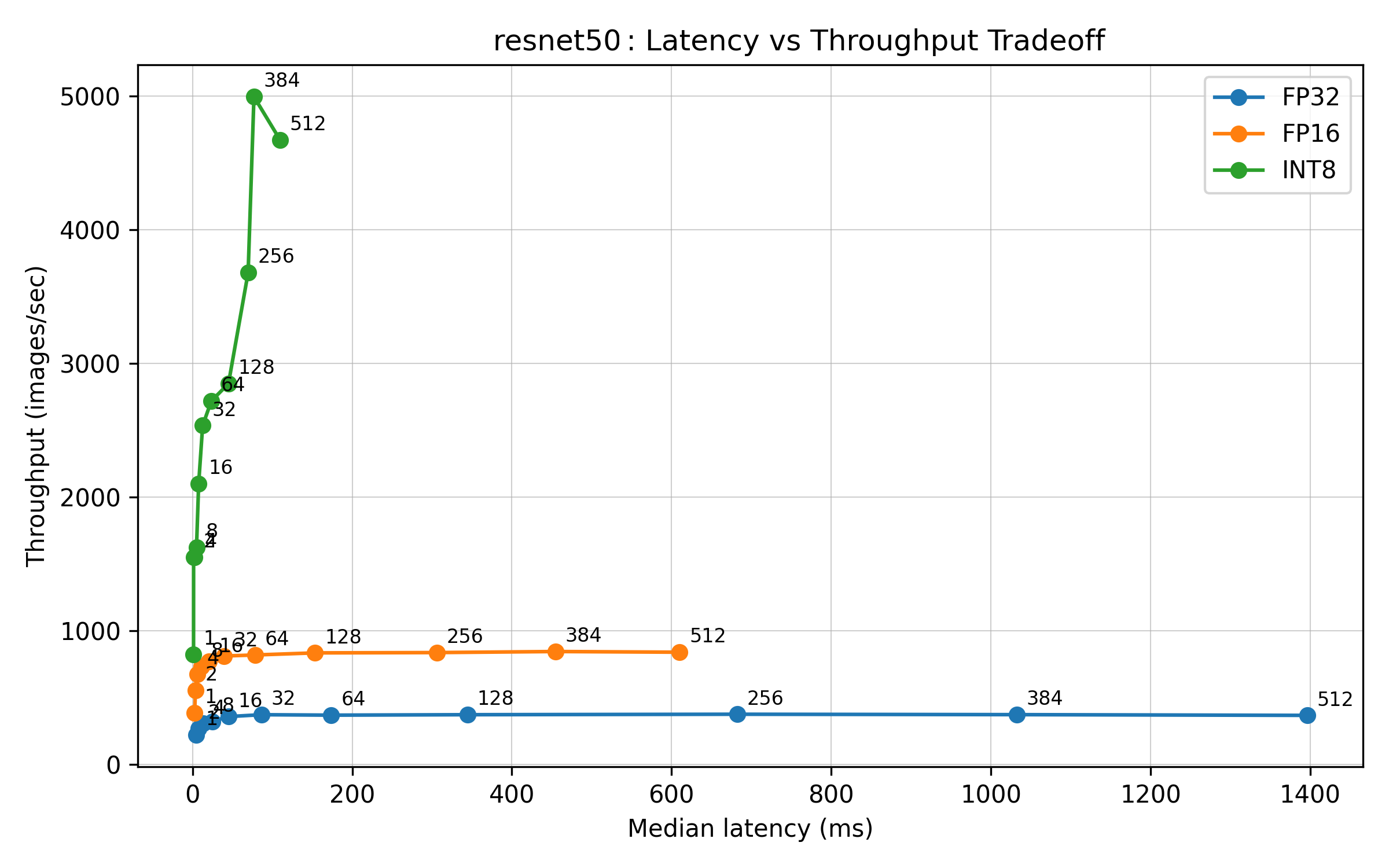}
        \caption{ResNet50}
        \label{fig:r50_t4_pareto}
    \end{subfigure}
    \hfill
    \begin{subfigure}[b]{0.32\textwidth}
        \centering
        \includegraphics[width=\textwidth]{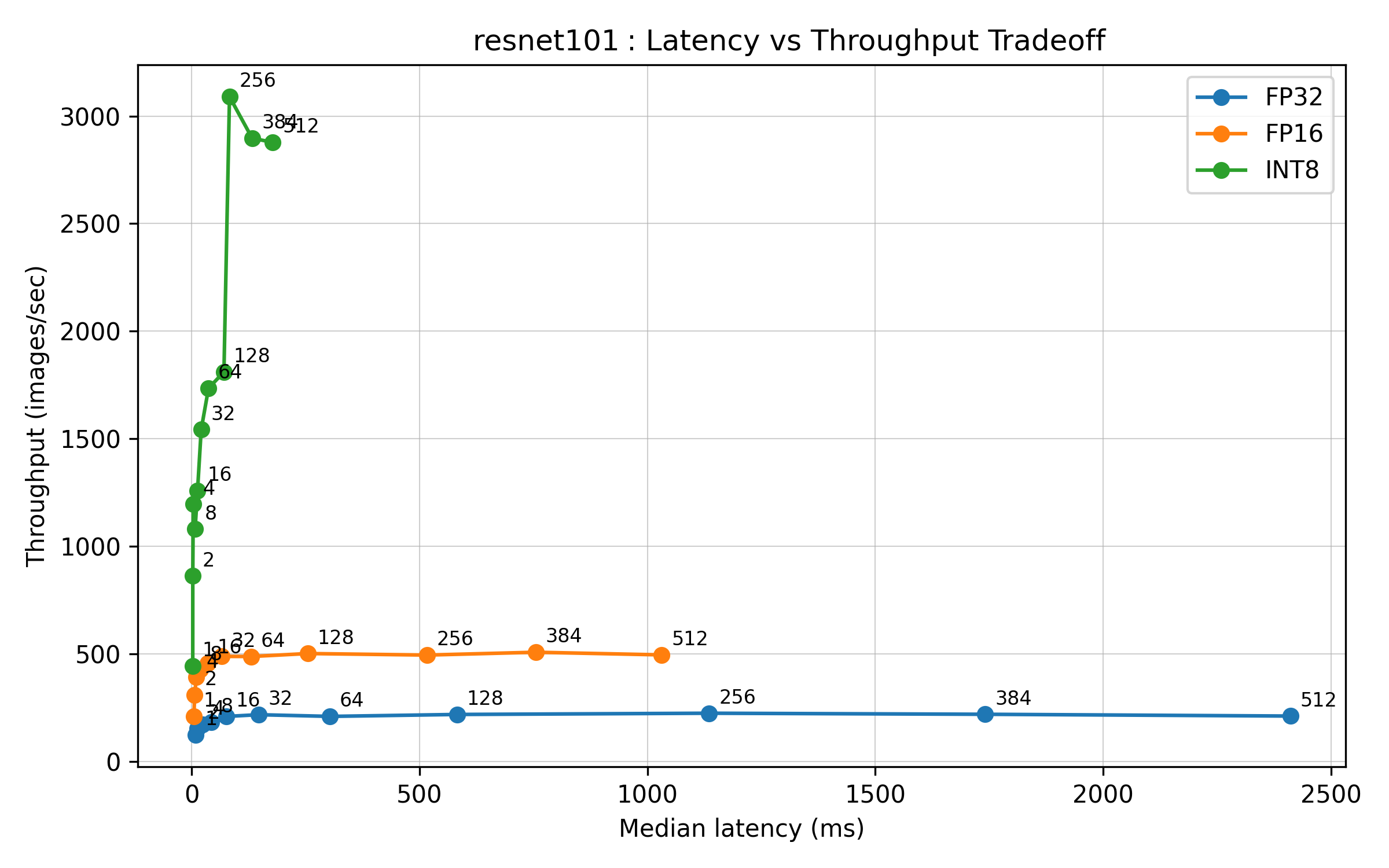}
        \caption{ResNet101}
        \label{fig:r101_t4_pareto}
    \end{subfigure}

    \vspace{12pt}


    \begin{subfigure}[b]{0.32\textwidth}
        \centering
        \includegraphics[width=\textwidth]{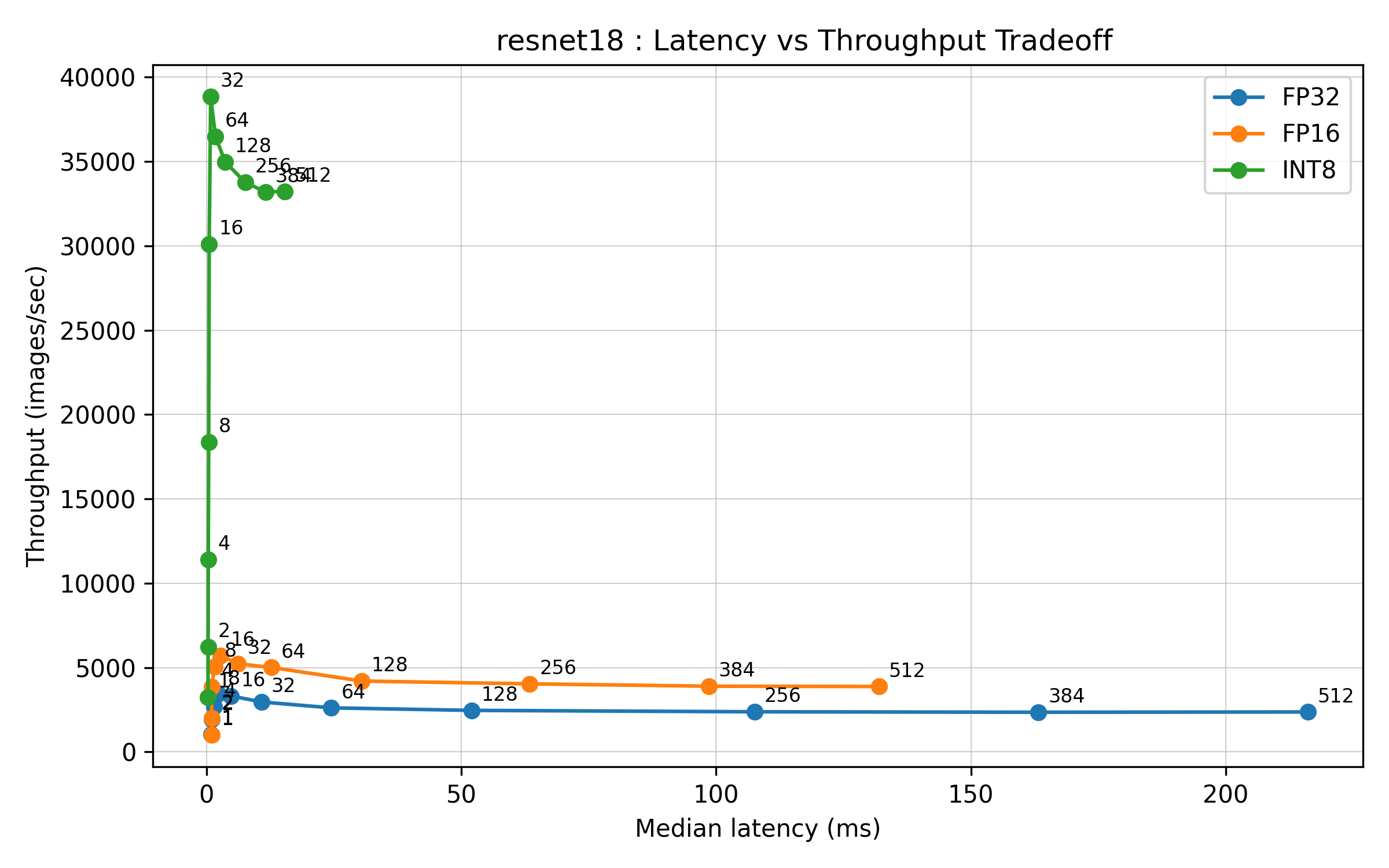}
        \caption{ResNet18}
        \label{fig:r18_l4_pareto}
    \end{subfigure}
    \hfill
    \begin{subfigure}[b]{0.32\textwidth}
        \centering
        \includegraphics[width=\textwidth]{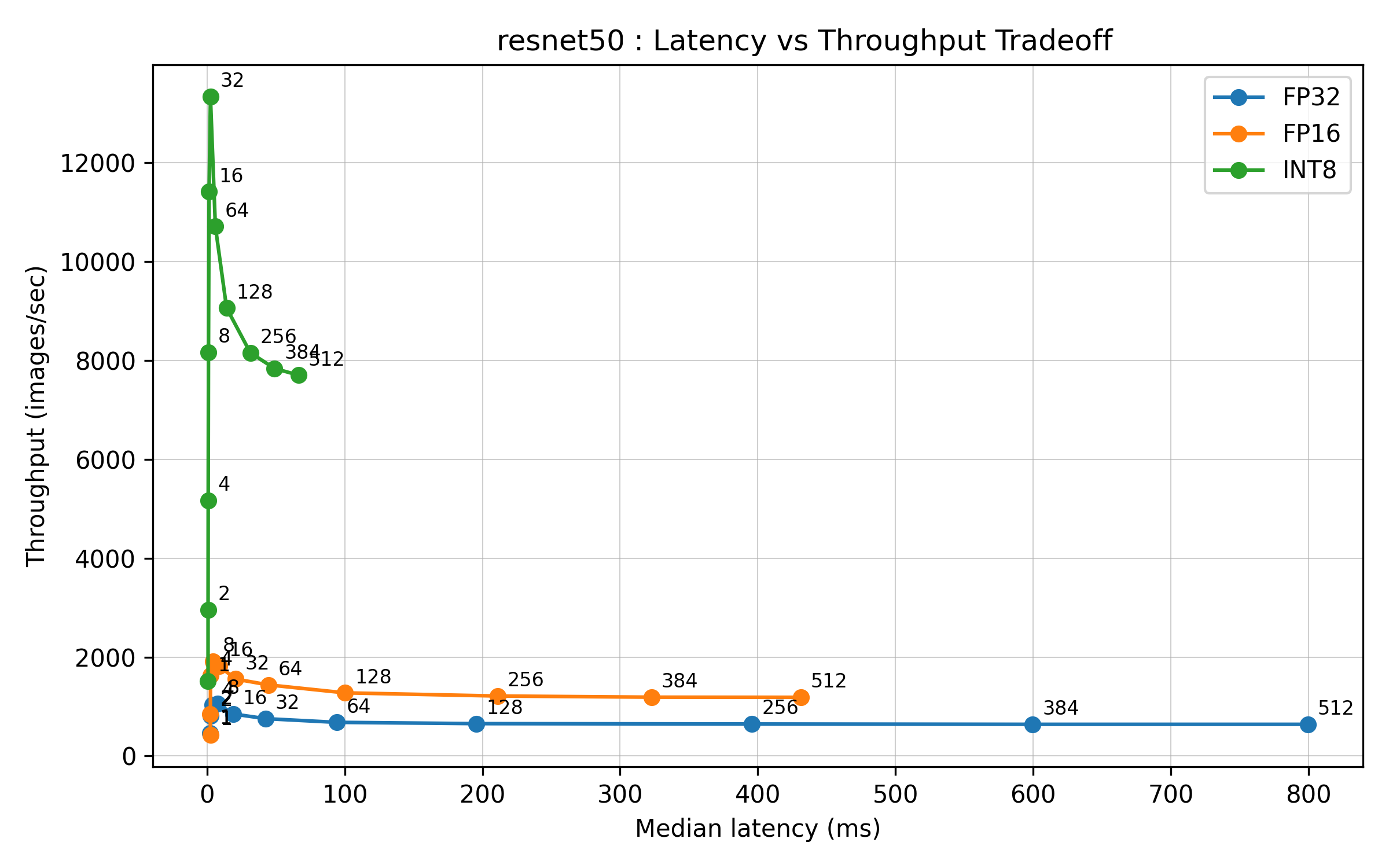}
        \caption{ResNet50}
        \label{fig:r50_l4_pareto}
    \end{subfigure}
    \hfill
    \begin{subfigure}[b]{0.32\textwidth}
        \centering
        \includegraphics[width=\textwidth]{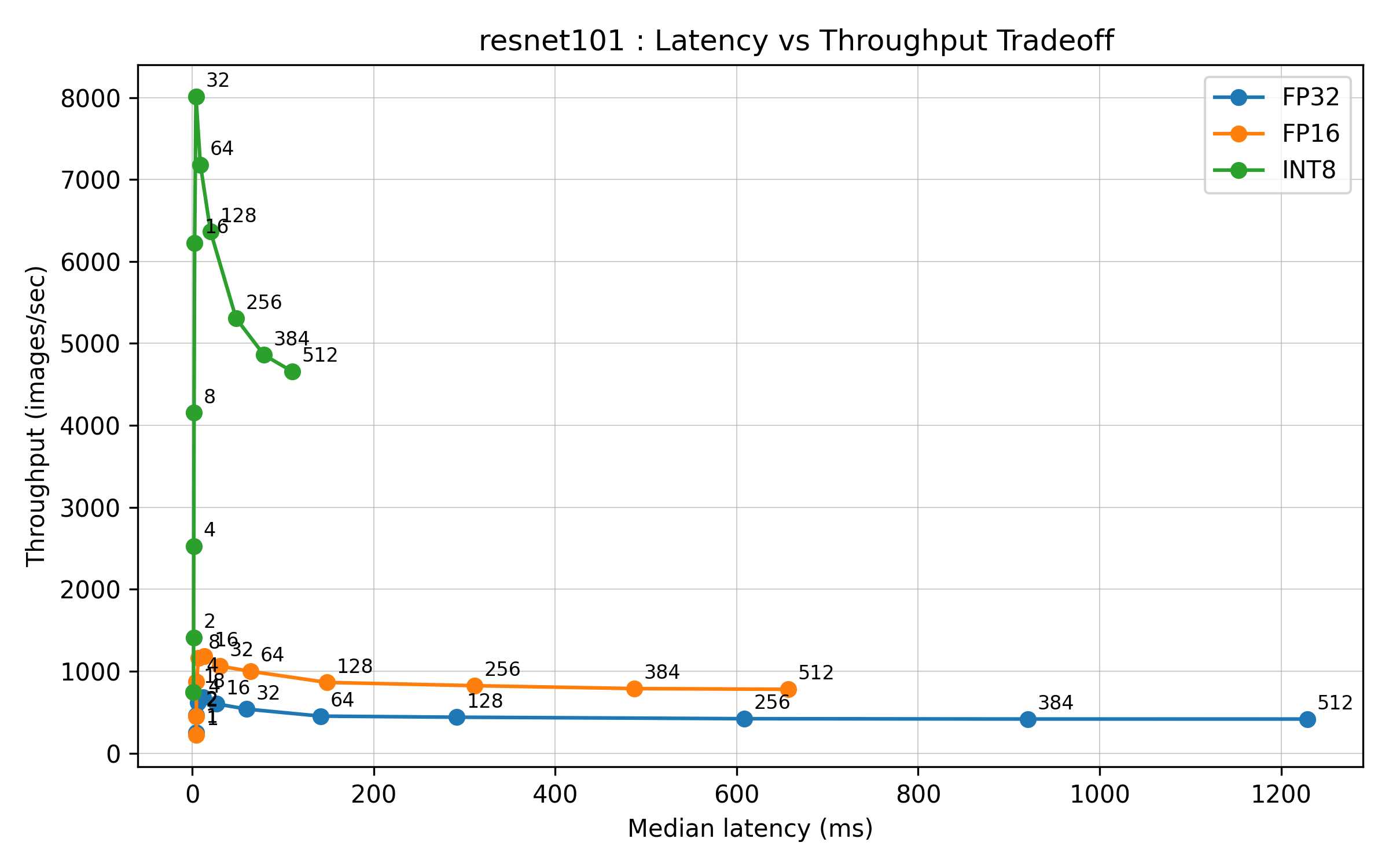}
        \caption{ResNet101}
        \label{fig:r101_l4_pareto}
    \end{subfigure}

	\caption{
	Latency vs throughput Pareto tradeoff across batch sizes.
	Top row shows NVIDIA T4 results and bottom row shows NVIDIA L4 results.
	Each point represents a batch size configuration.
	Moving right increases latency while moving up increases throughput.
	INT8 achieves the best throughput--latency tradeoff, followed by FP16 and FP32.
	L4 consistently dominates T4 across most operating regions,
	providing higher throughput for comparable latency levels.
	}
    \label{fig:pareto_t4_vs_l4}

\end{figure*}

\subsection{Tail Latency Behavior Across T4 and L4}

Tail latency, measured using P99 response time, captures worst-case inference performance and provides insight into system predictability under varying workload sizes. Across both NVIDIA T4 and NVIDIA L4 platforms, P99 latency increases with batch size for all evaluated models (ResNet18, ResNet50, and ResNet101) and precision modes (FP32, FP16, and INT8). This increase reflects the larger activation working set and longer kernel execution times associated with higher batch sizes.

Across both GPUs, a consistent ordering of precision modes is observed. FP32 exhibits the highest P99 latency, FP16 provides a substantial reduction, and INT8 consistently achieves the lowest tail latency. For example, at larger batch sizes (B=512), ResNet18 P99 latency on T4 reaches approximately 400 ms in FP32, compared to roughly 220 ms on L4, representing an improvement of about 45\%. Under FP16 precision, ResNet18 P99 latency decreases from roughly 200 ms on T4 to approximately 130 ms on L4, corresponding to an improvement of about 35\%. INT8 provides the lowest tail latency on both platforms, with L4 achieving approximately 15--20 ms at large batch sizes compared to roughly 50--60 ms on T4, representing more than a 60\% reduction.

For ResNet50, the numerical differences between architectures become more pronounced as model complexity increases. At B=512, FP32 P99 latency on T4 approaches approximately 1350 ms, whereas L4 achieves roughly 800 ms, corresponding to an improvement of about 40\%. FP16 shows a similar trend, with T4 reaching approximately 600 ms compared to about 430 ms on L4. INT8 again demonstrates the strongest improvement, with T4 reaching approximately 160--170 ms compared to roughly 60--70 ms on L4.

For the deepest evaluated model, ResNet101, tail latency differences further widen due to increased arithmetic intensity. At larger batch sizes, FP32 P99 latency on T4 exceeds approximately 2200 ms, while L4 achieves roughly 1200 ms, representing an improvement of about 45\%. FP16 latency decreases from approximately 1000 ms on T4 to around 650 ms on L4. INT8 continues to show the lowest tail latency, with T4 reaching approximately 180--200 ms compared to roughly 100--110 ms on L4.

Overall, L4 demonstrates consistently lower P99 latency across all evaluated workloads and precision modes relative to T4. Observed reductions typically range from 35\% to 50\% for FP32 and FP16 precision modes, and from approximately 40\% to 65\% for INT8. These improvements reflect architectural enhancements in the L4 GPU, including increased compute throughput, improved Tensor Core efficiency, higher memory bandwidth, and larger cache capacity. Reduced precision further improves tail-latency stability by lowering computational cost per inference and reducing variability in execution time. INT8 consistently provides the strongest tail-latency performance across both platforms, making it particularly well suited for latency-sensitive inference deployments requiring predictable response times and tight service-level objectives.

\subsection{Latency Comparison Across Precision Modes}

Figure~\ref{fig:latency_summary_t4_l4} summarizes median latency across
models and precision modes for both T4 and L4 GPUs.

Across all models, INT8 achieves the lowest latency due to reduced numerical precision
and optimized Tensor Core execution. FP16 provides intermediate latency,
while FP32 shows the highest latency due to increased computational cost.

Latency increases with model depth, with ResNet101 showing the highest inference time.

Comparing architectures, L4 demonstrates consistently lower latency than T4,
indicating improved compute efficiency and memory subsystem performance.

\begin{figure*}[t]
    \centering


    \begin{subfigure}[b]{0.32\textwidth}
        \centering
        \includegraphics[width=\textwidth]{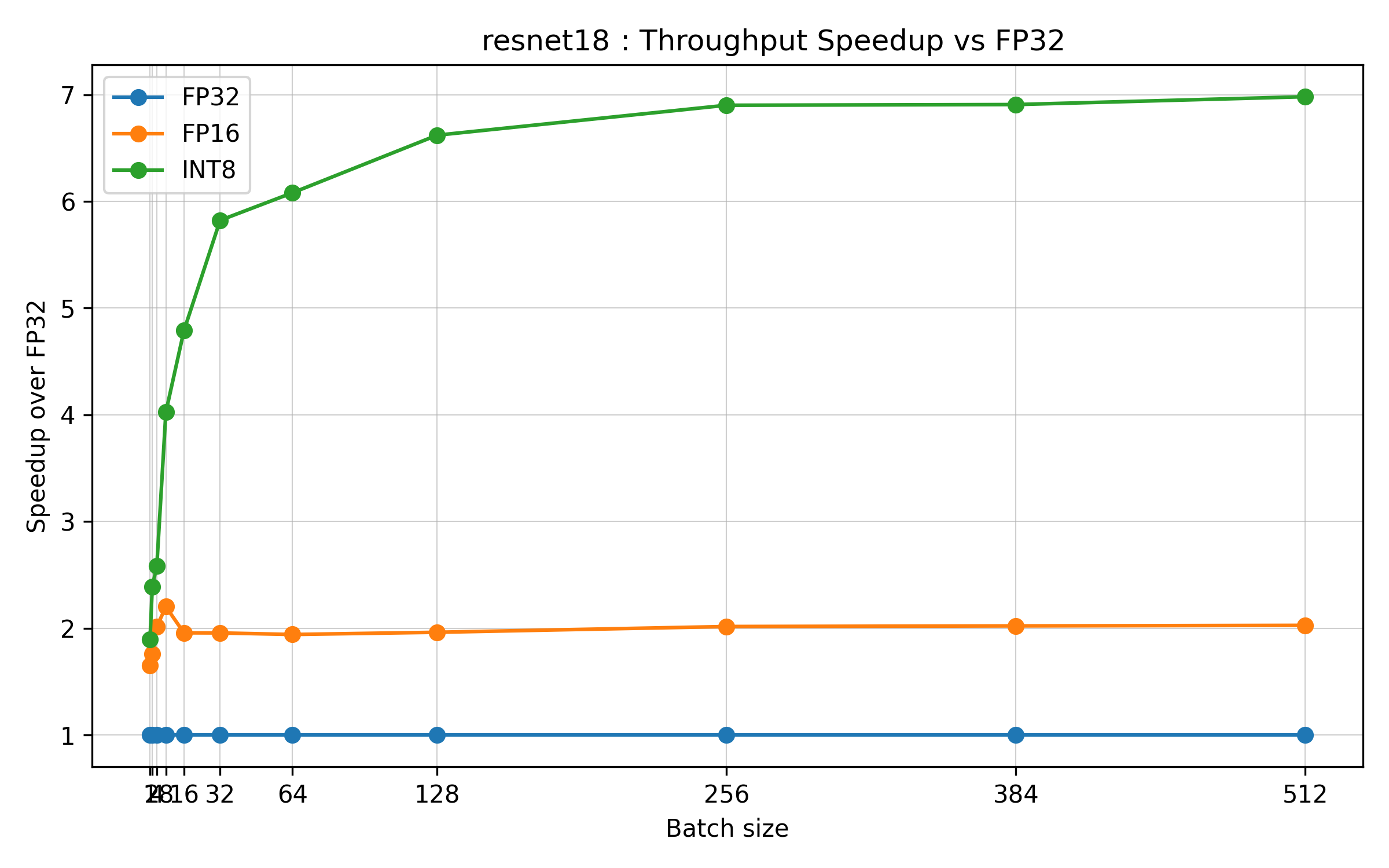}
        \caption{ResNet18}
        \label{fig:r18_t4_speedup}
    \end{subfigure}
    \hfill
    \begin{subfigure}[b]{0.32\textwidth}
        \centering
        \includegraphics[width=\textwidth]{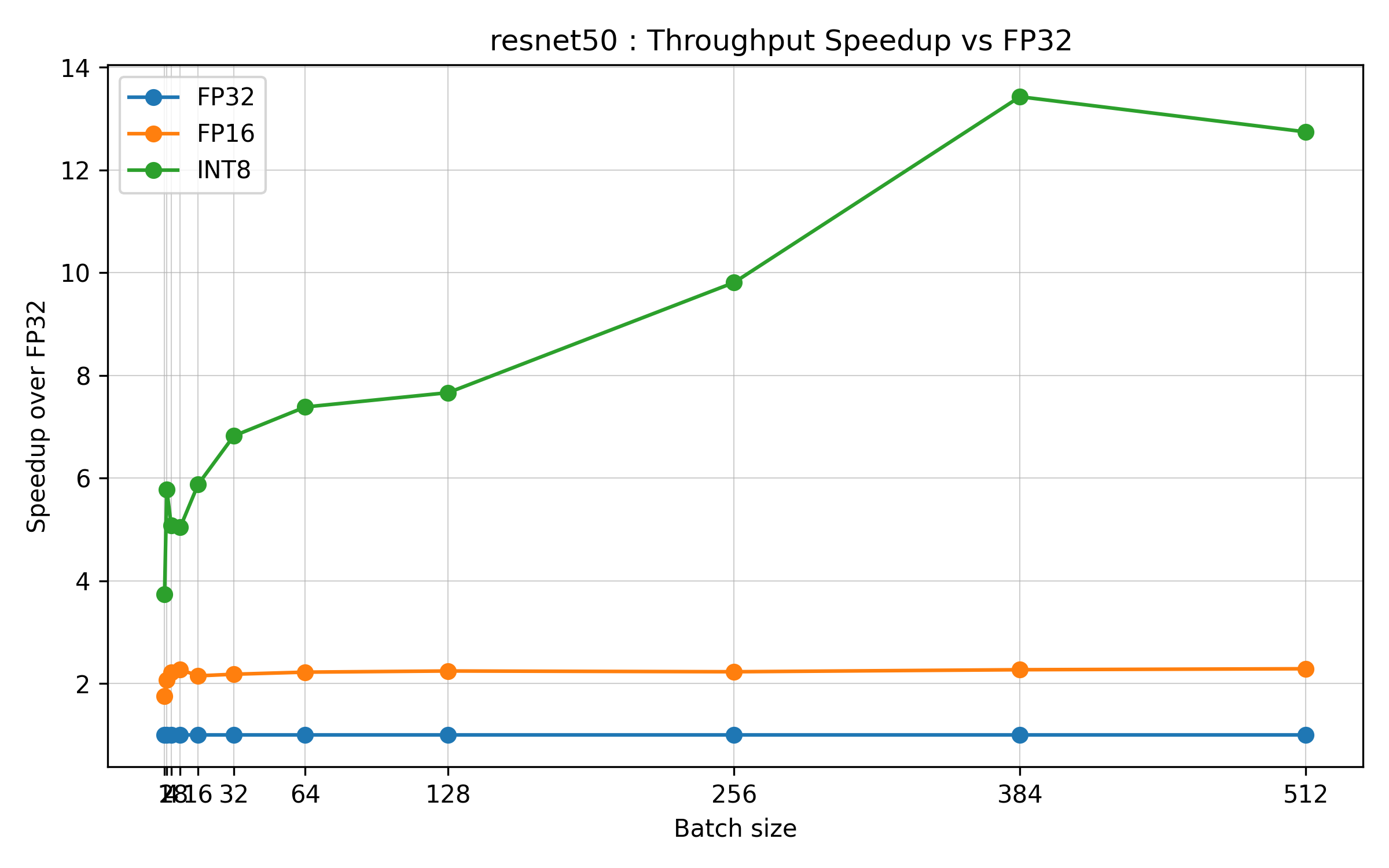}
        \caption{ResNet50}
        \label{fig:r50_t4_speedup}
    \end{subfigure}
    \hfill
    \begin{subfigure}[b]{0.32\textwidth}
        \centering
        \includegraphics[width=\textwidth]{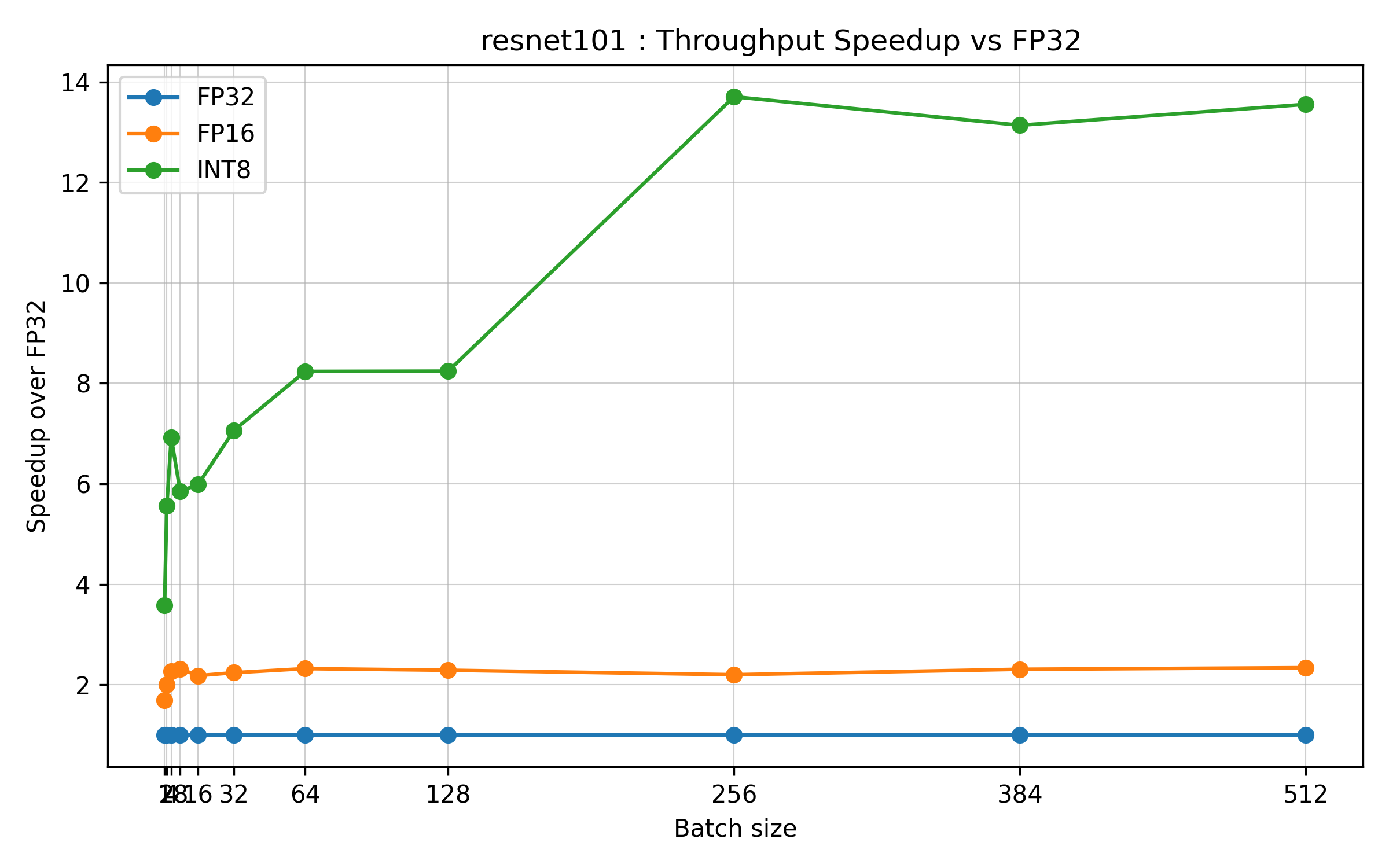}
        \caption{ResNet101}
        \label{fig:r101_t4_speedup}
    \end{subfigure}

    \vspace{12pt}


    \begin{subfigure}[b]{0.32\textwidth}
        \centering
        \includegraphics[width=\textwidth]{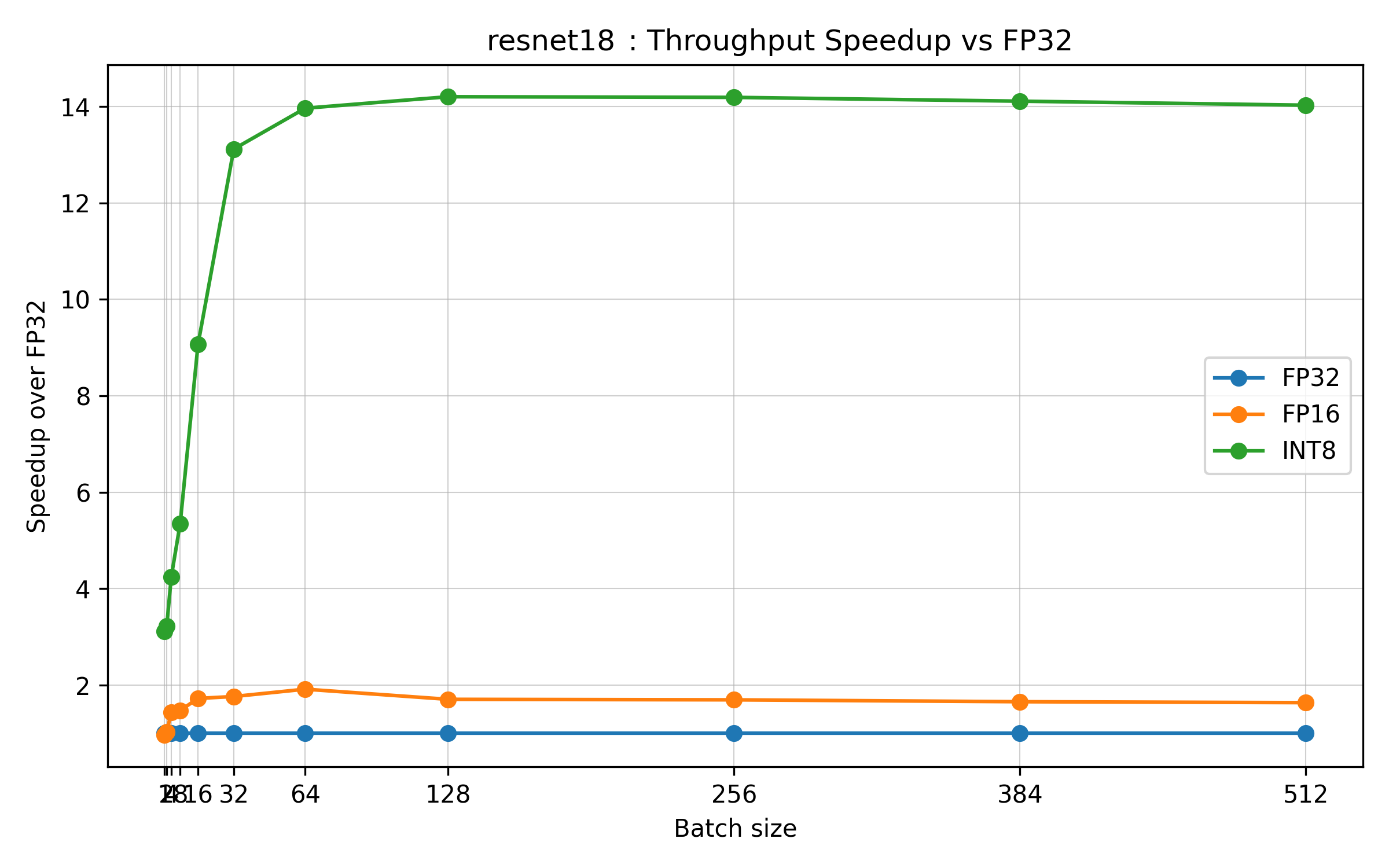}
        \caption{ResNet18}
        \label{fig:r18_l4_speedup}
    \end{subfigure}
    \hfill
    \begin{subfigure}[b]{0.32\textwidth}
        \centering
        \includegraphics[width=\textwidth]{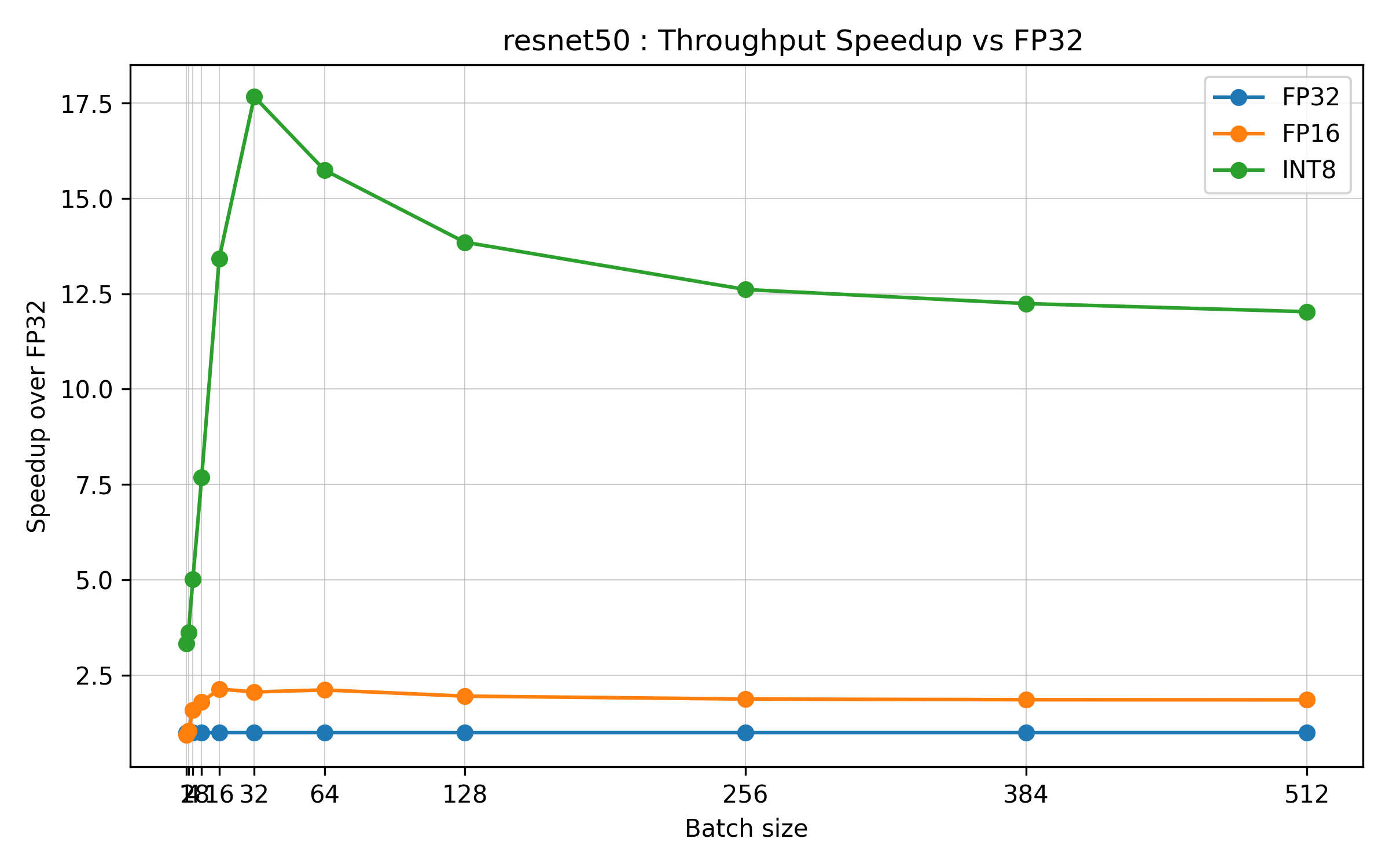}
        \caption{ResNet50}
        \label{fig:r50_l4_speedup}
    \end{subfigure}
    \hfill
    \begin{subfigure}[b]{0.32\textwidth}
        \centering
        \includegraphics[width=\textwidth]{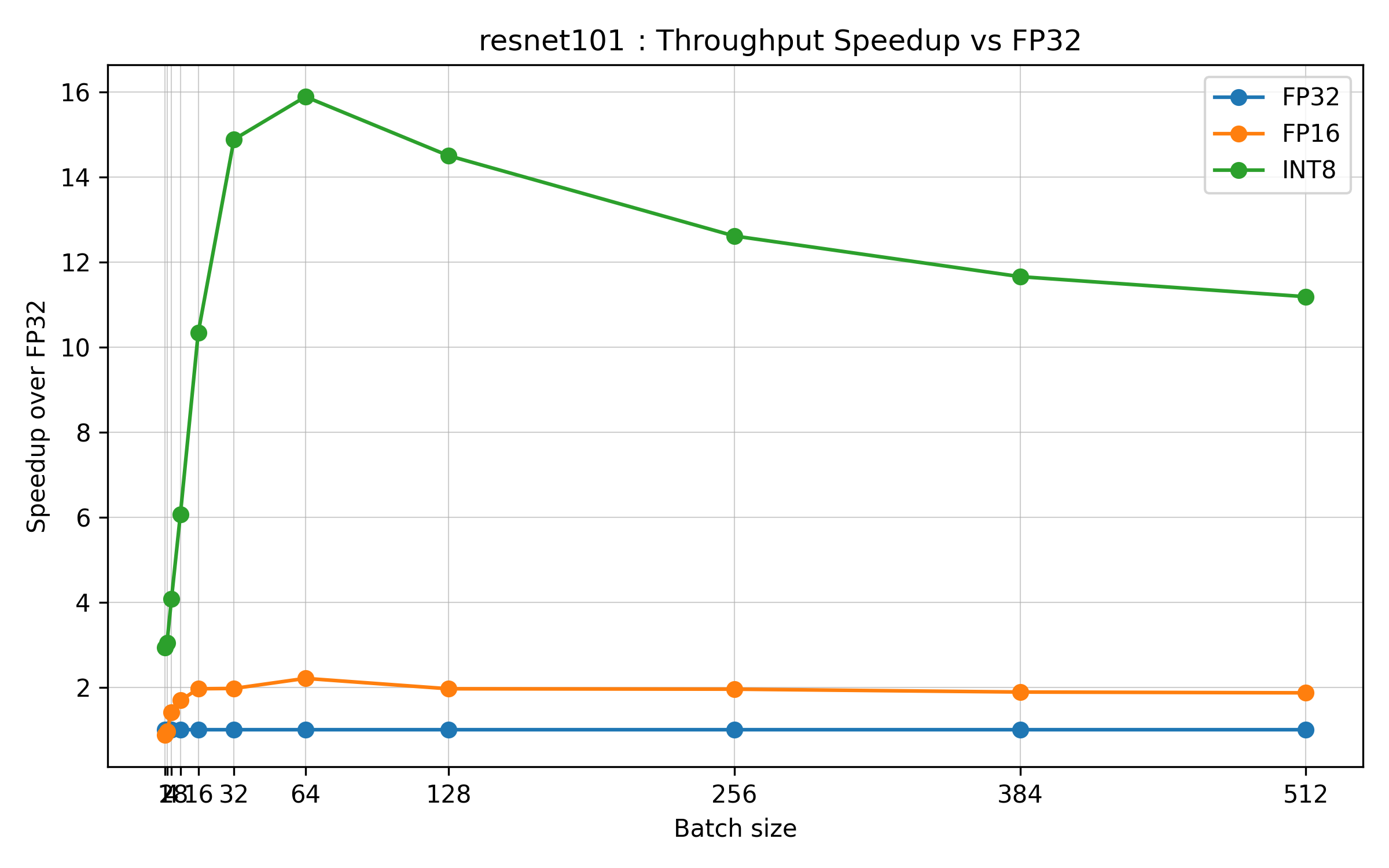}
        \caption{ResNet101}
        \label{fig:r101_l4_speedup}
    \end{subfigure}

    \caption{
    Throughput speedup relative to FP32 baseline across batch sizes.
    Top row shows NVIDIA T4 results, bottom row shows NVIDIA L4 results.
    FP16 provides moderate speedup over FP32, while INT8 achieves significantly
    higher speedup due to reduced numerical precision and improved Tensor Core utilization.
    Speedup increases with batch size until reaching a stable region where GPU compute
    resources are fully utilized. L4 achieves higher peak speedup compared to T4,
    reflecting architectural improvements in modern inference-optimized GPUs.
    }

    \label{fig:speedup_t4_vs_l4}

\end{figure*}

\subsection{GPU Memory Utilization (NVML)}

Figure~\ref{fig:nvml_memory_t4_vs_l4} shows GPU memory utilization measured
using NVIDIA Management Library (NVML) across batch sizes for both T4 and L4 GPUs.

Across all configurations, GPU memory usage is primarily determined by model size
and numerical precision, with relatively small increases as batch size grows.
This indicates that model weights and framework runtime allocations dominate total
memory usage, while activation memory contributes incrementally as batch size increases.

\textbf{T4 memory behavior.}
On NVIDIA T4, FP32 precision shows the highest memory allocation across all models.
ResNet18 requires approximately 4.9 GB of memory, ResNet50 approximately 7.6 GB,
and ResNet101 approximately 6.1 GB.

FP16 reduces memory requirements significantly.
ResNet18 requires approximately 3.1 GB, ResNet50 approximately 5.2 GB,
and ResNet101 approximately 5.2 GB.

INT8 TensorRT demonstrates the lowest memory footprint.
ResNet18 requires approximately 1.95 GB at small batch sizes, increasing slightly
to approximately 2.2 GB at $B=512$.
ResNet50 increases from approximately 2.65 GB to 2.9 GB across the same range,
while ResNet101 increases from approximately 2.65 GB to approximately 2.95 GB.

Memory growth across batch sizes is modest.
For example, ResNet50 INT8 memory increases by less than 300 MB when batch size
increases from 1 to 512, indicating that activation storage contributes only a
small fraction of total memory usage.

\textbf{L4 memory behavior.}
NVIDIA L4 shows slightly higher baseline memory allocation compared to T4.
FP32 memory usage reaches approximately 5.6 GB for ResNet18, approximately 8.0 GB
for ResNet50, and approximately 7.9 GB for ResNet101.

FP16 memory usage on L4 remains consistent with expected precision scaling.
ResNet18 requires approximately 3.0 GB, ResNet50 approximately 5.3 GB,
and ResNet101 approximately 5.3 GB.

INT8 TensorRT again demonstrates the lowest memory utilization.
ResNet18 requires approximately 2.75 GB at small batch sizes and increases slightly
to approximately 3.05 GB at $B=512$.
ResNet50 increases from approximately 3.45 GB to approximately 3.75 GB,
while ResNet101 increases from approximately 3.45 GB to approximately 3.8 GB.

Several important memory utilization trends emerge from the experimental results. Memory usage remains largely independent of batch size across the evaluated configurations, confirming that model parameters dominate total GPU memory consumption. This behavior indicates that the majority of allocated GPU memory is associated with static model weights and framework-level runtime buffers rather than batch-dependent activation scaling.

INT8 precision consistently reduces the overall memory footprint by approximately 2--3$\times$ compared to FP32, enabling larger batch sizes within fixed GPU memory constraints. Reduced numerical precision decreases the storage requirements for model parameters and intermediate activations, allowing more efficient utilization of available device memory while supporting higher inference throughput.

The NVIDIA L4 shows slightly higher baseline memory allocation (approximately 300--500 MB) compared to the NVIDIA T4. This difference is likely attributable to variations in CUDA runtime workspace allocation, kernel scheduling buffers, or driver-level optimization strategies that influence initial memory reservation behavior.

As expected, larger models require proportionally more GPU memory due to increased parameter counts and deeper network structures. For example, FP32 memory usage increases from approximately 4.9 GB for ResNet18 to approximately 8.0 GB for ResNet50 on L4, reflecting the additional storage required for convolutional weights and intermediate feature maps.

Overall, the results indicate that precision reduction improves both computational efficiency and memory utilization. Lower precision enables larger batch sizes without exceeding GPU memory limits, contributing directly to the throughput improvements observed in Section~\ref{sec:throughput_scaling}.

\subsection{Latency–Throughput Tradeoff (Pareto Analysis)}

Figure~\ref{fig:pareto_t4_vs_l4} illustrates the tradeoff between median latency
and throughput across batch sizes for ResNet18, ResNet50, and ResNet101 on both
T4 and L4 GPUs.

Each point represents a batch size configuration, showing how throughput improves
as latency increases due to larger batches enabling greater parallel execution.
Operating points located toward the upper-left region of each plot represent more
efficient configurations, achieving higher throughput with lower response time.

\textbf{T4 Pareto behavior.}
On NVIDIA T4, FP32 configurations show limited throughput improvement as latency
increases. For example, ResNet18 FP32 throughput increases from approximately
700 images/sec at low latency ($\sim$6 ms) to approximately 1289 images/sec at
high latency ($\sim$400 ms), indicating diminishing returns from increasing batch size.

FP16 improves the tradeoff by increasing throughput at moderate latency levels.
ResNet18 FP16 achieves approximately 2569 images/sec at approximately 200 ms latency,
providing roughly 2$\times$ higher throughput compared to FP32 at similar latency.

INT8 TensorRT provides substantially better tradeoff characteristics.
ResNet18 INT8 achieves approximately 8837 images/sec at approximately 60 ms latency,
demonstrating both significantly higher throughput and lower latency compared to FP32.

Similar trends are observed for larger models.
For ResNet50, INT8 reaches approximately 5066 images/sec at approximately 110 ms latency,
while FP32 achieves only approximately 382 images/sec at approximately 700 ms latency.
For ResNet101, INT8 achieves approximately 3125 images/sec at approximately 170 ms latency,
compared to approximately 226 images/sec at approximately 2400 ms latency for FP32.

\textbf{L4 Pareto behavior.}
NVIDIA L4 demonstrates substantial improvement across the entire Pareto frontier.
Across all models, L4 achieves higher throughput at significantly lower latency
compared to T4.

For ResNet18, L4 INT8 achieves approximately 38,932 images/sec at approximately
15 ms latency, representing more than a 4$\times$ throughput improvement over
T4 INT8 while simultaneously reducing latency by approximately 4$\times$.

ResNet50 L4 INT8 achieves approximately 13,388 images/sec at approximately
65 ms latency, compared to approximately 5066 images/sec at approximately 110 ms latency
on T4.

For ResNet101, L4 INT8 achieves approximately 8026 images/sec at approximately
105 ms latency, compared to approximately 3125 images/sec at approximately 170 ms latency
on T4.

Several important patterns emerge from the Pareto curves. INT8 precision consistently forms the Pareto-optimal frontier across all evaluated models, achieving both higher throughput and lower latency compared to FP16 and FP32 configurations. This indicates that reduced numerical precision improves computational efficiency while maintaining favorable latency characteristics, allowing more inference operations to be processed per unit time.

FP16 provides intermediate performance improvements, reducing latency by approximately 1.5--2$\times$ compared to FP32 while increasing throughput by approximately 2--4$\times$. These improvements reflect the benefits of reduced precision arithmetic while maintaining higher numerical resolution than INT8.

Across all workloads, the NVIDIA L4 shifts the Pareto frontier upward and to the left relative to the NVIDIA T4, indicating simultaneous improvements in both throughput and latency. This shift demonstrates the architectural advantages of the L4, including improved Tensor Core performance, memory bandwidth efficiency, and enhanced parallel execution capability.

As expected, larger batch sizes improve throughput but increase latency, demonstrating the inherent tradeoff between response time and processing efficiency in GPU inference systems. However, the L4 achieves near-peak throughput at relatively small batch sizes ($B=16$--$32$), enabling high performance without requiring large latency increases. This characteristic is particularly beneficial for real-time inference environments where strict latency constraints limit the use of very large batch sizes.

These Pareto curves illustrate that optimal batch size selection depends on workload requirements. Latency-sensitive applications may operate at smaller batch sizes to minimize response time, while throughput-oriented batch processing systems may benefit from larger batch sizes that maximize computational efficiency. Overall, the results show that modern GPU architectures significantly improve the latency--throughput tradeoff, enabling higher inference performance without proportional increases in response time.

\subsection{Speedup Relative to FP32 Baseline}

Figure~\ref{fig:speedup_t4_vs_l4} shows throughput speedup relative to FP32
across batch sizes for ResNet18, ResNet50, and ResNet101 on both T4 and L4 GPUs.

Across all models, reduced precision consistently improves performance relative
to FP32. FP16 provides moderate speedup due to reduced numerical precision and
efficient Tensor Core utilization, while INT8 demonstrates substantially larger
performance gains by enabling increased parallel execution and reduced memory
traffic.

\textbf{T4 speedup behavior.}
On NVIDIA T4, FP16 provides approximately 1.8$\times$–2.2$\times$ speedup
across most batch sizes.
For example, ResNet18 FP16 achieves approximately 2.0$\times$ speedup at large
batch sizes ($B \geq 128$), while ResNet50 and ResNet101 show similar gains
near 2.1$\times$–2.3$\times$.

INT8 demonstrates significantly larger improvements.
For ResNet18, INT8 speedup increases rapidly from approximately 2.5$\times$
at small batch sizes to approximately 7$\times$ at larger batch sizes.
ResNet50 achieves peak speedup near 13$\times$ relative to FP32,
while ResNet101 achieves approximately 13–14$\times$ speedup at moderate
batch sizes ($B=256$–512).

Speedup increases with batch size until the GPU reaches effective utilization.
Beyond moderate batch sizes, speedup stabilizes as hardware resources become
fully saturated.

\textbf{L4 speedup behavior.}
NVIDIA L4 demonstrates higher speedup across most configurations compared to T4,
reflecting improvements in Tensor Core throughput, memory bandwidth,
and architectural parallelism.

FP16 on L4 provides consistent speedup of approximately 1.6$\times$–2.1$\times$
across models, similar to T4 but achieved at smaller batch sizes due to improved
hardware efficiency.

INT8 provides the largest gains.
For ResNet18, INT8 speedup reaches approximately 14$\times$ relative to FP32,
remaining stable across larger batch sizes.
ResNet50 achieves peak speedup near 17–18$\times$ at moderate batch sizes
($B=32$–64), while ResNet101 achieves approximately 15–16$\times$ speedup.

Compared to T4, L4 achieves similar or higher peak speedup at significantly
smaller batch sizes, indicating improved architectural efficiency and better
utilization of Tensor Core resources.

Several important speedup trends emerge when comparing reduced precision modes against the FP32 baseline. FP16 consistently provides approximately 2$\times$ speedup across all evaluated models, demonstrating the efficiency of half-precision computation on Tensor Core architectures. Reduced precision enables higher arithmetic throughput and more efficient utilization of GPU compute resources while maintaining numerical stability suitable for inference workloads.

INT8 provides the largest performance improvement, reaching up to 14$\times$ speedup on the NVIDIA T4 and up to 18$\times$ speedup on the NVIDIA L4. These results highlight the substantial computational advantages enabled by lower numerical precision, where increased arithmetic density allows more operations to be executed per cycle.

Speedup increases rapidly between batch sizes 1--32 as GPU utilization improves and parallel execution resources become more effectively utilized. Beyond this range, speedup stabilizes as compute resources approach saturation, indicating that additional batching provides diminishing relative performance gains.

Across most configurations, the NVIDIA L4 achieves comparable or higher speedup at smaller batch sizes compared to the NVIDIA T4, indicating improved compute efficiency and reduced dependence on aggressive batching strategies. Larger models benefit significantly from reduced precision because increased computational complexity amplifies the advantages of Tensor Core acceleration \cite{micikevicius2018mixedprecision}.

Overall, the results demonstrate that reduced precision substantially improves inference efficiency relative to FP32 baselines. Modern architectures such as L4 further amplify these benefits by achieving higher speedup at smaller batch sizes, improving both throughput scalability and latency--throughput tradeoff characteristics.

\section{Conclusion}
\label{sec:conclusion}

This paper presented DEEP-GAP, a systematic evaluation of GPU inference performance
across modern datacenter-oriented architectures using NVIDIA T4 and NVIDIA L4 GPUs.
By extending the GDEV-AI benchmarking methodology to GPU environments, we analyzed
throughput, latency, tail latency, memory utilization, and precision-based scaling
across multiple ResNet models.

Our results demonstrate that reduced numerical precision plays a critical role in
maximizing inference efficiency. FP16 consistently improves throughput over FP32,
while INT8 TensorRT achieves the highest performance gains due to reduced data width
and increased Tensor Core utilization. Across all evaluated models, INT8 provides
the most favorable latency-throughput tradeoff, enabling significantly higher
processing capacity within similar latency constraints.

Comparative analysis between T4 and L4 reveals substantial architectural improvements
in newer GPU generations. L4 achieves significantly higher throughput and improved
scaling efficiency across batch sizes, while maintaining stable latency behavior.
Peak throughput improvements of up to 58$\times$ relative to CPU execution highlight
the impact of specialized hardware acceleration for deep learning inference workloads.

Our findings show that optimal batch sizes typically occur within moderate ranges,
where GPU compute resources are fully utilized without incurring excessive memory
bandwidth pressure or runtime overhead. Beyond this region, throughput stabilizes
or slightly declines, indicating a shift from compute-bound to memory-bound execution.

Overall, the results provide practical guidance for selecting precision modes and
batch sizes when deploying inference workloads in production datacenter environments.
The methodology and findings presented in DEEP-GAP contribute to a clearer understanding
of how modern GPU architectures translate theoretical compute capability into
real-world inference performance.

\section{Future Work}

While this study evaluates single-tenant inference performance under controlled
conditions, future work will extend DEEP-GAP to multi-tenant environments where
multiple workloads share the same GPU infrastructure.

In real-world datacenter deployments, GPUs are commonly shared across heterogeneous
inference services with differing latency and throughput requirements. Future work
will investigate scheduling strategies, resource sharing mechanisms, and
quality-of-service (QoS) considerations that influence performance predictability
under concurrent workloads.

This direction extends DEEP-GAP toward understanding how modern GPU architectures
behave under realistic shared deployment conditions, providing insights for efficient
resource allocation in large-scale AI systems.

\end{document}